\documentclass[hyper,letterpaper,notoc]{JHEP3}
\input{epsf}
\usepackage{amsfonts}
\usepackage{amsbsy}
\usepackage{amsmath}
\usepackage{amssymb}
\usepackage{epsfig}
\usepackage{cite}
\usepackage{epstopdf}
\usepackage{url}

\newdimen\tableauside\tableauside=1.0ex
\newdimen\tableaurule\tableaurule=0.4pt
\newdimen\tableaustep
\def\phantomhrule#1{\hbox{\vbox to0pt{\hrule height\tableaurule width#1\vss}}}
\def\phantomvrule#1{\vbox{\hbox to0pt{\vrule width\tableaurule height#1\hss}}}
\def\sqr{\vbox{
  \phantomhrule\tableaustep
  \hbox{\phantomvrule\tableaustep\kern\tableaustep\phantomvrule\tableaustep}%
  \hbox{\vbox{\phantomhrule\tableauside}\kern-\tableaurule}}}
\def\squares#1{\hbox{\count0=#1\noindent\loop\sqr
  \advance\count0 by-1 \ifnum\count0>0\repeat}}
\def\tableau#1{\vcenter{\offinterlineskip
  \tableaustep=\tableauside\advance\tableaustep by-\tableaurule
  \kern\normallineskip\hbox
    {\kern\normallineskip\vbox
      {\gettableau#1 0 }%
     \kern\normallineskip\kern\tableaurule}
  \kern\normallineskip\kern\tableaurule}}
\def\gettableau#1 {\ifnum#1=0\let\next=\null\else
  \squares{#1}\let\next=\gettableau\fi\next}

\title{Refined Black Hole Ensembles and Topological Strings}

\author{Mina Aganagic$^{1,2}$ and Kevin Schaeffer$^1$\\
\\
${}^1$ Center for Theoretical Physics, 
University of California, Berkeley, CA 94720, USA
\\
${}^2$ Department of Mathematics,
University of California, Berkeley, CA 94720, USA}

\abstract{We formulate a refined version of the Ooguri-Strominger-Vafa (OSV) conjecture. The OSV conjecture that $Z_{BH} = |Z_{top}|^2$ relates the BPS black hole partition function to the topological string partition function $Z_{top}$. In the refined conjecture, $Z_{BH}$ is the partition function of BPS black holes counted with spin, or more precisely the protected spin character. $Z_{top}$ becomes the partition function of the refined topological string, which is itself an index.
Both the original and the refined conjecture are examples of large $N$ duality in the 't Hooft sense. 
The refined conjecture applies to non-compact Calabi-Yau manifolds only, so the black holes are really BPS particles with large entropy, of order $N^2$. The refined OSV conjecture states that the refined BPS partition function has a large $N$ dual which is captured by the refined topological string. 
We provide evidence that the conjecture holds by studying local Calabi-Yau threefolds consisting of line bundles over a genus $g$ Riemann surface. 
We show that the refined topological string partition function on these geometries is computed by a two-dimensional TQFT. We also study the refined black hole partition function arising from $N$ D4 branes on the Calabi-Yau, and argue that it reduces to a $(q,t)$-deformed version of two-dimensional  $SU(N)$ Yang-Mills.  Finally, we show that in the large $N$ limit this theory factorizes to the square of the refined topological string in accordance with the refined OSV conjecture. }

\begin{document}


\section{Introduction}
The Ooguri-Strominger-Vafa (OSV) conjecture gives a beautiful relation between the partition function of four-dimensional BPS black holes in a type IIA string theory compactified on a Calabi-Yau and the topological A-model string partition function\cite{Ooguri:2004zv}.  Consider the BPS black hole partition function, \(Z_{BH}\), in a mixed ensemble given by fixing the magnetic charge, \(p_{\Lambda}\), and summing over the electric charge with electric potential, \(\phi_{\Lambda}\). 
The OSV conjecture relates the exact microscopic entropy of the black hole, captured by $Z_{BH}$ to the macroscopic entropy, computed in terms of the topological string 
\begin{equation}
Z_{\textrm{BH}}(p^{\Lambda},\phi^{\Lambda}) \sim \vert Z_{\textrm{top}}(X^{\Lambda})\vert ^{2} \label{eqn:osvoriginal}
\end{equation}
where \(X^{\Lambda} = p^{\Lambda} + \frac{i}{\pi}\phi^{\Lambda}\).  This reproduces the Bekenstein-Hawking entropy/area law to the leading order, but otherwise it computes quantum gravitational corrections to it. The entropy on the left is a supersymmetric index.  The topological string on the right computes F-terms in the four-dimensional low energy effective action of the IIA string theory on the Calabi-Yau.  But from Wald's formula these F-terms are precisely the corrections to the Bekenstein-Hawking black hole entropy. The OSV relation is an example of gauge/gravity duality, where the gauge theory is the theory on the D-branes comprising the black hole. This aspect of the correspondence was emphasized in  \cite{Ooguri:2004zv, Vafa:2004qa, Aganagic:2004js ,Ooguri:2005vr, Dijkgraaf:2005bp, Beasley:2006us, Sen:2008yk, Sen:2008vm, Dabholkar:2010uh, Sen:2011ba, Dabholkar:2011ec}.\footnote{The large $N$ dual of a black hole with fixed both magnetic and electric charges is naturally the real version of the topological string, recently studied  in
\cite{Vafa:2012fi}. Its partition function has both the holomorphic $Z_{top}$ and the anti-holomorphic piece ${\bar Z}_{top}$.} Note that  both sides of (\ref{eqn:osvoriginal}) depend only on the K\"ahler moduli and not the complex structure moduli - on the right, this is a well-known property of A-model topological strings, and on the left this is a consequence of computing a well-behaved BPS index.

It is natural to ask if there are meaningful ways to generalize the conjecture (\ref{eqn:osvoriginal}). The most general possible black hole partition function would include a chemical potential for angular momentum.  This partition function, also known as a spin character, has been extensively studied in the context of motivic wall crossing in four-dimensional \(\mathcal{N}=2\) field theory \cite{KSoriginal, Dimofte:2009tm, Dimofte:2009bv, Cecotti:2009uf, Gaiotto:2010be} and supergravity \cite{Andriyash:2010qv, Andriyash:2010yf}.\footnote{Rotating single-centered black holes in four dimensions cannot be supersymmetric.  The black holes that one is studying here are multicentered configurations that carry intrinsic angular momentum in their electromagnetic fields. Note that despite the fact that these are multi-centered, they still correspond to bound states \cite{Denef:2000nb}.} For compact Calabi-Yau manifolds, this spin character will depend sensitively on both the K\"ahler and complex structure of the manifold.  On a {\it noncompact} Calabi-Yau, however, the situation improves. We can form a {\it protected spin character} by utilizing the \(SU(2)_{R}\)-symmetry of four-dimensional \(\mathcal{N}=2\) theory \cite{Gaiotto:2010be}.  The protected spin character is a genuine index that only depends on the K\"ahler moduli and is constant except at real codimension-one walls of marginal stability.  Therefore, this protected spin character gives a well-behaved and computable definition for the refined BPS black hole partition function $Z_{\textrm{ref BH}}(p^{\Lambda},\phi^{\Lambda},y)$, depending on one extra parameter $y$ to keep track of the spin.

There is also a natural candidate for what may replace the topological string partition function. The refined topological string is a one-parameter deformation of topological string theory. Just like the refined black-hole partition function, the refined topological string partition function $Z_{\textrm{ref top}}(X^I,  y)$ also makes sense only in the non-compact limit, and also utilizes the $SU(2)_R$ symmetry to be defined. The refined topological string is defined as an index in M-theory \cite{Hollowood:2003cv, Iqbal:2007ii}. There are several equivalent ways to compute the index, either by counting  spinning M2-branes \cite{Iqbal:2007ii}, or alternatively, as the refined BPS index for D2 and D0-branes bound to a single D6 brane\cite{Iqbal:2003ds , Dijkgraaf:2006um, Dimofte:2009bv}, to name two. Yet another way to compute the index is as the Nekrasov partition function of the five dimensional gauge theory that arises from M-theory at low energies.\footnote{Alternatively, it has a B-model formulation, at least in some cases, in terms of \(\beta\)-deformed matrix models \cite{Dijkgraaf:2009pc, Cheng:2010yw,Aganagic:2011mi}.} 

Having found a generalization of both $Z_{BH}$ and $Z_{top}$, defined in the same circumstances, and depending on the same set of parameters, it is natural to
conjecture that there is in fact a refinement of the OSV conjecture that relates the two:

\begin{equation}
Z_{\textrm{ref BH}}(p^{\Lambda},\phi^{\Lambda},y) = \vert Z_{\textrm{ref\;top}}(k^{\Lambda},\epsilon_{1},\epsilon_{2})\vert^{2}.
\end{equation}
Despite the fact that the ingredients for the conjecture fit naturally, one would still like to have a {rationale} for why the conjecture should hold.
Because the Calabi-Yau is non-compact, the BPS black holes we are discussing are really BPS particles with large entropy. While one can imagine the system arising by taking a limit where we take the mass of the black-hole to infinity, at the same rate as we take the Planck mass to infinity so that entropy stays finite, the $SU(2)_R$ symmetry we are using makes sense only in the non-compact limit. Correspondingly some of the justifications for the OSV conjecture, for example those based on Wald's formula, may no longer be sound in the refined setting, since the supergravity solution is singular.

However, as we mentioned, the original OSV relation is fundamentally a large $N$ duality.  For the BPS states that came from a theory on $N$ D-branes, we always get an $SU(N)$ gauge theory describing the particles. In the 't Hooft large $N$ limit of this theory one expects to get a string theory, whether or not there are dynamical gravitons in the theory. Indeed, many famous examples of large $N$ duality are of this kind, see for example \cite{Gopakumar:1998ki}.
In particular, the duality should still hold even for the non-compact Calabi-Yau; it is merely difficult to check beyond the protected quantities. 
Finally, from the perspective of large $N$ duality, once we modify one side of the correspondence, the duality, at least in principle, { fixes} what the other side { has to be}. Furthermore, there is a way to understand directly from the large $N$ duality ${\it why}$ refining the black hole ensemble has to correspond to a refinement of the topological string on the other side.

To precisely test the refined OSV conjecture, one would like to compute both sides of the relation and compare explicitly.  For the unrefined OSV conjecture, this was done for local, non-compact Calabi-Yau manifolds in \cite{Vafa:2004qa, Aganagic:2004js, Aganagic:2005wn}, and agreement was found.  These local Calabi-Yaus can be thought of as the limit of compact Calabi-Yaus in the neighborhood of a shrinking two- (or four-) cycle.  In this limit, gravity decouples so we need to be precise about what we mean by the black hole partition function.  We must require that the D-branes forming the BPS black hole wrap a cycle that becomes noncompact in this limit so that their entropy remains nonzero.  In this limit, the black hole partition function is simply computed by a Witten index on the noncompact brane worldvolume. In the paper, we will run a similar test in the refined case. 
To this end, we will develop techniques to solve for the refined partition functions on both sides of the conjecture.  Remarkably, the refined OSV conjecture passes the tests just as well as the original OSV.\footnote{In subsequent work, \cite{Vafa:2004qa, Dijkgraaf:2005bp, Dabholkar:2005by, Dabholkar:2005dt, Shih:2005he,  Aganagic:2006je, Denef:2007vg } several aspects of the conjecture were clarified. For one, while the relation holds to all orders in perturbation theory, the exact relation requires summing over nonperturbative corrections to the macroscopic entropy, taking the form of ``baby universes'' \cite{Dijkgraaf:2005bp, Aganagic:2006je}.  Second, at the perturbative level the right side of equation \ref{eqn:osvoriginal} should include a summation over \(\phi \to \phi + 2\pi i n\) so that the periodicities of \(\phi\) match on both sides.  Additionally, the right side generically contains an additional measure factor of \(g_{top}^{-2}e^{-K}\), which is natural from the viewpoint of K\"ahler quantization since \(Z_{top}\) transforms as a wavefunction \cite{Verlinde:2004ck, Ooguri:2005vr, Denef:2007vg}.  Such a measure factor did not appear in the unrefined local curve examples of \cite{Vafa:2004qa, Aganagic:2004js}.  One way to understand this is to observe that these geometries do not have a holomorphic anomaly, which implies that \(Z_{top}\) actually transforms as a function rather than a wavefunction -- therefore the additional measure factors from K\"ahler quantization are absent.  In this paper when we study the refined OSV relation on these local geometries, we will again find that no measure factors appear.}  This provides strong evidence in support of our conjecture.

The paper is organized as follows. In section 2, we review the ingredients of the original OSV conjecture for both compact and non-compact Calabi-Yaus.
In section 3, we motivate the refined OSV conjecture first from the perspective of the \(AdS_{2}/CFT_{1}\) correspondence, and second by studying the wall crossing of \(D4\) branes splitting into \(D6\) -\(\overline{D6}\) bound states.  In the noncompact setting these arguments are necessarily heuristic, but we believe they capture the correct physics. 
In section 4, we explain how the refined topological string on Calabi-Yau manifolds of the form \(X =\mathcal{L}_{1}\oplus \mathcal{L}_{2} \to \Sigma_{g}\) can be completely solved by a two-dimensional topological quantum field theory (TQFT).  We then use this TQFT to compute refined topological string amplitudes for the above geometries.  We also show that our results agree precisely with the five-dimensional Nekrasov partition function of \(U(1)\) gauge theories with \(g\) adjoints that are engineered by these Calabi-Yaus.  
In section 5, we study the black hole side of the correspondence.  Mathematically, the refined partition function for D4/D2/D0-branes computes the \(\chi_{y}\) genus of the relevant instanton moduli spaces.  As originally suggested in \cite{Witten:2011zz}, this can be thought of as a categorification of the euler characteristic invariants computed by the \(\mathcal{N}=4\) Vafa-Witten theory.  We then specialize to \(D4\) branes wrapping the geometry \(C_{4} = (\mathcal{L}_{1} \to \Sigma_{g})\) and study the refined BPS partition function of bound states with \(D2\) and \(D0\)-branes.  We propose that this partition function is computed by a \((q,t)\)-deformation of two-dimensional Yang-Mills, which is closely related to the refined Chern-Simons theory studied in \cite{Aganagic:2011sg, Gadde:2011uv, Aganagic:2012au}.  Wrapping branes on the geometry \(C = (\mathcal{O}(-1) \to \mathbb{P}^{1})\), we show that \((q,t)\)-deformed Yang Mills precisely reproduces a mathematical result of Yoshioka and Nakajima for the \(\chi_{y}\) genus of instanton moduli space \cite{Yoshioka:1996pd, Nakajima:2003uh}.  
In section 6 we connect the black hole and topological string perspectives by studying the large N limit of the \((q,t)\)-deformed Yang Mills theory.  We find that the theory factorizes to all orders in \(1/N\) into two copies of the refined topological string partition function.  This gives a nontrivial check of our refined OSV conjecture.
Finally, in section 7 we explain an alternative way to compute the refined black hole partition function on \(\mathcal{O}(-1) \to \mathbb{P}^{1}\).  The refined bound states are counted by using the semi-primitive refined wall-crossing formula \cite{Dimofte:2009bv}, thus giving a refined extension of the techniques used in \cite{Nishinaka:2010qk, Nishinaka:2010fh}.

\section{The OSV Conjecture: Unrefined and Refined \label{sec:osv}}
We start by reviewing the remarkable conjecture of Ooguri, Strominger, and Vafa (OSV) connecting four-dimensional BPS black holes with topological strings \cite{Ooguri:2004zv}.  Consider IIA string theory compactified on a Calabi-Yau, \(X\), with four-dimensional black holes arising from D-branes wrapping holomorphic cycles in \(X\).  In terms of four-dimensional gauge fields, the \(D0\) and \(D2\) branes are electrically charged while the \(D4\) and \(D6\) branes are magnetically charged.

The object of interest for the OSV conjecture is the mixed black hole partition function given by fixing the magnetic charges and summing over electric charges with chemical potentials,
\begin{equation}
Z_{BH}(P_{6},P_{4};\phi_{2},\phi_{0}) = \sum_{Q_{2},Q_{0}}\Omega(P_{6},P_{4},Q_{2},Q_{0})e^{-\phi_{2}Q_{2}-\phi_{0}Q_{0}}
\end{equation}
where we have denoted \(D6\) charges by \(P_{6}\), \(D4\) charges by \(P_{4}\), \(D2\) charges by \(Q_{2}\), and \(D0\) charges by \(Q_{0}\).  Here \(\phi_{2}\) and \(\phi_{0}\) are chemical potentials associated to the electrically charged D-branes.  \(\Omega(P,Q)\) is computed by the Witten index in the corresponding charge sector,
\begin{equation}
\Omega(P_{6},P_{4},Q_{2},Q_{0}) = \textrm{Tr}_{\mathcal{H}_{P,Q}}(-1)^{F}
\end{equation}
and only receives contributions from BPS black holes.

The OSV conjecture states that this mixed black hole partition function is equal to the square of the A-model topological string partition function on \(X\),
\begin{equation}
Z_{BH}(P_{6},P_{4};\phi_{2},\phi_{0}) = \vert Z_{top}(g_{top},k)\vert^{2}
\end{equation}
where \(k\) is the complexified K\"ahler form on \(X\).  The projective coordinates on moduli space are given by,
\begin{equation}
X_{I} = P_{I} + i \frac{\phi_{I}}{\pi}
\end{equation}
This implies that the string coupling constant and K\"ahler moduli are determined by the magnetic charges and electric potentials,
\begin{eqnarray}
g_{top} & = & \frac{4\pi i}{X_{0}} = \frac{4\pi i}{P_{6}+i\frac{\phi_{0}}{\pi}} \\
k_{I} & = & 2\pi i \frac{X_{I}}{X_{0}} = \frac{1}{2}g_{top}\Big(P_{4,I}+\frac{i\phi_{2,I}}{\pi}\Big)
\end{eqnarray}
Here the real part of \(X_{I}\) is fixed by the attractor mechanism which determines the near-horizon Calabi-Yau moduli in terms of the black hole charge.  Both sides of the relation should be considered as expansions in \(1/Q\) where \(Q\) is the total graviphoton charge of the black hole.  From the change of variables, we have a perturbative expansion in \(g_{top}\) if either \(P_{6}\) or \(\phi_{0}\) is large.  We will usually set \(P_{6}=0\) so it is natural to take both \(\phi_{0}\) and \(P_{4}\) to infinity is such a way that \(g_{top}\) becomes small and the K\"ahler form remains constant.

One way to further understand the OSV relation is by inverting it,
\begin{equation}
\Omega(P_{6},P_{4},Q_{2},Q_{0}) = \int d\phi_{0}d\phi_{2} e^{Q_{2}\phi_{2}+Q_{0}\phi_{0}} \vert Z_{top}\vert^{2}
\end{equation}
so that black hole degeneracies are formally computed by the topological string.  It is known that because the topological string partition function obeys the holomorphic anomaly equations \cite{Bershadsky:1993cx}, it transforms as a wavefunction under changes of polarization on the Calabi-Yau moduli space \cite{Witten:1993ed}.  Thus, from quantum mechanics the appearance of \(\vert Z_{top} \vert^{2}\) is very natural.  In fact, this interpretation implies that \(\Omega(P,Q)\) is the Wigner quasi-probability function on \((P,Q)\) phase space \cite{Ooguri:2004zv}.

The original OSV conjecture focused on the case of compact Calabi-Yau manifolds, so that the wrapped D-branes correspond to black holes in four-dimensional \(\mathcal{N}=2\) supergravity.  However, as explored in \cite{Vafa:2004qa, Aganagic:2004js, Aganagic:2005dh, Dijkgraaf:2005bp, Aganagic:2005wn, Aganagic:2006je}, it is interesting to study the OSV conjecture for local Calabi-Yau manifolds which can be thought of as the decompactification limit of the compact case.  In this limit gravity decouples which means that the four dimensional planck mass goes to infinity.  Since the Bekenstein-Hawking entropy of a black hole is proportional to \(M_{BH}^{2}/M_{Pl}^{2}\), to obtain a finite entropy in this limit we should also take \(M_{BH}\to\infty\).  This can be accomplished simply by wrapping \(D4\) or \(D6\)-branes on cycles that become non-compact in this limit.  The precise OSV relation remains the same in this limit, except that now the black hole partition function is naturally computed by a partition function on the worldvolume of the noncompact \(D4\) or \(D6\)-branes.

The advantage of taking this limit is that both sides of the OSV relation are exactly solvable, leading to a highly nontrivial check of the conjecture.  The conjecture has been tested perturbatively to all orders in \cite{Vafa:2004qa, Aganagic:2004js, Aganagic:2005wn}, and non-perturbative corrections in the form of baby universes have been computed in \cite{Dijkgraaf:2005bp, Aganagic:2006je}.

\subsection{Refining the Conjecture}
Now that we have reviewed the OSV conjecture, a natural question to ask is whether the black hole degeneracy computed by \(\Omega(P,Q)\) is the most general index that counts four-dimensional BPS black holes.  In fact, we could include information about spin by replacing the Witten index, 
\begin{equation}
\textrm{Tr}_{\mathcal{H}_{BPS}}(-1)^{F}
\end{equation}
by the spin character,
\begin{equation}
\textrm{Tr}_{\mathcal{H}_{BPS}}(-1)^{F}\exp(-2\gamma J_{3})
\end{equation}
where \(J_{3}\) is the three-dimensional generator of rotations and \(\gamma\) is the conjugate chemical potential.\footnote{Rotating single-centered black holes in four dimensions cannot be supersymmetric.  The black holes that we are studying here are multicentered configurations that carry intrinsic angular momentum in their electromagnetic fields. Note that despite the fact that these are multi-centered, they still typically correspond to bound states \cite{Denef:2000nb}.}  These spin-dependent BPS traces and their wall-crossing behavior have been studied extensively in the context of \(\mathcal{N}=2\) field theory \cite{KSoriginal, Dimofte:2009bv, Dimofte:2009tm, Cecotti:2009uf, Gaiotto:2010be} and supergravity \cite{Andriyash:2010qv, Andriyash:2010yf}.  However, this trace has the drawback of not being an index, which means that it will be sensitive to both the complex and K\"ahler moduli.  

If we consider a local Calabi-Yau manifold by taking the gravity decoupling limit, there is a preserved \(SU(2)\) R-symmetry that appears.  As explained in \cite{Gaiotto:2010be}, this can be used to form a protected spin character that is a genuine index, and only depends on the K\"ahler moduli through wall-crossing,

\begin{equation}
\textrm{Tr}_{\mathcal{H}_{P,Q}}(-1)^{2J_{3}}e^{-2\gamma(J_{3}-R)} = \sum_{J_{3},R}\Omega(P,Q;J_{3},R)e^{-2\gamma(J_{3}-R)}
\end{equation}
Now we can form the mixed ensemble of black holes counted with spin where, as in the ordinary case, we fix the magnetic charge and sum over the electric charge,

\begin{equation}
Z_{\textrm{ref BH}}(P_{6},P_{4};\phi_{2},\phi_{0};\gamma) = \sum_{Q_{2},Q_{0},J_{3},R}\Omega(P_{6},P_{4},Q_{2},Q_{0};J_{3},R)e^{-2\gamma (J_{3}-R)-\phi_{2}Q_{2}-\phi_{0}Q_{0}}
\end{equation}
We will refer to this as the refined black hole partition function.

We would like to know whether there exists a generalization of the topological string whose square is equal to \(Z_{\textrm{ref BH}}\).  A natural candidate for this one-parameter deformation is well-known, and is given by the refined topological string. 

Recall the definition of the refined topological string as the index of M-theory, depending on Kahler moduli $k$ and two additional parameters $\epsilon_1$ and ${\epsilon_2}$. The refined topological string partition function \cite{Hollowood:2003cv, Iqbal:2007ii, Aganagic:2011mi}  on a Calabi-Yau $X$ is given by computing the index of M-theory on the geometry,
\begin{equation}
(X\times TN \times S^{1})_{\epsilon_1,\epsilon_2}
\end{equation}
where \(TN\) denotes the Taub-NUT spacetime, and upon going around the \(S^{1}\) the Taub-NUT is twisted by,
\begin{eqnarray}
(z_{1}\, , \, z_2 )  \qquad \to \qquad (q\, z_{1} , \,t^{-1}\,z_{2} )
\end{eqnarray}
where
$$ q=e^{-\epsilon_{1}}\qquad  t=e^{-\epsilon_{2}}.
$$
In addition, we must include an R-symmetry twist to preserve supersymmetry.  This twist is implemented by a geometric Killing vector on the non-compact Calabi-Yau, \(X\).  Note that, in our notation, the unrefined limit is $\epsilon_1=\epsilon_2$, unlike in much of the literature, where one typically defines $\epsilon_2$ with a different sign. The partition function of M-theory in this geometry is computing the index of the resulting theory on $TN\times S^1$,

\begin{equation}
Z_{ref\; top}(\epsilon_1,\epsilon_2;k) = \textrm{Tr}(-1)^{2S_{1}+2{S_2}}q^{S_{1}-{R}}t^{{R}-S_{2}}e^{- k_I Q^I_2} \label{eqn:refindex}
\end{equation}
where \(S_{1}\) and \(S_{2}\) are the spins in the \(z_{1}\) and \(z_{2}\) directions, respectively, and \({R}\) is the R-charge of the state.  We have schematically indicated that the partition function depends on the Kahler moduli $k$, via the M2 brane contributions to the index. Note that although the trace is over all states, only BPS states will make a contribution. The index can be computed in several different ways: by counting spinning M2-branes \cite{Hollowood:2003cv, Iqbal:2007ii} or as the Omega-deformed instanton partition function \cite{Nekrasov:2002qd}. In analogy with the unrefined case \cite{Iqbal:2003ds}, the refined topological string can also be written in terms of the refined Donaldson-Thomas invariants that compute the BPS protected spin character for D2 and D0-branes bound to a single D6 brane\cite{Dimofte:2009bv}. Given that the refined topological string also counts BPS particles with spin and only depends on the K\"ahler moduli of \(X\), it is should be related to the refined black hole partition function. 

The index is related to the ordinary topological string partition function is we set $\epsilon_1=\epsilon_2 = g_s$, and reduce on the thermal $S^1$ to IIA. Then, we get IIA string theory on $X\times TN$, whose partition function is the same as the ordinary topological string partition function, where $g_s$ is the topological string coupling constant  \cite{Dijkgraaf:2006um}. In particular, M2 branes wrapping holomorphic curves in $X$ and the thermal $S^1$  become the worldsheet instantons of the topological string.  For $\epsilon_1\neq \epsilon_2$, the theory has no known worldsheet formulation, at the moment.

There is yet another way to view the partition function (\ref{eqn:refindex}), which will be useful for us. This corresponds to the TST dual formulation, where instead, we go down to IIA string theory on the $S^1$ in the Taub-Nut space \cite{Dijkgraaf:2006um}. This turns the Taub-Nut space into a single D6 brane wrapping $X\times S^1$.
 In this case, the refined topological string partition function has the interpretation as the refined spin character, counting the bound states of the D6 brane on $X$ with D0 and D2 branes. In terms of the $SO(4)=SU(2)_{\ell }\times SU(2)_r$ rotation symmetry of the Taub-Nut space,  the D0 brane charge $Q_0$ is the $2J_3^{\ell}$ component of the $SU(2)_{\ell}$ spin in M-theory, the $SO(3)=SU(2)$ rotation symmetry in IIA is identified, under the dimensional reduction, with the $SU(2)_r$ symmetry in M-theory, while the $SU(2)_R$ R-symmetry is manifestly the same in both IIA and M-theory. In particular, the refinement is associated with the diagonal $SU(2)_d \subset SU(2)_r\times SU(2)_R$ spin\cite{Nekrasov:2003rj}. This allows us to rewrite \ref{eqn:refindex} as
 
 \begin{equation}
Z_{ref\; top}(\epsilon_1,\epsilon_2;k) = \textrm{Tr}(-1)^{2J_3}\; e^{\frac{\epsilon_1 +\epsilon_2}{2}Q_0}\; e^{\frac{\epsilon_1 -\epsilon_2}{2}(2J_3 - 2R)}e^{k_I Q^I_2} \label{eqn:refindexD}.
\end{equation}
In writing this, we used the fact that $2J_3^{\ell} = S_1-S_2$,  $2J_3^{r}=S_1+S_2$, which is obvious from the way the $SU(2)_{\ell}\times SU(2)_r$ acts on the coordinates $z_1,z_2$ of the Taub-Nut space, and furthermore, as we just reviewed, that $Q_0=2J^{\ell}_3$ and $2J_3=2J_3^r$.

This leads us to propose the refined OSV conjecture relating the protected spin character of black holes to the refined topological string\footnote{For an alternate proposal relating the Nekrasov partition function to non-supersymmetric extremal black holes, see \cite{Saraikin:2007jc}.},

\begin{equation}
Z_{\textrm{ref BH}}(P_{6},P_{4};\phi_{2},\phi_{0};\gamma) = \vert Z_{\textrm{ref top}}(\epsilon_{1},\epsilon_{2},k)\vert^{2} \label{eqn:refosv}
\end{equation}
To complete the conjecture, we propose that the variables are related by,
\begin{eqnarray}
k_{I} & = & \frac{2\pi i (\frac{i\phi_{2,I}}{\pi} + \beta P_{4,I})}{\frac{i\phi_{0}}{\pi} + \beta P_{6}} \nonumber \\
\epsilon_{1} & = & \frac{4\pi i C}{i\frac{\phi_{0}}{\pi}+ \beta P_{6}} \label{eqn:osvchange} \\
\epsilon_{2} & = & \frac{4\pi i C \beta}{i\frac{\phi_{0}}{\pi}+ \beta P_{6}} \nonumber
\end{eqnarray}
where we have defined the variable \(\beta \equiv 1 - \frac{\gamma}{2\pi i}\), and we have included an additional constant \(C\).\footnote{As explained in \cite{Ooguri:2004zv}, an arbitrary constant $C$ is needed in the compact case due to the fact that $X_I$ are not functions on the moduli space, but sections of a line bundle. In the non-compact case, which we study, this degree of freedom is fixed. In the unrefined case, it is typically set to $1$. Our choice of the refined value is such that it reduces to $1$ when we set $\epsilon_{1,2}$ to be equal.} The most natural choice for \(C\), as we explain in  section \ref{sec:motiv} is 
\begin{equation}
C = \frac{2\epsilon_{1}}{\epsilon_{1} + \epsilon_{2}}
\end{equation}
so that when \(P_{6}=0\) we have,
\begin{equation}
\phi_{0} = \frac{8\pi^{2}}{\epsilon_{1} + \epsilon_{2}}
\end{equation}
for the D0 brane chemical potential, and moreover
$$
{\gamma} ={2\pi i} \frac{\epsilon_{1} - \epsilon_{2}}{\epsilon_{1} + \epsilon_{2}}
$$
for the spin chemical potential.

%



In the next section we will motivate the conjecture, and explain the origin of the change of variables. Note that, in the specific example that we study in sections \ref{sec:tqft}-\ref{sec:factor}, we will find that one gets a slightly different effective value of \(C\),  for the reason we will explain (having to do with a shift in the zero of the spin for the D0 branes).
\section{Motivating the Refined Conjecture \label{sec:motiv}}

As explained in \cite{Ooguri:2004zv, Vafa:2004qa, Ooguri:2005vr, Dijkgraaf:2005bp, Beasley:2006us, Sen:2008yk, Sen:2008vm, Dabholkar:2010uh, Sen:2011ba, Dabholkar:2011ec}, the OSV conjecture is an instance of large $N$ duality. In this case, the gauge theory is the $SU(N)$ gauge theory on the $N$ D-branes wrapping cycles of the Calabi-Yau manifold $X$ and comprising the black holes. The large $N$ dual of this theory is a string theory in the back-reacted geometry. In the full physical string theory, the near-horizon geometry of a BPS black hole in four dimensions takes the form
\begin{equation}
AdS_{2} \times S^{2} \times X.
\end{equation}
The OSV conjecture deals with supersymmetric sub-sectors of the theory. On the black-hole side, we consider the Witten index of the theory on $N$ D-branes; this is typically a partition function of a topological $SU(N)$ gauge theory in one dimension less. On the large $N$ dual side, the partition function ends up depending only on F-terms in the low energy effective action, which are captured by the topological string partition function.
The OSV conjecture can also be thought of as a consequence of large $N$ duality in the topological setting alone. 
The 't Hooft large $N$ duality, relating a $SU(N)$ gauge theory to a string theory is a very general phenomenon. It should encompass any $SU(N)$ gauge theory, including topological ones, and requires a string theory on the dual side, though not one containing dynamical gravity.
In particular, in \cite{Vafa:2004qa, Aganagic:2004js} it was shown that OSV conjecture holds even for non-compact Calabi-Yau manifolds. In the physical version of the theories studied there,  the Planck mass is infinite, and the the black holes horizon will have zero area, making it difficult to study the large $N$ duality in the full physical theory.

In the refined context, we have to take the Calabi-Yau to be non-compact, since the R-symmetry which is necessary to compute the  protected index exists only in that case. However, the theory on $N$ D-branes is still a $SU(N)$ gauge theory, with large entropy at large $N$.  $Z_{\textrm{ref BH}}$ is simply a one parameter deformation of the ordinary black-hole partition function $Z_{\textrm{BH}}$. The dual description of the theory at large $N$ has to be a string theory, on general grounds, and moreover, a suitable one-parameter deformation of the topological string theory. Now, we will explain why this one-parameter deformation is the refined topological string.

The statement of the refined OSV conjecture is that we can refine both sides of the OSV duality, by keeping track of the ${J_3-R}$ charge. Why this should be true is most tranparent in yet another way to understand the OSV, namely, using wall crossing \cite{Denef:2007vg}.  In this case, we do not take the near-horizon limit but instead we study general D4/D2/D0-brane bound states.  We can then perform \(TST\)-duality on this partition function so that it is dominated by ``polar'' states. Generically, these polar states can be made to decay by varying the background K\"ahler parameters.  Along a real co-dimension one wall, the state will decay into a D6/D4/D2/D0 state and a \(\overline{D6}\)/D4/D2/D0-state.  
By a chain of dualities (lifting to M-theory, then reducing on a different circle), the  \(AdS_{2} \times S^{2} \times X\) geometry (see \cite{Beasley:2006us, Denef:2007vg} for details) can be related to IIA string theory on $X$, with a \(D6-\overline{D6}\) pair.

Further, from the primitive wall-crossing formula we know that the degeneracies will factorize,
\begin{equation}
\Omega(D4+\ldots) \sim \Omega(D6+\ldots)\Omega(\overline{D6}+\ldots)
\end{equation}
Now the key observation is that the degeneracies of D6/D4/D2/D0 brane bound states are precisely computed by Donaldson-Thomas invariants, which are further identified with the topological string.  On the topological string side, the S-duality used is precisely what relates the D6 brane partition function to the topological string \cite{Nekrasov:2004js}, as we reviewed in the previous section.
Therefore, we can identify the D6-brane bound states with \(Z_{top}\) and the \(\overline{D6}\)-brane bound states with \(\overline{Z}_{top}\).  Therefore, the semiprimitive wall-crossing formula gives precisely the factorization expected from OSV,
\begin{equation}
Z_{BH} \sim \vert Z_{top} \vert^{2}
\end{equation}

In the refined setting,  the argument goes through in precisely the same way as in the unrefined case; we simply replace, on both sides of the duality, the Witten index of the D4 brane and the D6 branes by the protected spin character. Then we simply use the refined primitive wall-crossing formula which also factorizes. On the black hole side, the protected spin character of the D4 branes is the refined black hole partition function $Z_{\textrm{ref BH}}$, and on the topological string side, the protected spin character of the D6 branes is the topological string partition function -- moreover, we get both $Z_{\textrm{top ref}}$ and ${\overline Z}_{\textrm{top \;ref}}$ from the D6 branes and the $\overline{D6}$ branes, and thus
\begin{equation}
Z_{\textrm{ref BH}} \sim \vert Z_{\textrm{ref top}} \vert^{2}.
\end{equation}
This argument makes it obvious that the OSV conjecture should hold in the refined setting, as we conjectured. 

The one subtlety in this argument is that on a noncompact Calabi-Yau geometry, there is actually no place in moduli space where the \(D4\)-branes can be made to split into \(D6-\overline{D6}\) constituents, since D6 and \(\overline{D6}\) branes will always have opposite central charges because of the noncompactness of the Calabi-Yau.  However, this issue was already present in the unrefined case, where it did not affect the validity of the conjecture, as was shown in \cite{Vafa:2004qa,Aganagic:2004js}. Thus, there is no reason to think it would affect our refined conjecture either. Thus, we believe that this \(D6\)-\(\overline{D6}\) decomposition captures the correct physics of D4/D2/D0-brane bound states in both the unrefined and the refined case, and this moreover leads to a refined OSV formula. 

In the rest of this section, we will provide further support for the conjecture, and explain the identification of the parameters we gave previously.

\subsection{Refined OSV and The Wave Function on the Moduli Space}

The refined topological string partition function is a wave function on the moduli space, just like in the ordinary topological string case \cite{Krefl:2010fm, Krefl:2010jb, Huang:2011qx, Aganagic:2011mi}. The quantum mechanics on the moduli space played a central role in understanding the original conjecture, and the same is true in the refined case.
In this respect, there only two differences between the refined and the unrefined topological string thing: for one, the effective value of the Plank's constant of the theory $g_s^2$, becomes $\epsilon_1\epsilon_2$ (recall that, in the unrefined case, $\epsilon_1$ and $\epsilon_2$ coincide), 
$$
g_s^2\qquad  \rightarrow \qquad  g_s^2=\epsilon_1 \epsilon_2.
$$
Secondly, the wave function that the topological string partition function computes changes:  in the refined case, this wave function depends on the additional parameter $\beta =\epsilon_2/\epsilon_1$. 

For this discussion of the quantum mechanics, it is useful to switch to the mirror perspective and study the refined B-model \cite{Aganagic:2011mi}. The refined B-model only depends on the complex structure moduli space, which can be parametrized by the holomorphic three form \(\Omega \in H_{3}(X)\).  We can choose a symplectic basis for \(H_{3}(X)\) such that \(A_{I}\cap B^{J} = \delta^{J}_{I}\), and define coordinates,
\begin{equation}
X_{I} = \int_{A_{I}}\Omega, \qquad\qquad F^{J} = \int_{B^{J}}\Omega
\end{equation}
From special geometry, we know that classically these variables are not independent and that there exists a prepotential, \(F^{(0)}\), such that,
\begin{equation}
F^{J} = \frac{\partial F^{(0)}}{\partial X_{J}}
\end{equation}
But now it is important to recognize that this prepotential is the genus zero contribution of the refined topological string,
\begin{equation}
Z_{\textrm{ref top}} = \exp\Big(\frac{1}{\epsilon_{1}\epsilon_{2}}F\Big) = \exp\Bigg(\frac{1}{\epsilon_{1}\epsilon_{2}}F^{(0)} + \ldots\Bigg)
\end{equation}
Therefore in the full quantum theory, we can represent \(F^{J}\) as the operator \(F^{J} = \epsilon_{1}\epsilon_{2}\frac{\partial}{\partial X_{J}}\) and this leads to the commutation relations,
\begin{equation}
\lbrack F^{J}, X_{I} \rbrack = \epsilon_{1}\epsilon_{2} \delta^{J}_{I}
\end{equation}
We could have applied this same reasoning to the conjugated theory, \(\overline{Z}_{\textrm{ref top}}\) which gives,
\begin{equation}
\lbrack \overline{F}^{J}, \overline{X}_{I} \rbrack = \epsilon_{1}\epsilon_{2} \delta^{J}_{I}
\end{equation}
and finally all of the barred variables commute with all of the unbarred variables.
Note that, in this case, because the Calabi-Yau is non-compact, the moduli space is always governed by the rigid special geometry of the ${\cal N}=2$ field theory, rather than the local special geometry of ${\cal N}=2$ supergravity.

Now consider formally inverting the refined OSV relation,
\begin{equation}\label{eq-overlap}
\Omega(P_{6},P_{4},Q_{2},Q_{0};\gamma) = \int d\phi_{0}d\phi_{2} e^{Q_{2}\phi_{2}+Q_{0}\phi_{0}} \vert Z_{\textrm{ref top}}\vert^{2}
\end{equation}
where
\begin{equation}
\Omega(P_{6},P_{4},Q_{2},Q_{0};\gamma) = \sum_{J_{3},R}\Omega(P_{6},P_{4},Q_{2},Q_{0};J_{3},R)e^{-2\gamma(J_{3}-R)}.
\end{equation}
We say that this is a formal inversion, since in the non-compact case $\phi_0$ is always just a parameter, so in particular, it does not really make sense integrating over it. This aside, note that for the relation such as (\ref{eq-overlap}) to make sense, it {\it has to be the case} that 
$Z_{\textrm{ref top}}$ is indeed a wave function on the moduli space. This is because while the left hand side is independent of the choice of polarization, i.e. the choice of basis of $A$- and $B$-cycles, for an arbitrary function on the moduli space, the right hand side {\it would} depend on such a choice, and the conjecture would not have a chance to hold. Because $Z_{\textrm{ref top}}$ is a wave function, while all the terms on the right hand side depend on the choice of polarization, the integral does not depend on such a choice. 

More precisely, for this to hold, one has to have the following commutation relations. We follow the reasoning in \cite{Ooguri:2004zv}, and formally introduce magnetic potentials, \(\chi^{I}\) in addition to the electric potential that we have already used.  In the refined black hole partition function, we cannot specify both the electric charge and the electric potential at the same time, so they must have nontrivial commutation relations.  Similarly, we require that the new magnetic potentials are conjugate to the magnetic charges.  Therefore, we find
\begin{eqnarray}
\lbrack \phi_{I}, Q^{J} \rbrack & = & \lbrack P_{I}, \chi^{J} \rbrack = \frac{i\pi}{2}\delta^{J}_{I} \nonumber \\
\lbrack \phi_{I}, P_{J} \rbrack & = & \lbrack \chi^{I}, Q^{J} \rbrack = 0 \\
\lbrack Q^{I}, P_{J} \rbrack & = & \lbrack \chi^{I}, \phi_{J} \rbrack = 0 \nonumber
\end{eqnarray}
where for convenience we have included an extra normalization factor above.  For the right hand side of \ref{eq-overlap} to be invariant under symplectic transformations one needs the black hole commutation relations and the topological string commutation relations, to be consistent. This requires,
\begin{eqnarray}
X_{I} & = & C'\epsilon_{2} P_{I} + i\frac{\epsilon_{1}}{C'}\frac{\phi_{I}}{\pi} \label{eqn:xf} \\
F^{I} & = & C'\epsilon_{2} Q^{I} + i\frac{\epsilon_{1}}{C'}\frac{\chi^{I}}{\pi} \nonumber
\end{eqnarray}
for some arbitrary constant, \(C'\).  Note that is in prefect agreement with the refined OSV change of variables in equation \ref{eqn:osvchange} upon fixing the constant to \(C'=1\). In fact, these relations were our main motivation for the change of variables we proposed in section 3, as a part of our conjecture. Notice that \(\Omega(P,Q;\gamma)\) still has the interpretation as a Wigner quasi-probability distribution on phase space, just as it did in \cite{Ooguri:2004zv}, but now it depends on the additional auxillary parameter, \(\gamma\).

To understand the identification of $\epsilon_1$ and $\epsilon_2$  with the D0 brane chemical potential $\phi_0$ and the spin fugacity $\gamma$, we can use the wall crossing derivation, which forces the identification of parameters.
The only subtlety is that, to relate the refined topological string to the black-hole ensemble, we need to perform the TST duality. The TST duality relates this to the chemical potentials before and after as follows:

\begin{equation}\label{Sduality}
\phi_{0}\;\; \to \;\;\phi_{0}'=\frac{4\pi^{2}}{\phi_{0}}, \qquad \phi_{2} \;\; \to\;\; \phi_{2}'=2\pi i \frac{\phi_{2}}{\phi_{0}}, \qquad \gamma \;\;\to \;\;\gamma'=2\pi i \frac{\gamma}{\phi_{0}}
\end{equation}
The derivation of this is presented in appendix \ref{sec:sduality}.

As we reviewed in the previous section, the chemical potential for the D0 branes bound to the D6 brane is 

\begin{equation}
\phi'_{0} = \frac{\epsilon_{1} + \epsilon_{2}}{2},
\end{equation}
and the spin is captured by
\begin{equation}
\gamma' = \frac{\epsilon_{1} - \epsilon_{2}}{2},
\end{equation}
For this to be consistent with TST duality, the chemical potentials in the black hole ensemble need to be
\begin{equation}
\phi_{0} = \frac{4\pi^{2}}{\epsilon_{1} + \epsilon_{2}},
\end{equation}
for the D0 brane charge, and moreover the spin needs to be captured by
\begin{equation}
\gamma = {2 \pi i}\frac{\epsilon_{1} - \epsilon_{2}}{\epsilon_{1} + \epsilon_{2}}
\end{equation}
just as we gave in the previous section. In particular, $\gamma/2\pi i = 1-\beta$.

The rest of this paper is devoted to testing our conjecture. There is a class of geometries where both sides of duality are computable explicitly.
These correspond to Calabi-Yau manifolds that are complex line bundles over a Riemann surface. 
After developing the necessary tools to precisely compute both sides of the refined OSV formula, we show that the refined OSV conjecture holds true perturbatively to all orders for these geometries.

\section{Refined Topological String on 
${\mathcal L}_1\oplus {\mathcal L}_1 \rightarrow \Sigma $ \label{sec:tqft}}

For some simple Calabi-Yau manifolds, the refined topological string partititon function  is exactly computable by cutting the Calabi-Yau into simple pieces, and sewing them back together.
The open-string version of this index was computed explicitly on simple geometries in \cite{Aganagic:2011sg} and used to solve the refined Chern-Simons theory completely.  We will follow a similar approach here in the closed string case, for Calabi-Yau manifolds of the form
$$
{\mathcal L}_1\oplus {\mathcal L}_1 \rightarrow \Sigma
$$
To obtain a Calabi-Yau manifold, the degrees of the line bundles must satisfy the property,
\begin{equation}
\textrm{deg}(\mathcal{L}_{1}) + \textrm{deg}(\mathcal{L}_{2}) = -\chi(\Sigma) = 2g-2
\end{equation}
The key idea is to chop up our geometries by introducing stacks of infinitely many M5 brane/anti-brane pairs wrapping Lagrangian three-cycles as in the original topological vertex \cite{Aganagic:2003db}.\footnote{In the refined setting, we must choose whether to wrap these M5 branes on the \(z_{1}\) or \(z_{2}\) plane of the Taub-NUT space.  This gives two types of refined A-branes, which can be denoted as \(q\)-branes and \(t\)-branes.  At each boundary of our geometry we can place either type of brane, leading to different choices of basis for each Hilbert space.  In this paper we will not need this rich structure, and we will implicitly place \(q\)-branes at each puncture.  We refer the reader to \cite{AS2, ASVertex} for details on general \(q\)/\(t\)-brane amplitudes.}  Then the computation of the refined index on these chopped geometries reduces to counting M2 branes ending on these M5 branes.  In this paper, we simply explain the structure of the TQFT, and refer the reader to \cite{ASVertex} for the details of computing these amplitudes by counting M2-brane contributions.  

\subsection{A TQFT for the Refined Topological String}

The basic building blocks of the TQFT are given by the annulus (A), cap (C), and pant (P) geometries. Since degrees of bundles and euler characteristics add upon gluing, this gives a way of building up more complicated bundles over Riemann surfaces.

\begin{figure}[htp]
\centering
\includegraphics[scale=0.80]{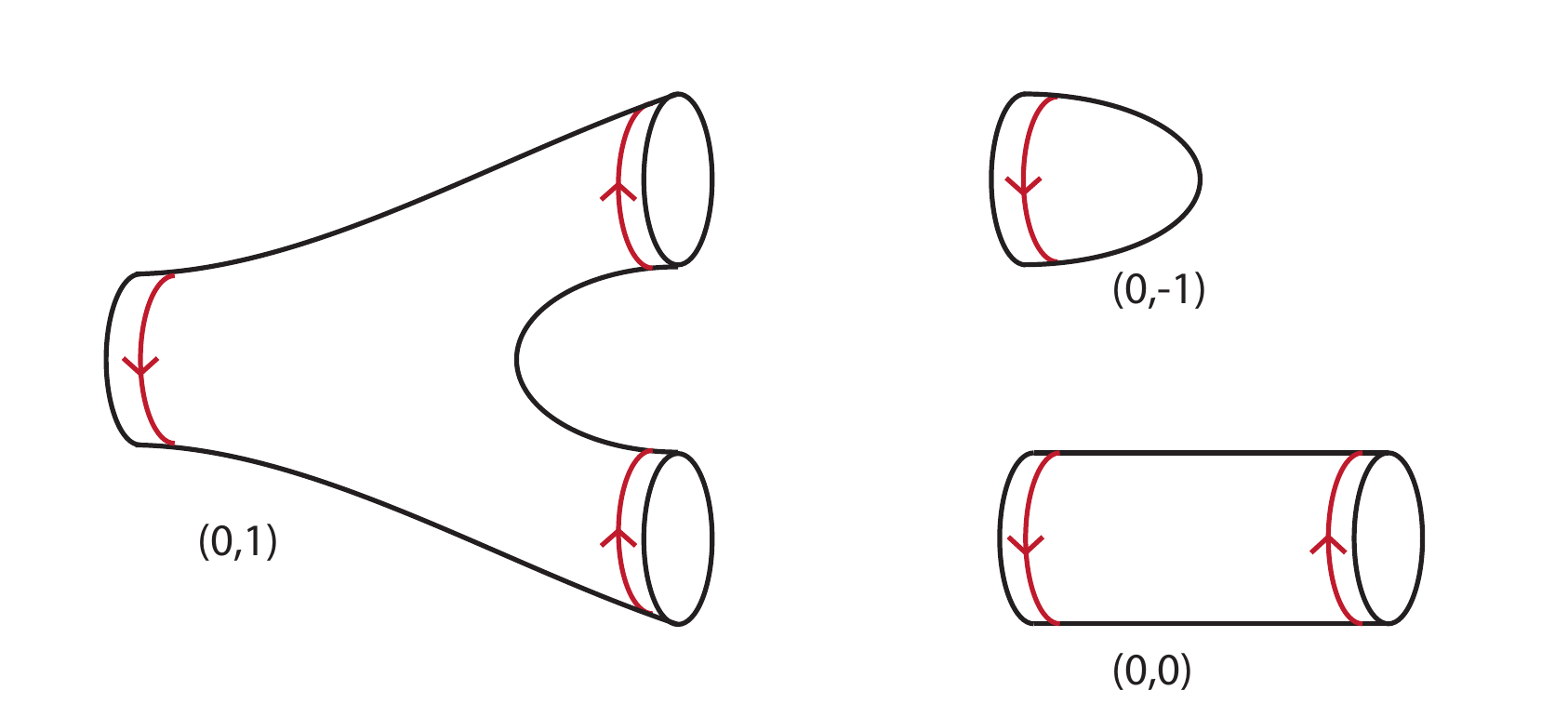}
\caption{The building blocks of the refined TQFT are the pant, cap, and annulus geometries, along with complex bundles of degree $(d_{1},d_{2})$ over each Riemann surface. \label{fig:tqftpieces}}
\end{figure}

We start by considering the simplest geometry, which is the annulus (shown in Figure \ref{fig:tqftpieces}) with two trivial complex line bundles over it given by \(A^{(0,0)} = \mathbb{C}^{*} \times \mathbb{C}^{2}\).  
\begin{eqnarray}
Z_{\textrm{ref top}}(A^{(0,0)}) & = & \sum_{R}\frac{1}{g_{R}(q,t)}M_{R}(U;q,t)M_{R}(V;q,t)
\end{eqnarray}
Here we are summing over all \(U(\infty)\) representations \(R\), and \(M_{R}(U;q,t)\) is the associated Macdonald polynomial which reduces on setting \(q=t\) to the simpler \(\textrm{Tr}_{R}(U)\).  The Macdonald metric, \(g_{R}\) computes the inner product of a Macdonald polynomial with itself and is given by,\footnote{Note we are including additional (q/t) factors in our definition of $g_{R}$ and \(M_{R}(U)\) compared to the standard definitions in \cite{macdonald_hall}.  The advantage of these factors is that they restore the symmetry, \(g_{R}(q,t) = g_{R}(q^{-1},t^{-1})\) (see also \cite{Awata:2008ed} for a similar shift).  We refer the reader to Appendix \ref{sec:macs} for more details on our Macdonald polynomial conventions.}
\begin{eqnarray}
g_{R}(q,t) & = & (t/q)^{\vert R \vert/2}\prod_{(i,j)\in R}\frac{1-t^{R^{T}_{j}-i}q^{R_{i}-j+1}}{1-t^{R^{T}_{j}-i+1}q^{R_{i}-j}} \label{eqn:uinftymetric} \\
& = & \prod_{(i,j)\in R}\frac{t^{\frac{R^{T}_{j}-i}{2}}q^{\frac{R_{i}-j+1}{2}}-t^{-\frac{R^{T}_{j}-i}{2}}q^{-\frac{R_{i}-j+1}{2}}}{t^{\frac{R^{T}_{j}-i+1}{2}}q^{\frac{R_{i}-j}{2}} - t^{-\frac{R^{T}_{j}-i+1}{2}}q^{-\frac{R_{i}-j}{2}}} \label{eqn:infmetric}
\end{eqnarray}
Since we are working with \(U(\infty)\) representations, this metric is the \(N\to\infty\) limit of the ordinary \(SU(N)\) Macdonald metric.  We refer the reader to Appendix \ref{sec:macs} for our Macdonald polynomial conventions.

Having discussed the simplest geometry, we should now explain how building blocks are glued together.  Recall that ordinarily when we want to glue two boundaries together, we should set their holonomies to be equal except that the boundaries should have opposite orientation.  This orientation reversal simply flips one of the holonomies from \(U\) to \(U^{-1}\).  Finally, to glue together the boundaries we must integrate over the Hilbert space at the boundaries.  

This is also true in the refined setting, except that the integration measure is deformed to the one natural for Macdonald polynomials.  If we denote the eigenvalues of \(U\) by \(e^{u_{i}}\), then the Macdonald measure is given by,
\begin{equation}
\Delta(U;q,t) = \prod_{m=0}^{\infty}\prod_{i\neq j}\frac{\Big(1-q^{m}e^{u_{i}-u_{j}}\Big)}{\Big(1-tq^{m}e^{u_{i}-u_{j}}\Big)}
\end{equation}
Then gluing two boundaries gives,
\begin{equation}
\int du_{i}\Delta(U;q,t)M_{R_{1}}(U)M_{R_{2}}(U^{-1}) = g_{R_{1}}(q,t)\delta_{R_{1}R_{2}}
\end{equation}
where \(g_{R}\) is the Macdonald metric for infinitely many variables introduced above, which is to be contrasted with the finite N Macdonald metric which we will define below in equation \ref{eqn:metricbeta}.  Thus, the \(M_{R}\) give an orthogonal but not orthonormal basis for the boundary Hilbert space.  Although we could remove the explicit metric factors \(g_{R}\) by choosing a different normalization for \(M_{R}\), it will actually be more convenient in this paper to keep them.

As a simple consistency check, gluing two annuli with trivial bundles should give back the original annulus amplitude.  But this is clearly true, since the two annuli contribute a total factor of \(g_{R}^{-2}\) while the gluing process contributes a factor of \(g_{R}\) so that the resulting amplitude is equal to the original annulus amplitude.

Now that we have explained gluing, we also want to know how to introduce nontrivial bundles.  Note that since \(\chi(A)=0\), any choice of line bundles over the annulus must satisfy \(\textrm{deg}(\mathcal{L}_{1}) = -\textrm{deg}(\mathcal{L}_{2})\).  The simplest nontrivial choice is the geometry \(A^{(1,-1)}\).  This geometry can be alternatively understood as implementing a change in framing, which has been studied for the refined topological vertex in \cite{Iqbal:2007ii}.  The resulting amplitude is given by,
\begin{equation}
Z_{\textrm{ref top}}(A^{(1,-1)}) =  \sum_{R}\frac{1}{g_{R}(q,t)}q^{\frac{1}{2}\vert\vert R \vert\vert^{2}}t^{-\frac{1}{2}\vert\vert R^{T} \vert \vert^{2}} M_{R}(U;q,t)M_{R}(V;q,t)
\end{equation}
where \(\vert \vert R \vert \vert^{2}= \sum_{i}R_{i}^{2}\) and \(\vert \vert R^{T}_{i} \vert \vert^{2} = \sum_{i} (R^{T}_{i})^{2} = \sum (2i-1)R_{i}\).

Next we study the cap geometries (shown in Figure \ref{fig:tqftpieces}), which are given by two complex line bundles over the disc.  Since the euler characteristic of the disc is equal to \(1\), the degrees of the line bundles must satisfy \(\textrm{deg}(\mathcal{L}_{1}) + \textrm{deg}(\mathcal{L}_{2}) = -1\).  In practice, it suffices to determine the \((0,-1)\) amplitudes, since the rest can be obtained by gluing.

It is helpful to notice that this geometry is equivalent to \(\mathbb{C}^{3}\) with a stack of branes inserted on one leg of the vertex.  Thus the cap amplitude can be computed by the refined topological vertex amplitude, \(C_{R\cdot\cdot}\),  with branes on the \(q\)-leg or from refined Chern-Simons \cite{AS2}.  The result is,
\begin{eqnarray}
Z(C^{(0,-1)}) & = & \sum_{R}\frac{1}{g_{R}}\textrm{dim}_{q,t}(R)M_{R}(U;q,t)
\end{eqnarray}
where we have defined the \((q,t)\)-dimension of a representation \(R\) by,
\begin{eqnarray}
\textrm{dim}_{q,t}(R) & = & (q/t)^{\frac{1}{2}\vert R \vert}M_{R}(t^{\rho};q,t) \label{eqn:uinftyqtdim} \\
& = & q^{\frac{1}{4}\vert\vert R \vert\vert^{2}}t^{-\frac{1}{4}\vert\vert R^{T} \vert \vert^{2}} \prod_{\tableau{1}\in R}\Big(q^{\frac{a(\tableau{1})}{2}}t^{\frac{l(\tableau{1})+1}{2}}-q^{-\frac{a(\tableau{1})}{2}}t^{-\frac{l(\tableau{1})+1}{2}}\Big)^{-1} \nonumber
\end{eqnarray}
where \((\rho)_{i}=-i+1/2\) is the \(U(\infty)\) Weyl vector, while 
\begin{eqnarray}
a(\tableau{1}) & = & R_{i}-j \label{eqn:armlength} \\
l(\tableau{1}) & = & R^{T}_{j}-i \nonumber
\end{eqnarray}
are the arm- and leg-lengths, respectively, of a box in the Young Tableau of \(R\).  The \((q,t)\)-dimension can be understood as a \((q,t)\)-deformation of the dimension of the symmetric group representation specified by \(R\).\footnote{Our notation differs slightly from the notation used for the unrefined case in \cite{Aganagic:2004js}.  In the limit \(t=q\), our \((q,t)\)-dimension is related to their \(d_{q}(R)\) by, \(\textrm{dim}_{q,q}(R) = s_{R}(q^{\rho}) = q^{\frac{1}{4}\kappa_{R}}d_{q}(R)\).}  To obtain the cap with a different choice of line bundles we can simply glue on the \(A^{(1,-1)}\) or \(A^{(-1,1)}\) annuli.

Finally, we must specify the three-punctured sphere amplitude (see Figure \ref{fig:tqftpieces}), which we refer to as the ``pant.''   Since the three-punctured sphere has the euler characteristic \(\chi=-1\), the degree of the line bundles must add to one in this case.  To compute this amplitude it is helpful to recall some general properties that our TQFT must satisfy.  Since it computes the refined topological A-model, the TQFT must be independent of complex structure moduli.  Specifically this means that the amplitude for a Riemann surface should not depend on how it is formed by gluing simpler geometries.  For this to be true, the pant amplitude should be symmetric in the three punctures, and thus should be diagonal in the Macdonald basis,
\begin{equation}
Z_{\textrm{ref top}}(P^{(0,1)}) = \sum_{R}P_{R}\;M_{R}(U_{1})M_{R}(U_{2})M_{R}(U_{3})
\end{equation}

Now it is helpful to recognize that the pant, cap, and annulus are not all independent.  We can form the annulus by capping off one of the punctures of the pant.  By consistency and using the fact that \(Z(P^{(0,1)})\) is diagonal, we can solve for the pant amplitude,
\begin{equation}
Z_{\textrm{ref top}}(P^{(0,1)}) =  \sum_{R}\frac{1}{g_{R}\textrm{dim}_{q,t}(R)}M_{R}(U_{1})M_{R}(U_{2})M_{R}(U_{3})
\end{equation}

So far we have described the structure of the A-model on these geometries as a TQFT, but it is important to remember that the theory is not purely topological since it depends on the K\"ahler moduli.  For Calabi-Yaus of the form \(\mathcal{L}_{1}\oplus \mathcal{L}_{2} \to \Sigma_{g}\), there is only one K\"ahler modulus, \(k\), that measures the area of the Riemann surface.  In fact, as is familiar from the topological vertex \cite{Aganagic:2003db}, the partition function depends on this modulus only by introducing a term, \(e^{-k\vert R \vert}\) in the sum over representations.

Altogether, we have given the necessary data to solve the theory completely.  As an application of these results, we can study geometries of the form \(\mathcal{L}_{1}\oplus \mathcal{L}_{2} \to \Sigma_{g}\) where \(\Sigma_{g}\) is a genus \(g\) Riemann surface.  For this to be a Calabi-Yau manifold we must have \(\textrm{deg}(\mathcal{L}_{1}) = 2g-2+p\) and \(\textrm{deg}(\mathcal{L}_{2}) = -p\).  Then the refined amplitude on this geometry is given by,
\begin{equation}
Z_{\textrm{ref top}}^{(g,p)}(q,t) = \sum_{R}\Bigg(\frac{\textrm{dim}_{q,t}(R)^{2}}{g_{R}}\Bigg)^{1-g}q^{\frac{(2g-2+p)}{2}\vert\vert R \vert \vert^{2}}t^{-\frac{(2g-2+p)}{2}\vert\vert R^{T} \vert \vert^{2}}Q^{\vert R \vert} \label{eqn:reftopgeneral}
\end{equation}
where we have defined the exponentiated K\"ahler modulus as \(Q=e^{-k}\).  It can be checked that this has the expected symmetry,
\begin{equation}
Z_{\textrm{ref top}}^{(g,p)}(q,t) = Z_{\textrm{ref top}}^{(g,p)}(t^{-1},q^{-1})
\end{equation}
which implies that the Gopakumar-Vafa invariants  come in complete multiplets of $SU(2)_\ell$ (as was the case in the unrefined limit).  However, the amplitude is not symmetric under the exchange \(q\leftrightarrow t\). This implies that the Gopakumar-Vafa invariants for these Calabi-Yaus do not come in full representations of \(SU(2)_{r}\), but only carry \(U(1)_{r}\subset SU(2)_r\) charge. The BPS states come from quantizing the moduli space of curves in $X$, together with the $U(1)$ bundle on them. The $SU(2)_r$ spin content comes from cohomologies of the moduli of the curve itself, while the $SU(2)_{\ell}$ comes from the bundle. In the present case,  the curve is $\Sigma$ itself. Its moduli space is in general non-compact, as typically one of the two line bundles over $\Sigma$ has positive degree. Correspondingly, the Lefshetz $SU(2)_r$ action on the cohomologies of the moduli space does not have to result in complete multiplets -- there can be contributions that escape to infinity. 
(See Appendix \ref{sec:gv} for some sample computations of Gopakumar-Vafa invariants for these geometries.) In fact, the only case when the moduli space is compact is when $\Sigma = {\mathbb P}^1$, and both line bundles are ${\cal O}(-1)$. It is easy to see that in this case the amplitude does in fact have the \(q\leftrightarrow t\) symmetry as well. 

In addition, the amplitude is not symmetric under exchange of the two line bundles, which is equivalent to taking \(p \to 2-2g-p\).  This tells us that one of the line bundles is distinguished from the other in the refined setting.  In fact, this arises because the index in equation \ref{eqn:refindex} includes an \(R\)-symmetry twist that rotates a specific bundle in the noncompact Calabi-Yau (for more details see \cite{AS2}). Equivalently, as will be explained in section \ref{sec:factor}, these refined topological string amplitudes can be obtained by taking the large N limit of \(D4\) branes wrapping one of the bundles.  In the unrefined case, the large N limit does not retain information about which bundle the \(D4\) branes wrapped, but in the refined case this choice has an effect on the closed string amplitude.  This is related to the above observation since in both the \(D4\) construction and in the closed string construction we must choose an R-symmetry rotation to preserve supersymmetry. However, these symmetries are not completely lost since the amplitude is symmetric under simultaneously exchanging \(q\leftrightarrow t\) and exchanging the bundles,
\begin{equation}
Z_{\textrm{ref top}}^{(g,p)}(q,t) = Z_{\textrm{ref top}}^{(g,2-2g-p)}(t,q)
\end{equation}
As we will explain in section \ref{sec:u1engineering}, \(p\) specifies the five-dimensional Chern-Simons coupling of the geometrically engineered gauge theory.  In \cite{Awata:2008ed}, it was similarly observed that for geometries that engineer five-dimensional \(SU(N)\) gauge theories, the refined topological string is only symmetric under the simultaneous exchange of \(q\leftrightarrow t\) and inverting the Chern-Simons level, \(k\to -k\).

So far, we have computed all the non-trivial contributions to the refined topological string.  However, we should also include by hand the additional pieces that appear at genus zero and one.  In the unrefined case, these arise from constant maps.  In the refined case, for geometries that engineer five-dimensional gauge theories, these contributions arise from the classical prepotential and the one-loop determinant of the instanton partition function.  These degree zero pieces take the form,
\begin{equation}
Z_{0}(q,t) = \Big(M(q,t)M(t,q)\Big)^{\chi/4}\exp\Bigg(\frac{1}{\epsilon_{1}\epsilon_{2}} \frac{a k^{3}}{6}+\frac{\epsilon_{2}}{\epsilon_{1}} b \frac{k}{24}\Bigg) \label{eqn:degzero}
\end{equation}
where \(M(q,t)\) is the refined MacMahon function,
\begin{equation}
M(q,t) = \prod_{i,j=1}^{\infty}\Big(1-t^{i}q^{j-1}\Big)
\end{equation}
and \(\chi\) is the euler characteristic of the Calabi-Yau, while \(a\) is related to the triple intersection of the K\"ahler class and \(b\) is the second Chern class of the Calabi-Yau.  These numbers are ambiguous because of the non-compactness of our geometries, but it was argued in \cite{Aganagic:2004js} that for the connection with black holes the natural values are,
\begin{equation}
\chi = 2-2g, \qquad \qquad a = - \frac{1}{p(p+2g-2)}, \qquad \qquad b = \frac{p+2g-2}{p}
\end{equation}

Note, that we have split the MacMahon function into two pieces, related by interchanging \(q\) and \(t\).  This split naturally appears in section \ref{sec:factor}, when making the connection with the refined black hole partition function.  A similar splitting was recently observed for the motivic Donaldson-Thomas invariants of the conifold in \cite{Morrison:2011rz}.

In section \ref{sec:u1engineering}, we will give further evidence that this refined amplitude is the correct one by comparing it with the equivariant instanton partition function of the geometrically engineered five-dimensional field theory.  Before doing so, however, it will be helpful to discuss one final aspect of the TQFT that arises when D-branes are included in the fiber of the complex line bundles.

\subsection{Branes in the Fiber \label{sec:fiberbranes}}
So far we have solved for the refined string on bundles over closed Riemann surfaces and Riemann surfaces with boundaries.  These boundaries naturally end on branes wrapping an \(S^{1}\) in the base and two dimensions in the fiber.  However, for understanding the refined OSV conjecture, it will be helpful to also consider introducing branes in the fiber.  For unrefined topological strings, this was studied in \cite{Aganagic:2004js}, and our analysis will follow a similar approach.

We consider a lagrangian brane at a point, \(z\), in the base Riemann surface, \(\Sigma_{g}\).  Since the brane is local in the base, we only need to study a neighborhood of \(z\).  Thus it is natural to introduce branes in the base that chop up the geometry into a disc, \(D\), containing \(z\), and its complement, \(\Sigma \setminus D\). The full amplitude is given by,
\begin{equation}
Z = \sum_{R, Q}Z_{R}(\Sigma \setminus D)Z_{RQ}(D)M_{Q}(V;q,t)
\end{equation}
where \(V\) is the holonomy around the branes in the fiber.  The amplitude on the complement, \(\Sigma\setminus D\), can be solved by gluing using the amplitudes in the previous section, but we still need to solve for the disc amplitude with two sets of branes.

\begin{figure}[htp]
\centering
\includegraphics[scale=1.00]{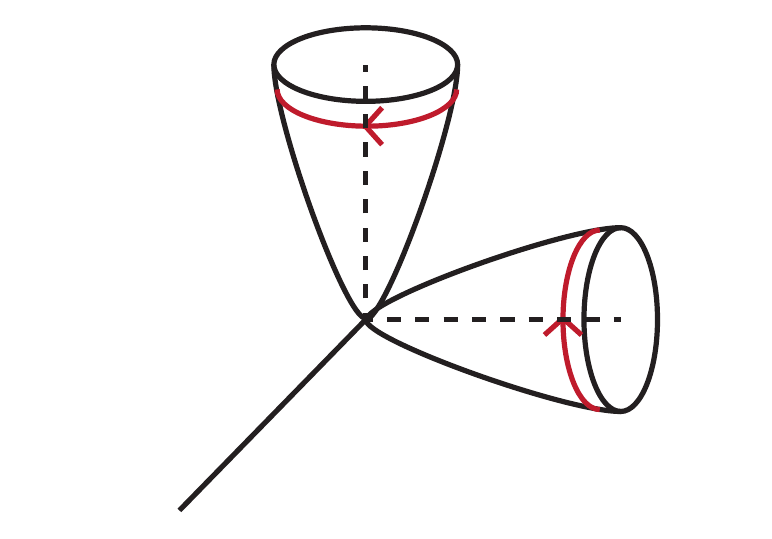}
\caption{The full cap amplitude with branes in both the fiber and the base. \label{fig:smat}}
\end{figure}

This can be accomplished by noticing that \(D\) has the topology of \(\mathbb{C}^{3}\) with the base and fiber branes on two legs of the vertex.  Thus, the full cap amplitude,
\begin{equation}
Z(D) = \sum_{R,Q} Z_{RQ}(D)M_{R}(U)M_{Q}(V)
\end{equation}
is simply computed by the refined topological vertex amplitude with two stacks of branes on different legs, as shown in Figure \ref{fig:smat}.  

Alternatively, as will be explained in \cite{AS2}, this amplitude can be solved by following the refined Chern-Simons theory through a geometric transition.  From this perspective, \(Z_{RQ}\) is computed by the large N limit of the refined Chern-Simons S-matrix,
\begin{eqnarray}
W_{RQ} & = & \lim_{N\to\infty}t^{-\frac{N(\vert R\vert + \vert Q \vert)}{2}}(q/t)^{\frac{\vert R \vert + \vert Q \vert}{4}}\frac{S_{RQ}(q,t;N)}{S_{00}(q,t;N)} \\
& = & (q/t)^{\frac{\vert R \vert + \vert Q \vert}{2}}M_{R}(t^{\rho})M_{Q}(t^{\rho}q^{R})
\end{eqnarray}
where we have used the symmetrized definition of the infinite-variable Macdonald polynomials in Appendix \ref{sec:macs}.
By including the appropriate metric factors, we obtain,
\begin{equation}
Z_{RQ} = \frac{1}{g_{R}g_{Q}}W_{RQ}
\end{equation}
Note that if we set the representation of the fiber brane to be trivial, \(Q=0\), then this geometry is the same as the cap (C) that we studied above.  This is consistent with the fact that \(W_{R0} = \textrm{dim}_{q,t}(R)\).

As an example, take the geometry, \(\mathcal{O}(2g-2+p)\oplus \mathcal{O}(-p) \to \Sigma_{g}\) with branes in the fiber over \(h\) points.  Then the refined amplitude is given by,

\begin{eqnarray}
Z_{\textrm{ref top}}^{(g,p,h)}(q,t) = \sum_{R, R_{1}, \cdots, R_{h}}& & \frac{g_{R}^{g-1}}{W_{R0}^{2g-2+h}}\frac{W_{RR_{1}}\cdots W_{RR_{h}}}{g_{R_{1}}\cdots g_{R_{h}}}q^{\frac{(2g-2+p)}{2}\vert\vert R \vert \vert^{2}}t^{-\frac{(2g-2+p)}{2}\vert\vert R^{T} \vert \vert^{2}}Q^{\vert R \vert} \nonumber \\
& & \cdot M_{R_{1}}(V_{1})\cdots M_{R_{h}}(V_{h})
\end{eqnarray}

It is also useful to understand how an anti-brane can be introduced that wraps the fiber.  Recall that in the unrefined topological string, converting a brane into an anti-brane corresponds to taking,
\begin{equation}
s_{R}(U) \to (-1)^{\vert R \vert}s_{R^{T}}(U)
\end{equation}
where \(s_{R}(U)\) is the Schur function.  The analogue of this reversal in the refined setting corresponds to taking,
\begin{equation}
M_{R}(U;q,t) \to \iota M_{R}(U;q,t)
\end{equation}
where \(\iota\) is defined by how it acts on power sums, \(p_{n}(x)\),
\begin{equation}
\iota(p_{n}) = -p_{n}
\end{equation}
We refer the reader to \cite{AS2, ASVertex} for more details on this construction.  This implies that the disc amplitude with an anti-brane in the fiber is given by,
\begin{equation}
\widetilde{Z}(D) = \sum_{R,Q}\frac{1}{g_{R}g_{Q}}W_{RQ}M_{R}(U)\iota M_{Q}(V)
\end{equation}
If we want to rewrite this in the \(M_{Q}(V)\) basis, this can be done by using a generalized Cauchy identity (see Appendex \ref{sec:macs}),
\begin{eqnarray}
\widetilde{Z}(D) & = & \sum_{R,Q}\frac{1}{g_{R}g_{Q}}(q/t)^{\frac{\vert R \vert + \vert Q \vert}{2}}M_{R}(t^{\rho})M_{Q}(t^{\rho}q^{R})M_{R}(U)\iota M_{Q}(V) \\
& = & \sum_{R}\frac{1}{g_{R}}M_{R}(t^{\rho})M_{R}(U)\sum_{Q}\frac{1}{g_{Q}}(q/t)^{\frac{\vert R \vert + \vert Q \vert}{2}}M_{Q}(t^{\rho}q^{R})\iota M_{Q}(V) \\
& = & \sum_{R}\frac{1}{g_{R}}M_{R}(t^{\rho})M_{R}(U)\sum_{Q}\frac{1}{g_{Q}}(q/t)^{\frac{\vert R \vert + \vert Q \vert}{2}} \iota M_{Q}(t^{\rho}q^{R}) M_{Q}(V) \\
& = & \sum_{R,Q}\frac{1}{g_{R}g_{Q}}\widetilde{W}_{RQ}M_{R}(U)M_{Q}(V)
\end{eqnarray}
where we have defined \(\widetilde{W}_{RQ}\) by,
\begin{equation}
\widetilde{W}_{RQ} = (q/t)^{\frac{\vert R \vert + \vert Q \vert}{2}}M_{R}(t^{\rho})\iota M_{Q}(t^{\rho}q^{R})
\end{equation}
This amplitude will be particularly important for studying the genus \(g=0\) OSV conjecture in section \ref{sec:genuszero}.

It is important to note that the fiber brane has a modulus, \(k_{f}\).  If we take \(u\) to be a coordinate for one of the fibers, then the fiber brane sits at \(\vert u \vert^{2}=const\).  As is standard, this real modulus combines with the holonomy to form the complexified K\"ahler parameter, \(k_{f}\).  Including this modulus simply modifies the partition function as,

\begin{equation}
M_{R}(U) \to e^{-k_{f}\vert R \vert}M_{R}(U)
\end{equation}
This modulus will appear in section \ref{sec:factor} when we discuss the ``ghost branes'' that appear in tests of the refined OSV conjecture.

\subsection{Refined Topological Strings on $\mathcal{L}_{1}\oplus\mathcal{L}_{2}\to \Sigma_{g}$ and $5d$ $U(1)$ Gauge Theories\label{sec:u1engineering}}
Now that we have defined a TQFT that computes refined topological string amplitudes, we would like to verify our proposal.  A simple check is that the Gopakumar-Vafa invariants are integers.  We have verified this in general, and we present a few examples in Appendix \ref{sec:gv}.

We can perform a much stronger check of our proposal by using geometric engineering.  We consider M-theory on the Calabi-Yau, \(X=\mathcal{O}(p)\oplus\mathcal{O}(2g-2-p)\to \Sigma_{g}\).  This is known to engineer five dimensional \(U(1)\) gauge theory with \(g\) hypermultiplets in the adjoint representation, and with a level \(k_{CS}=1-g-p\) five-dimensional Chern-Simons term turned on \cite{Katz:1996ht, Chuang:2010ii, Chuang:2012dv}.\footnote{The motivic Donaldson-Thomas invariants of these geometries were also studied recently in \cite{Chuang:2010ii, Chuang:2012dv}.  In general, the motivic invariants of a Calabi-Yau, \(X\), will differ from the refined invariants that we compute in this paper.  Motivic invariants depend on the motive of X and thus are sensitive to its complex structure.  In contrast, our refined invariants are computed by a supersymmetric index which makes them invariant under complex structure deformations.  These differences are reflected in the connection with geometric engineering.  In \cite{Chuang:2010ii, Chuang:2012dv}, the motivic invariants for these geometries were related to the instanton partition function with the adjoint mass equal to \(m=(\epsilon_{1}-\epsilon_{2})/2 \leftrightarrow \widetilde{y}=\sqrt{q/t}\).  Our refined invariants are identified with the different parameter choice, \(m=0 \leftrightarrow \widetilde{y}=1\).  We thank Emanuel Diaconescu for helpful discussions on this point.}

%

Now we consider the K-theoretic equivariant instanton partition function for these theories.\footnote{Ordinarily, such counting would not be sensible because \(U(1)\) instantons are singular and because the adjoint representation of \(U(1)\) is trivial.  However, this instanton counting is performed by turning on background noncommutativity which both resolves \(U(1)\) instantons and causes fields in the \(U(1)\) adjoint representation to transform nontrivially.} The original index in \cite{Nekrasov:2002qd} that computes the K-theoretic instanton partition function is exactly the same index that we have used to compute the refined A-model in equation \ref{eqn:refindex}, so the two partition functions must agree.

As explained in \cite{Chuang:2012dv}, the instanton partition function for this five-dimensional field theory is given by,
\begin{eqnarray}
Z_{U(1)}^{g,k_{CS}}(q,t,\widetilde{Q}) & = & \sum_{\mu}\prod_{\tableau{1}\in \mu}\Big(q^{-l(\tableau{1})-1/2}t^{a(\tableau{1})+1/2}\Big)^{k_{CS}}\Big(1-q^{-l(\tableau{1})}t^{-a(\tableau{1})-1}\Big)^{g-1} \nonumber \\
& & \cdot\Big(1-q^{l(\tableau{1})+1}t^{a(\tableau{1})}\Big)^{g-1}(t/q)^{\frac{(g-1)\vert \mu \vert}{2}} \widetilde{Q}^{\vert \mu\vert}
\end{eqnarray}
where \(q\) and \(t^{-1}\) are the equivariant parameters rotating the \(z_{1}\) and \(z_{2}\) planes respectively.  The sum is over all Young Tableaux, \(\mu\), and the arm and leg length of a box in such a tableau (defined in equation \ref{eqn:armlength}) are denoted by \(a(\tableau{1})\) and \(l(\tableau{1})\), respectively.  

By using the definitions of the metric and $(q,t)$-dimension in equations \ref{eqn:uinftymetric} and \ref{eqn:uinftyqtdim}, we can rewrite the equivariant instanton partition function as,

\begin{eqnarray}
Z_{U(1)}^{g,k_{CS}}(q,t,\widetilde{Q}) & = & \sum_{\mu}\Bigg(\frac{\textrm{dim}_{q,t}(\mu)^{2}}{g(\mu)}\Bigg)^{1-g} \Big(q^{-\frac{1}{2}\vert\vert \mu \vert \vert^{2}}t^{\frac{1}{2}\vert\vert \mu^{T} \vert \vert^{2}}\Big)^{k_{CS}+1-g}(-1)^{\vert R\vert}\widetilde{Q}^{\vert \mu \vert}
\end{eqnarray}
But now it is clear that this agrees with the refined topological string partition function of equation \ref{eqn:reftopgeneral},
\begin{equation}
Z_{U(1)}^{g,k_{CS}}(q,t,\widetilde{Q}) = Z_{\textrm{ref top}}^{g,p}(q,t,Q)
\end{equation}
upon making the change of variables,
\begin{eqnarray}
\widetilde{Q} & = & (-1)^{g-1}Q \nonumber \\
k_{CS} & = & 1-g-p \nonumber
\end{eqnarray}
This verifies in general our proposed refinement of the Bryan-Pandharipande TQFT for arbitrary line bundles over a Riemann surface.

\section{Refined Black Hole Entropy \label{sec:branes}}
In this section we study BPS bound states of $N$ D4 branes wrapping a four-cycle inside a Calabi-Yau, and carrying D2 and D0 brane charge.  We start by explaining that the refined counting of D4/D2/D0-brane BPS bound states computes the \(\chi_{y}\)-genus of the moduli space of instantons on the four-cycle wrapped by the D4 branes.  We then specialize to the case of interest in this paper -- IIA string theory compactified to four-dimensions on the class of Calabi-Yau manifolds, \(X\), that consist of two complex line bundles over a Riemann surface, and show how to compute the $\chi_y$ genus in the examples that arise there.  Finally, as a check of our results in this section,
we compare our answers in the case when the four-cycle is ${\cal O}(-1)\rightarrow {\mathbb P}^1$ 
against a direct computation of the cohomologies of the moduli space of instantons, by Yoshioka and Nakajima in \cite{Yoshioka:1996pd, Nakajima:2003uh}, and find a perfect agreement.

In the unrefined case, the black hole partition function is the index
\begin{equation}
Z_{BH} = \textrm{Tr}_{\mathcal{H}_{BPS}}(-1)^{F}e^{-\phi_{2}Q_{2}}e^{-\phi_{0}Q_{0}} \label{eqn:urtrace}
\end{equation}
where \(Q_{0}\) and \(Q_{2}\) are the D0 and D2 charges, while \(\phi_{0}\) and \(\phi_{2}\) are the respective chemical potentials.
Since we are working in the large volume limit, we can identify D0/D2/D4 bound states with nontrivial \(U(N)\) bundles, \(V\), over \(C_{4}\).  The D-brane charges and Chern classes of this bundle are related by,
\begin{equation}
Q_{2} = c_{1}(V), \qquad \qquad Q_{0} = ch_{2}(V)
\end{equation}
Therefore, calculating degeneracies will reduce to field theoretic computations on the D4-brane worldvolume.  
Since the D4 brane wraps \(\mathbb{R}_{t} \times C_{4}\), we can associate to \(C_{4}\) a Hilbert space\(\mathcal{H}\), which is graded by D2/D0-brane charge, angular momentum, \(J_{3}\), and R-charge, \(R\).   Now we would like to compute the BPS degeneracies as a trace over this entire Hilbert space.  This can be done easily by using the Witten index, since non-BPS contributions will cancel out.  Therefore we must simply compute the D4-brane path integral on \(S^{1} \times C_{4}\),
\begin{equation}
Z_{BH} = \textrm{Tr}_{\mathcal{H}}(-1)^{F}e^{-\phi_{2}c_{1}}e^{-\phi_{0}ch_{2}}
\end{equation}

Since the D-branes are wrapping a curved geometry, the gauge theory is topologically twisted along \(C_{4}\)\cite{Bershadsky:1995qy}.  In our case, \(\mathcal{H}_{BPS}\) is equal to the cohomology of instanton moduli space for the corresponding topological sector.  Therefore, computing the Witten index reduces to computing the euler characteristic, \(\chi(\mathcal{M})\), for the moduli space of instantons on \(C_{4}\).


We have presented this computation entirely from a five-dimensional perspective because this approach will easily generalize to the refined setting.  However, in the unrefined case we could also reduce on the \(S^{1}\) and study the four-dimensional gauge theory.  This leads to four-dimensional topologically twisted \(\mathcal{N}=4\) Yang-Mills \cite{Vafa:1994tf} on \(C_{4}\) with the observables,
\begin{equation}
S = \frac{\phi_{0}}{8\pi^{2}}\int {\rm Tr\,} F\wedge F + \frac{\phi_{2}}{2\pi}\int {\rm Tr\,} F \wedge \omega_{\Sigma}
\end{equation}
inserted into the action.  Here, \(\omega_{\Sigma}\) is the K\"ahler class of the Riemann surface, \(\Sigma_{g}\).  Since this is the Vafa-Witten \cite{Vafa:1994tf} twist of \(\mathcal{N}=4\), the four-dimensional perspective explains why we are computing the euler characteristic of instanton moduli space.

Now that we have discussed the unrefined case, we would like to count BPS states while keeping information about angular momentum and R-charge.  As explained in section \ref{sec:osv}, our goal is to compute the protected spin character of D2 and D0-branes bound to the D4-branes,
\begin{equation}
Z_{BH} = \textrm{Tr}_{\mathcal{H}_{BPS}}(-1)^{2J_{3}}y^{J_{3}-R}e^{-\phi_{2}Q_{2}}e^{-\phi_{0}Q_{0}} \label{eqn:rtrace}
\end{equation}
where we have used the variable, \(y\equiv e^{-2\gamma}\).  Since this is an index, it only receives contributions from BPS states. This means we can extend the trace to be over the full D4-brane Hilbert space ${\cal H}$,
\begin{equation}
Z_{BH} = \textrm{Tr}_{\mathcal{H}}(-1)^{2J_{3}}y^{J_{3}-R}e^{-\phi_{2}Q_{2}}e^{-\phi_{0}Q_{0}} \label{eqn:rtrace}.
\end{equation}
This also means that $Z_{BH}$ can be computed by the five-dimensional path integral and will be invariant under small deformations.\footnote{One should contrast this with the most general trace,
$$
Z_{BH} = \textrm{Tr}_{\mathcal{H}_{BPS}}(-x)^{J_{3}+R}(-y)^{J_{3}-R}e^{\phi_{2}Q_{2}-\phi_{0}Q_{0}} \label{eqn:double}
$$
where \(J_{3}\) is the generator of the \(Spin(3)\) rotation group in the (3+1)-dimensional spacetime, and \(R\) is the \(U(1)\) R-charge of the four-dimensional BPS states.  Unfortunately, this trace cannot be extended to the full Hilbert space, since non-BPS states will contribute nontrivially.  This means that the doubly-refined trace in equation \ref{eqn:double} cannot be computed by a five-dimensional path-integral, and is therefore analogous to the five-dimensional Khovanov-Rhozansky construction of \cite{Witten:2011zz}.}

To understand precisely what the protected spin character computes, it helps to remember that the Hilbert space \(\mathcal{H}_{BPS}\) can be identified with the cohomology of the moduli space \(\mathcal{M}\) of instantons on \(C_{4}\).  Once we fix the topological charges \(c_{1}\) and \(ch_{2}\), the most general geometric quantity that can be computed from the cohomology of \(\mathcal{M}\) is the Hodge polynomial,
\begin{equation}
e(\mathcal{M};x,y) = \sum_{p,q}(-1)^{p+q}x^{p}y^{q}\textrm{dim} H^{p,q}(\mathcal{M})
\end{equation} 
As explained in \cite{Diaconescu:2007bf}, the degrees, \((p,q)\) are related to the R-charge and spin by,
\begin{equation}
J_{3} = \frac{p+q}{2}, \qquad \qquad R = \frac{p-q}{2} \label{eqn:hodgespin}
\end{equation}
Therefore, the refined black hole partition function in equation \ref{eqn:rtrace} computes the generating function for the \(\chi_{y}\) genus of instanton moduli space,
\begin{eqnarray}
Z_{BH} & = & \sum_{c_{1},ch_{2}}e^{-\phi_{0}ch_{2}-\phi_{2}c_{1}}\sum_{p,q}(-1)^{p+q}y^{q}h^{p,q}(\mathcal{M}_{c_{1},ch_{2}}) \\
& = & \sum_{c_{1},ch_{2}}e^{-\phi_{0}ch_{2}-\phi_{2}c_{1}}\chi_{y}(\mathcal{M}_{c_{1},ch_{2}})
\end{eqnarray}

One more aspect of the protected spin character that we will need is its transformation properties under $S$-duality. In the unrefined case, the transformation properties are well known  \cite{Vafa:1994tf}. The partition function \ref{eqn:urtrace} transforms like a theta function, with modular parameter $\phi_0$.  We show in appendix \ref{sec:sduality} that in the refined case $S$-duality corresponds to replacing
\begin{equation}\label{Sduality}
\phi_{0} \to \frac{4\pi^{2}}{\phi_{0}}, \qquad \phi_{2} \to 2\pi i \frac{\phi_{2}}{\phi_{0}}, \qquad \gamma \to 2\pi i\frac{\gamma}{\phi_{0}}
\end{equation}
where the variable \(y\) in the \(\chi_{y}\) genus is related to \(\gamma\) by \(y = e^{-2\gamma}\).  In the rest of this section, we will show that, in the simple example of the family of Calabi-Yaus we have been studying, the $\chi_y$ genus of the instanton moduli space is computable explicitly in terms of a topological theory on the base Riemann surface $\Sigma$.
\subsection{D4 branes on ${\mathcal L}_1\oplus {\mathcal L}_2\rightarrow \Sigma$ \label{sec:d4x}}
In our local Calabi-Yau manifold,
$${\mathcal L}_1\oplus {\mathcal L}_2\rightarrow \Sigma_g$$
consider $N$ D4 branes wrapping the zero section of ${\mathcal L}_2$. The world-volume of the brane is
$${\cal D} =  ({\mathcal L}_1\to \Sigma_{g})$$
As before, we take ${\cal L}_1$ to have the first Chern class $-p$, so ${\mathcal L}_1$ is an $\mathcal{O}(-p)$ bundle over $\Sigma$.
In the unrefined case studied in\cite{Aganagic:2004js}, the partition function of the Vafa-Witten twisted \(\mathcal{N}=4\) $U(N)$ Yang-Mills on  ${\cal D}$
was shown to be computed by $q$-deformed two-dimensional bosonic Yang-Mills on \(\Sigma_{g}\). 
Roughly speaking, one can use localization on the fiber over the Riemann surface to reduce the four-dimensional theory down to a theory on $
\Sigma$. The basic observation is that one can use localization along the fiber of 
 \(\mathcal{O}(-p) \to \Sigma_{g} \) to reduce the four dimensional theory to a two dimensional theory on the fixed point set.
 This reduces the observables,
\begin{equation}
S =\frac{\phi_0}{8\pi^{2}}  \int {\rm Tr\,}  F\wedge F + \frac{\phi_2}{2\pi} \int {\rm Tr\,}  F \wedge \omega_{\Sigma}
\end{equation}
to
\begin{equation}
S = \frac{\phi_0}{4\pi^{2}} \int_{\Sigma_g}{\rm Tr\,}  \Phi \;  F + \frac{\phi_2}{2\pi} \int_{\Sigma_g} {\rm Tr\,} \Phi\; \omega_{\Sigma} - p \frac{\phi_0}{8\pi^{2}} \int_{\Sigma_g} \; {\rm Tr\,} \Phi^2 ,
\end{equation}
which is the action of the bosonic two dimensional Yang-Mills. Here, $\Phi$ is the holonomy of the circle at infinity of the $\mathbb{C}$ fiber -- the action becomes a boundary term. The last term reflects the topology of the fibration over $\Sigma_g$. The way it arises from four dimensions was explained in \cite{Aganagic:2004js}.  This is not quite the end of the story, as one has to be careful about the measure of the path integral. The fact that $\Phi$ comes from the holonomy around the $S^1$ turns it into a periodic variable --  this is why the theory is $q$-deformed 2d Yang-Mills, instead of ordinary 2d Yang Mills. In the limit where the \(D2\) brane chemical potential $\phi_2$ is turned off,  the \(q\)-deformed Yang-Mills on the above geometry reduces to an analytic continuation of ordinary Chern-Simons theory on a degree \(p\) \(S^{1}\) bundle over \(\Sigma_{g}\).  It is important to note that the ${\cal N}=4$ YM and Chern-Simons couplings are the same.

In this paper, we would like to solve for the corresponding refined amplitudes.  Since all of the arguments in the derivation so far were topological, the only thing that can change in a non-trivial way is the measure of the two-dimensional path integral.  While deriving the measure is straightforward in the unrefined case, it is more challenging in the refined theory. Instead, we will give pursue a different path. We will find another derivation of the fact that the Vafa-Witten partition function in this background is computed by $q$-deformed 2d Yang-Mills, in which understanding the deformation we need will be easy.
It will turn out that the theory we get is related to the refined Chern-Simons theory of \cite{Aganagic:2011sg, AS2}.

\subsection{From 4d ${\cal N}=4$ Yang-Mills to 2d $(q,t)$-deformed Yang-Mills \label{sec:n4qt}}
The idea of the derivation is to look at the same D4 brane background in a slightly different way -- by pairing the coordinates differently to get a Calabi-Yau four-fold instead. Doing so will make it manifest that, in the unrefined case, the theory we get is the same as Chern-Simons theory at $\phi_2=0$, or more generally, $2d$ $q$-deformed Yang Mills.

To begin with, we will consider the case \(p=0\) case, so the Calabi-Yau manifold is simply

$$ (\mathcal{O}(0) \oplus \mathcal{O}(2g-2) \to \Sigma_g ) = {\mathbb C} \times T^*\Sigma_g
$$
and the D4 branes wrap the divisor

$${\cal D} =  ({\cal O}(0)\to \Sigma_{g}) = {\mathbb C} \times \Sigma_g.$$

Thus, all together, the Vafa-Witten theory we are interested in, as we explained above, arises from studying the partition function of the $N$ D4-branes wrapping,
\begin{equation}
\mathbb{C} \times \Sigma_{g} \times S_t^{1} \label{Ob}
\end{equation}
in IIA string theory on,
\begin{equation}
\mathbb{C} \times T^{*}\Sigma_{g}\times \mathbb{R}^{3} \times S_t^{1} .\label{eqn:qspacetime}
\end{equation}
As we go around the temporal circle, \(S_t^{1}\), we compute the index,
\begin{equation}
Z = \textrm{Tr}(-1)^{F}e^{- \phi_0 Q_{0}}
\end{equation}
where \(Q_{0}\) is the \(D0\)-brane charge bound to the \(D4\)-branes.  We have temporarily set $\phi_2$ to zero. Our goal is to explain why this construction leads to the partition function of analytically continued Chern-Simons on \(S^{1} \times \Sigma_{g}\). Recall moreover that after a modular transformation, the partition function becomes manifestly equal to the partition function of Chern-Simons theory, with $q= e^{g_s} =e^{\frac{4\pi^2}{\phi_0}}$.

In \cite{Witten:2011zz}, analytically continued $SU(N)$ Chern-Simons theory on $S^1\times \Sigma_g$ was obtained from a string theory construction involving a stack of  $N$ D4 branes wrapping
\begin{equation}
  \mathbb{C} \times \Sigma_{g}\times S_t^{1}
\end{equation}
in IIA string theory on the Calabi-Yau fourfold $ T^{*}(\mathbb{C}\times \Sigma_{g}) $, or more precisely, on 
\begin{equation}
 T^{*}(\mathbb{C}\times \Sigma_{g}) \times \mathbb{R} \times S_t^{1} = \mathbb{C} \times T^{*} \Sigma_{g}  \times\mathbb{R}^{3} \times S_t^{1} \label{Wb}.
\end{equation}
At infinity of the D4 brane, we impose the D6 brane boundary conditions along
\begin{equation}
T^{*}(S^{1} \times \Sigma_{g}) \times \{0\} \times S^{1}_{t}
\end{equation}
We view the $\mathbb{C} \times \Sigma_{g}$ as the base of the cotangent space, and the D6 brane wraps the $S^1$ at the boundary of $\mathbb{C}$. 
The theory on the D4 branes (forgetting the temporal circle) on a Lagrangian cycle in a Calabi-Yau fourfold, is the Langlands twist of ${\cal N}=4$ $U(N)$ Yang-Mills on $ \mathbb{C} \times \Sigma_{g}$.
Upon going around the first \(S^{1}\), we compute the index on the \(D4\) brane worldvolume,
\begin{equation}
Z = \textrm{Tr}(-1)^{F}e^{-\phi_0' Q_{0}} \label{eqn:wittenindex}
\end{equation}
where $\phi_0' = \frac{4\pi^2}{\phi_0}=g_s$. Using string dualities, \cite{Witten:2011zz} argued that the theory on the D4 branes is $U(N)$ Chern-Simons theory, with $q=e^{g_s}.$ 

It may be surprising at first, but these two constructions are effectively the same. In the case studied in \cite{Witten:2011zz} one has $\mathcal{N}=4$ theory with the Langlands twist, while we are a priori interested in the Vafa-Witten twist. While the two are not the same on a generic four manifold $V$, if we take $V=\mathbb{C} \times \Sigma_{g}$, the difference disappears. We can argue that this is the case by recalling that topologically twisting merely implements the twisted version of supersymmetry imposed by the string background. As is manifest from \ref{Ob} and \ref{Wb}, the string backgrounds end up being the same in our case. The other apparent difference is that the Witten construction  involves \(D4\)-branes ending on a \(D6\)-brane at infinity.  In contrast, the first construction naively only involves \(D4\)-branes.  This discrepancy can be resolved by remembering that in the our setup, we still must impose boundary conditions at infinity on the noncompact \(\mathbb{C}\).  If we were to choose the boundary conditions $S$-dual to those of the \(D6\)-brane boundary conditions for the \(S^{1}\) at infinity, then the two setups agree. The S-duality is here simply to account for the fact that with D6 brane boundary conditions it is $q=e^{g_s}$ that keeps track of the instanton charge, where $q$ is parameter in terms of which the one naturally writes the Chern-Simons amplitudes -- while with the S-dual boundary conditions instead, it is $e^{-1/g_s}$  that keeps track of the instanton charge in the gauge theory on the four-manifold. 

As preparation for understanding the refined theory (and to ultimately make contact with the definition of refined Chern-Simons in \cite{Aganagic:2011sg}), it is helpful to also consider the unrefined setup in a slightly different geometry.  On very general grounds \cite{Witten:2011zz}, we expect the partition function of Langlands-twisted $SU(N)$ \(\mathcal{N}=4\) theory on a four-manifold \(V\), to be equal to the partition function of the $SU(N)$ Chern-Simons theory on the boundary \(\partial V\) of the four manifold.  The choice of the bulk geometry, \(V\), only potentially affects the integration contour of the analytically continued Chern-Simons partition function.  In the previous setup we studied \(V= \mathbb{C} \times \Sigma\), but instead we could choose \(V = \mathbb{R}_{+} \times S^{1} \times \Sigma\).  More explicitly we could take IIA string theory on the geometry,
\begin{equation}
 T^{*}({\mathbb C}^* \times \Sigma) \times \mathbb{R} \times S_{t}^{1} = \mathbb{C}^{*} \times T^{*} \Sigma  \times\mathbb{R}^{3} \times S_t^{1} \label{W2}
\end{equation}
with \(D4\) branes wrapping
\begin{equation}
\mathbb{R}_{+}\times S^{1} \times \Sigma\times S^{1}_{t}
\end{equation}
and with D6 brane boundary condition along,
\begin{equation}
\{0\} \times T^{*}(S^{1}\times \Sigma) \times \{0\} \times S^{1}_{t}.
\end{equation}
This is the more familiar realization of Chern-Simons that appears in the study of topological strings and in \cite{Witten:2011zz}.
%

Now we would like to understand the effect of refinement on these setups. Firstly, as we argued, the setups are effectively the same for the purposes of the index, so if we understand refinement in any one of these, we will have understood it in all the others as well. Moreover, note that already in the unrefined case, the first and the second (or third) setup, are related by $S$-duality. Thus, if, 
in the second and third constructions, we are computing the index 

\begin{equation}
Z_{ref} = \textrm{Tr}(-1)^{F}\exp\Bigg(-\phi_0'Q_{0} - 2\gamma'(J_{3}-R)\Bigg),
\end{equation}
where \(J_{3}\) rotates the \(\mathbb{R}^{3}\) spacetime and the R-symmetry acts geometrically by rotating the fiber of \(T^{*}\Sigma_{g}\), the index in the first setup is related to this by S-duality, \eqref{Sduality}

\begin{equation}\label{eq:smap}
\phi_{0}' \;\;\to\;\; \phi_0 = \frac{4\pi^{2}}{\phi_{0}'},  \qquad \gamma' \;\;\to\;\; \gamma = 2\pi i \frac{\gamma'}{\phi_{0}'}
\end{equation}
 and equals

\begin{equation}
Z_{ref} = \textrm{Tr}(-1)^{F}\exp\Bigg(-\phi_0Q_{0} - 2\gamma(J_{3}-R)\Bigg).
\end{equation}

This would in principle simply provide alternate setups to compute the index, but it would not save us the work of actually evaluating it. Fortunately, however, in the third setup, the index was already computed.  The problem of evaluating the refined index in this context was solved in \cite{Aganagic:2011sg, Aganagic:2012au}, in terms of the refined Chern-Simons theory on \(S^{1} \times \Sigma_{g}\). 
Since all three different setups, with the identification of parameters as in \ref{eq:smap} give rise to the same partition function, we conclude that the partition function in the first setup is simply the refined Chern-Simons partition function! 
For $\gamma'=0$, refined Chern-Simons becomes the same as ordinary Chern-Simons theory, analytically continued away from the integer level. 
As shown in \cite{Aganagic:2004js}, this in turn is the same as the 2d $q$-deformed Yang-Mills theory on $\Sigma$, upon reduction on the $S^1$ factor. Thus, we have derived the result of \cite{Aganagic:2004js}, by different means. Moreover, we have explained how to generalize it to the refined case.
We will explain below that there is a two dimensional theory theory related to refined Chern-Simons theory the same way the $q$-deformed Yang -Mills is related to the ordinary Chern-Simons theory; we will call this theory the $q,t$-deformed 2d Yang-Mills.

The only thing that remains to do is identify $\phi_0$ and $\gamma$, with the parameters $q,t$ that appear in the refined Chern-Simons partition function.  To do this, we need to back up slightly, and recall how the refined Chern-Simons theory was defined originally. The refined Chern-Simons partition function is defined as the index of M-theory in the background that arises by simply uplifting the third setup to M-theory. In this case, D6 branes lift to Taub-Nut space, and D4 branes lift to M5 branes. All together, we get M-theory on
\begin{equation}
T^*(S^1\times \Sigma)\times TN \times S^{1} ={\mathbb C}^* \times T^*\Sigma\times TN \times S^{1}
\end{equation}
with M5 branes on 
\begin{equation}
S^1\times \Sigma\times {\mathbb C} \times S^{1} 
\end{equation}
In addition, as we go around the $S^1$, the Taub-NUT is twisted by,
\begin{eqnarray}
(z_{1}\, , \, z_2 )  \qquad \to \qquad (q\, z_{1} , \,t^{-1}\,z_{2} ).
\end{eqnarray}
So, from this perspective, the index we are computing takes the form similar to that in equation \eqref{eqn:refindex}:

\begin{equation}
Z_{ref} = \textrm{Tr}(-1)^{F} q^{S_1 - R} t^{R-S_2} ,
\end{equation}
where $S_1$ rotates the complex plane $z_1$ wrapped by the M5 brane and $S_2$ generates the rotation of the $z_2$ plane, transverse to M5 brane and $R$ is another R-symmetry, coming from rotations of the fiber of $T^*\Sigma$. Moreover, let

$$
q=e^{-\epsilon_1} ,\qquad t=e^{-\epsilon_2}.
$$
We will see that these will turn out to be exactly the $\epsilon_1$ and the $\epsilon_2$ parameters that arize in the refined topological string.
If we reduce this to IIA, we get the third setup back with the definition of refined Chern-Simons we started the discussion with.  Naively, following arguments similar to those in section 2, one would have expected that we simply have $e^{-\phi_0'} = \sqrt{qt}=e^{-\frac{\epsilon_1+\epsilon_2}{2}}$ while $e^{\gamma' }= \sqrt{q/t}=e^{\frac{\epsilon_1-\epsilon_2}{2}}$. However, this would have treated $q$ and $t$ symmetrically, while  in refined Chern-Simons theory the symmetry is badly broken.
The fact that it is broken is very natural from the M-theory perspective, as on the M5 brane wrapping the $z_1$ plane $S_1$ corresponds to angular momentum, while $S_2$ is an R-symmetry, rotating the space transverse to the brane. Correspondingly, had we considered the M5 brane wrapping the $z_2$ plane instead, $S_2$ would have been the momentum on the brane.

To reconcile these two perspectives, one from M-theory with M5 branes and the other from IIA with D4 branes, we must understand how the choice of \(q/t\)-branes appears in the IIA geometry.  From the geometry of Taub-NUT space, it can be seen that this corresponds to whether the \(D4\)-brane runs along the positive or negative half-line in \(\mathbb{R}^{3}\).  More precisely, the choice of \(q/t\)-branes is translated into whether it fills,
\begin{equation}
\{0\} \times \{z>0\} \in \mathbb{R}^{3} \qquad \textrm{or} \qquad \{0\}\times \{z<0\} \in \mathbb{R}^{3}
\end{equation}

Now to see how this affects the identification of \(D0\)-brane charge, recall that D0-branes are magnetically dual to D6-branes.  In the presence of both kind of branes, the electromagnetic fields carry one unit of angular momentum along the vector connecting their positions.  But the D6-brane in our geometry is frozen at \(\{0\}\in \mathbb{R}^{3}\) and the D0-branes that bind to the \(D4\)-brane must sit at \(\{z>0\}\) or \(\{z<0\}\) depending on the type of refined brane.  Therefore, we find that the D0-brane always carries either \(+\frac{1}{2}\) unit of angular momentum, \(J_{3}\), or \(-\frac{1}{2}\) unit of angular momentum, depending on the type of \(D4\)-branes used.  When we compute the above trace, it is natural to only count the angular momentum that comes from other physics, and absorb this intrinsic angular momentum into the weighting of D0-brane charge.  From equation  this implies that each D0-brane is weighted by
\begin{equation}
\big(\sqrt{qt}\big)^{Q_{0}}\big(\sqrt{q/t}\big)^{Q_{0}} = q^{Q_{0}} \qquad \textrm{or} \qquad \big(\sqrt{qt}\big)^{Q_{0}}\big(\sqrt{q/t}\big)^{-Q_{0}} = t^{Q_{0}} 
\end{equation}
in perfect agreement with the M-theory perspective. Putting

$$ q=e^{-\epsilon_{1}}\qquad  t=e^{-\epsilon_{2}}.
$$
this implies we should identify (for the D4 brane ending from  \(\{z>0\}\)  on the D6 brane)

$$
\phi_0'=\epsilon_1', \qquad \gamma'=\frac{\epsilon_1-\epsilon_2}{2}.
$$
We can use this, and $S$-duality,  to identify the parameters $\phi_0$ and $\gamma$ in terms of $q$ and $t$, as:

$$
\phi_0=\frac {4 \pi ^2}{\epsilon_1}, \qquad \gamma= \pi i \frac{\epsilon_1-\epsilon_2}{\epsilon_1}
$$

So far our discussion has focused on the \(p=0\) case.  Once we consider nontrivial circle or line bundles over \(\Sigma\), the setups will differ since the Vafa-Witten and Langlands twists are not equivalent when \(p\neq 0\).  As discussed in \cite{Aganagic:2012au}, the framing factors in refined Chern-Simons can be understood as arising from a topological term.  We expect that such topological terms should be present, regardless of which setup we use.  Finally, we should remember that our goal is to count both \(D0\)-brane charge and \(D2\)-brane charge.  Therefore, we must reintroduce a term in the index, \(e^{-\phi_{2}Q_{2}}\).  This term is unaffected by refinement and takes precisely the same form as before.

\subsection{Refined Chern-Simons Theory and a  $(q,t)$-deformed Yang-Mills \label{sec:path}}

To summarize, the refined black hole partition function, 

\begin{equation}
Z_{ref} = \textrm{Tr}(-1)^{F}\exp\Bigg(-\phi_0Q_{0} - 2\gamma(J_{3}-R)\Bigg).
\end{equation}
corresponding to $N$ D4 branes on the divisor

$${\cal D} =  ({\cal O}(-p)\to \Sigma_{g})$$
is computed by refined $U(N)$ Chern-Simons theory on an $S^1$ bundle over $\Sigma_g$ of first Chern-Class $-p$, where the $q=e^{-\epsilon_1},t=e^{-\epsilon_2}$ parameters of refined Chen-Simons are related to $\phi_0$ and $\gamma$ as

\begin{equation}
\epsilon_1 = \frac{4\pi^2}{\phi_{0}} , \qquad \epsilon_2 =  {4\pi^2 \over \phi_0}(1-\frac{\gamma}{2\pi i}), \qquad \theta = \frac{2\pi\phi_{2}}{\phi_{0}} \label{eqn:cov}
\end{equation}

It will be useful for us to formulate the refined Chern-Simons theory as a two dimensional one, refining $q$-deformed  2d Yang-Mills, in particular since we still need to turn on $\phi_2$, the D2 brane chemical potential (we could have done this in the refined Chern-Simons theory as well using the natural contact structure, but we will not do that here). As we explained earlier, the topological terms in four dimensional action are unchanged by refinement; only the values of $\phi_0$, $\phi_2$ change.

\begin{equation}
S =\frac{\phi_0}{8 \pi^2}  \int {\rm Tr \, }F\wedge F + \frac{\phi_2}{2\pi} \int {\rm Tr \, }F \wedge \omega_{\Sigma}
\end{equation}
Namely,
localization on the ${\cal L}_1 = {\cal O}(-p)$ fiber, relates the 4d observables,
to 2d ones

\begin{equation}
S = \frac{\phi_0}{4\pi^{2}} \int_{\Sigma_g}{\rm Tr \, }  \Phi \;  F + \frac{\phi_2}{2\pi} \int_{\Sigma_g}{\rm Tr \, } \Phi\; \omega_{\Sigma} - p\frac{\phi_{0}}{8\pi^{2}} \int_{\Sigma_g} {\rm Tr \, }\Phi^2\; \omega_{\Sigma}. 
\end{equation}
Here, $\Phi$ is the holonomy of the four-dimensional gauge field around the circle at infinity of the ${\cal L}_1$ fiber.
We can also think of it as the holonomy of the Chern-Simons gauge field around the $S^1$. The origin of the last term, from the Chern-Simons perspective was reviewed in \cite{Aganagic:2012ne, Aganagic:2012au}. Here $\omega_{\Sigma}$ is a volume form on $\Sigma_g$, normalized to unit volume. The presence of this form in the action means that the 2d YM is invariant under area preserving diffeomorphisms only. 

Thus, we still get the action of the ordinary 2d Yang-Mills, but the measure has to be deformed -- both because of the periodicity of $\Phi$, and now because also of the $q,t$ dependence of the index. The measure factor was in fact the only difference between the refined and ordinary Chern-Simons theory, as well. Let ${\cal D}_{q,t} A$ be the refined Chern-Simons measure. This induces a measure on the gauge fields in two dimensions, but also on the holonomy. We will explain what the measure is in some detail later on, for now, let us leave it schematic. All together, the path integral of the theory is

$$
Z_{\textrm{ref BH}}=\int {\cal D}_{q,t} A\;{\cal D}_{q,t} \Phi\; \exp\Bigl( \frac{\phi_0}{4\pi^{2}} \int_{\Sigma_g}{\rm Tr \, }  \Phi \;  F + \frac{\phi_2}{2\pi} \int_{\Sigma_g}{\rm Tr \, } \Phi\; \omega_{\Sigma} - p\frac{\phi_{0}}{8\pi^{2}} \int_{\Sigma_g} {\rm Tr \, }\Phi^2\; \omega_{\Sigma}. \Bigr)
$$
We will refer to this theory as \((q,t)\)-deformed Yang-Mills theory.\footnote{This \((q,t)\)-deformed Yang-Mills theory has also appeared as a limit of the TQFT that computes the four-dimensional \(\mathcal{N}=2\) superconformal index in \cite{Gadde:2011uv}.  It is also worth noting that in the limit \(q\to 1, t \to 1\) this theory reduces to ordinary 2d Yang-Mills at zero coupling.}  

 We will use this path integral to derive the answer for the partition function, in the next subsection. For now, let us simply state the answer: For $N$ D4 branes wrapping a  degree \(-p\) complex line bundle fibered over a genus \(g\) Riemann surface \(\Sigma_{g}\), the resulting refined partition function for bound states with D2-D0 branes is given by,

\begin{equation}
Z_{\textrm{ref BH}}= \sum_{\mathcal{R}}\Bigg(\frac{(S_{\mathcal{R}0})^2}{G_{\mathcal{R}}}\Bigg)^{1-g}
 q^{\frac{p(\mathcal{R},\mathcal{R})}{2}}t^{p(\rho,\mathcal{R})}Q^{\sum_{i}\mathcal{R}_{i}} \label{eqn:qtym11}
 \end{equation}
where the sum is over representations, \(\mathcal{R}\), of \(U(N)\), \(\rho\) is the Weyl vector \((\rho)_{i}=\frac{N+1}{2}-i\), and \(Q=e^{-\phi_{2}}\) is the \(D2\)-brane chemical potential.  Here we have also used the definitions,
\begin{eqnarray}
S_{\mathcal{R}0} = S_{00}\textrm{dim}_{q,t}(\mathcal{R}) & = & \prod_{m=0}^{\beta-1}\prod_{1\leq i<j \leq N} \lbrack \mathcal{R}_{i} - \mathcal{R}_{j} + \beta(j-i)+m \rbrack_{q} \nonumber \\
G_{\mathcal{R}} & = & \prod_{m=0}^{\beta-1}\prod_{1 \leq i<j \leq N}\frac{ \lbrack \mathcal{R}_{i}-\mathcal{R}_{j}+\beta(j-i)+m\rbrack_{q} }{\lbrack \mathcal{R}_{i}-\mathcal{R}_{j}+\beta(j-i)-m\rbrack_{q}}
\end{eqnarray}
Note that \(G_{\mathcal{R}}\) is naturally the finite N version of the metric that appeared in equation \ref{eqn:infmetric}.


\subsection{A Path Integral for $(q,t)$-deformed Yang-Mills \label{sec:path}}

Let us now explain in more detain what the $q,t$-deformed 2d Yang -Mills is, and how to compute its partition function. 
The most straight-forward way to proceed is to note that, because the theory is essentially topological, we can simply formulate the theory on pieces of the Riemann surface, explain how to glue them together, and show that the answer is independent of the decomposition. This has essentially been done in \cite{Aganagic:2011sg, Aganagic:2012ne}, only from the 3d perspective of the refined Chern-Simons theory on $S^1$ fibration over the Riemann surface.  To avoid simply repeating the derivation of \cite{Aganagic:2011sg, Aganagic:2012ne}, we will instead compute the path integral directly. We will begin by recalling some of the results of  \cite{Aganagic:2004js}, where the \(U(N)\) \(q\)-deformed 2d Yang-Mills theory was studied. This section will not be entirely self contained, but will build on  \cite{Aganagic:2004js}.

In \cite{Aganagic:2004js}, it was shown that the path integral of the unrefined theory 
can be abelianized, so that we are left with \(U(1)^{N}\) gauge fields, \(A_{k}\), and \(N\) compact scalar fields, \(\phi_{k}\), which are the eigenvalues of $\Phi$.  Starting from the original integral:

$$
Z_{\textrm{ BH}}=\int {\cal D} A\;{\cal D} \Phi\; \exp\Bigl( \frac{1}{g_s} \int {\rm Tr \, }  \Phi \;  F + \frac{\theta}{g_s}  \int{\rm Tr \, }  \Phi\; \omega_{\Sigma} - \frac{p}{g_s} \int \;  {\rm Tr \, }  \Phi^2\; \omega_{\Sigma}. \Bigr)
$$
the abelianized version becomes

\begin{equation}
Z_{\textrm{qYM}}= \frac{1}{N!}\int^{'}\prod_{i}\mathcal{D}\phi_{i}\mathcal{D}A_{i} \Big(\Delta(\phi)\Big)^{1-g}\exp\Bigg(\sum_{i} {1\over g_s} \int_{\Sigma}d^{2}\sigma \Big(\frac{p}{2}\;\phi_{i}^{2}-{\theta} \;\phi_{i}\Big) -\frac{1}{g_s} \int_{\Sigma} F_{i}\phi_{i}\Bigg) \label{eqn:path}
\end{equation}
where we have used the measure factor,

\begin{equation}
\Delta(\phi) = \prod_{i\neq j}\Big(e^{\frac{\phi_{i}-\phi_{j}}{2}}-e^{\frac{\phi_{j}-\phi_{i}}{2}} \Big)
\end{equation}
and where \(\int^{'}\) indicates that the path integral omits those values of \(\phi\) for which \(\Delta(\phi)=0\).  As was argued in \cite{Aganagic:2004js}, this partition function is precisely equal to the black hole partition function of equation \ref{eqn:urtrace} with the identification,

\begin{equation}
\phi_{0} = \frac{4\pi^{2}}{g_s}, \qquad \qquad \phi_{2} = \frac{2\pi \theta}{g_s}
\end{equation}
Since this action is quadratic and very simple, the path integral of $q$-deformed Yang-Mills can be solved exactly.

As we discussed, in the refined case, the only thing that changes are the chemical potentials and the measure factors. The chemical potentials become

\begin{equation}
\phi_{0} = \frac{4\pi^{2}}{\epsilon_1}, \qquad \qquad \phi_{2} = \frac{2\pi \theta}{\epsilon_1}
\end{equation}
and $\gamma$ enters through the measure of the path integral that depends on both $\epsilon_1$ and $\epsilon_2$:

\begin{equation}
 \epsilon_2 =  {4\pi^2 \over \phi_0}(1-\frac{\gamma}{2\pi}) . 
\end{equation}
In the refined theory, the measure becomes 

\begin{equation}
\Delta(\phi) \to \Delta_{q,t}(\phi) = \prod_{m=0}^{\beta-1}\prod_{j\neq k}
\Big(q^{-{\frac{m}{2}}}e^{\frac{\phi_{i}-\phi_{j}}{2}}- q^{{\frac{m}{2}}}e^{\frac{\phi_{j}-\phi_{i}}{2}} \Big)
 \label{eqn:measure}
\end{equation}
where \(\beta=\epsilon_{2}/\epsilon_{1}\), and we have taken it to be a positive integer for computational convenience.
The path integral is thus given by:

\begin{equation}
Z_{qt\textrm{YM}}(\Sigma) = \frac{1}{N!}\int^{'}\prod_{i}\mathcal{D}\phi_{i}\mathcal{D}A_{i} \Big(\Delta_{q,t}(\phi)\Big)^{1-g}\exp\Bigg(\sum_{i} \int_{\Sigma}d^{2}\sigma\Big(\frac{p}{2\epsilon_{1}}\phi_{i}^{2}-\frac{\theta}{\epsilon_{1}} \phi_{i}\Big)-\frac{1}{\epsilon_{1}} \int_{\Sigma} F_{i}\phi_{i}\Bigg)
\end{equation}

Since this path integral is abelian, we can evaluate it explicitly following the approach of \cite{Blau:1993tv, Aganagic:2004js}.  We begin by evaluating the path integral over abelian gauge fields.  It is helpful to first change integration variables from \(\mathcal{D}A_{i}\) to \(\mathcal{D}F_{i}\).  However, we should only integrate over those two-forms \(F_{k}\) that are genuine bundles over \(\Sigma_{g}\), which means that we must impose,
\begin{equation}
\int_{\Sigma_{g}}F_{k} \in 2\pi \mathbb{Z}
\end{equation}
This can be accomplished by inserting a periodic delta function,
\begin{equation}
\sum_{\{n_{k}\}}\delta^{(N)}\Big(\int_{\Sigma_{g}}F_{k} - 2\pi n_{k} \Big) 
\end{equation}
which can be rewritten using Poisson resummation,
\begin{equation}
\sum_{\{n_{k}\}}\delta^{(N)}\Big(\int_{\Sigma_{g}}F_{k} - 2\pi n_{k} \Big) = \sum_{\{n_{k}\}}\exp\Big(in_{k}\int_{\Sigma_{g}}F_{k}\Big)
\end{equation}
Therefore the path integral takes the form,
\begin{eqnarray}
Z_{qt\textrm{YM}}(\Sigma)  = \frac{1}{N!}\int^{'} & & \prod_{i}\mathcal{D}\phi_{k}\mathcal{D}F_{k} \Big(\Delta_{q,t}(\phi)\Big)^{1-g} \\
& & \cdot\exp\Bigg(\sum_{i} \int_{\Sigma}d^{2}\sigma\Big(\frac{p}{2\epsilon_{1}}\phi_{k}^{2}-\frac{\theta}{\epsilon_{1}} \phi_{k}\Big)-\frac{1}{\epsilon_{1}} \int_{\Sigma} F_{k}(\phi_{k} - i\epsilon_{1}n_{k}) \Bigg) \nonumber
\end{eqnarray}
When performing the path integral over \(F_{k}\) we obtain a delta function from the last term, and the path integral over \(\phi\) only receives contributions from constant fields,
\begin{equation}
\phi_{k} = i\epsilon_{1}n_{k}
\end{equation}
Therefore, the path integral evaluates to,
\begin{equation}
Z_{qt\textrm{YM}}(\Sigma) = \frac{1}{N!}\sum_{\{n_{k}\}}'\Big(\Delta_{q,t}(i\epsilon_{1}n_{k})\Big)^{1-g}\exp\Big(-\frac{p\epsilon_{1}}{2}\sum_{k}n_{k}^{2}-i\theta \sum_{k} n_{k}\Big) \label{eqn:qtym1}
\end{equation}
where \(\sum'\) indicates that we should only sum over those field configurations where \(\Delta_{\epsilon_{1},\beta}(\phi)\) is nonzero.  From equation \ref{eqn:measure}, this means that we must impose \(n_{i} \neq n_{j}\), but we also require,
\begin{eqnarray}
n_{i} & \neq & n_{j} \pm 1 \nonumber \\
n_{i} & \neq & n_{j} \pm 2 \label{eqn:constraints} \\
& \vdots & \nonumber \\
n_{i} & \neq & n_{j} \pm (\beta-1) \nonumber 
\end{eqnarray}

It is also helpful to notice that both the refined measure and the action are invariant under Weyl reflections, so we can restrict the sum to the fundamental Weyl chamber so that \(n_{1}> n_{2} > \ldots > n_{N}\).  

We would like to use these observations to rewrite equation \ref{eqn:qtym1} as a sum over \(U(N)\) representations.  Recall that a \(U(N)\) representation is specified by highest weights satisfying \(\mathcal{R}_{1} \geq \mathcal{R}_{2} \geq \ldots \geq \mathcal{R}_{N}\).  Therefore, in order to satisfy the constraints of equation \ref{eqn:constraints}, we should shift each \(n_{k}\) by \(\beta k\).  For convenience, we can also shift all \(n_{k}\) by a constant amount.  This leads us to the identification,
 \begin{equation}
 n_{k} = \mathcal{R}_{k}+\beta \rho_{k}
 \end{equation}
where \(\rho_{k} = \frac{N+1}{2}-k\).  Then the partition function takes the form,
\begin{equation}
Z_{qt\textrm{YM}}(\Sigma) = \sum_{\mathcal{R}}\Big(\Delta_{q,t}(\mathcal{R})\Big)^{1-g} q^{\frac{p}{2}(\mathcal{R},\mathcal{R})}t^{p(\rho,\mathcal{R})}Q^{\vert \mathcal{R} \vert}
\end{equation}
where we have defined \(q=e^{-\epsilon_{1}}\), \(t=e^{-\beta\epsilon_{1}}\), and \(Q = e^{-i\theta}\), and where,
\begin{equation}
\Delta_{q,t}(\mathcal{R}) = \prod_{m=0}^{\beta-1}\prod_{i<j}\lbrack \mathcal{R}_{i}-\mathcal{R}_{j}+\beta(j-i)+m\rbrack_{q} \lbrack \mathcal{R}_{i}-\mathcal{R}_{j}+\beta(j-i)-m\rbrack_{q}
\end{equation}
where we have used the notation, \(\lbrack n \rbrack_{q} = q^{n/2}-q^{-n/2}\).  

We can now rewrite \(\Delta_{q,t}(\mathcal{R})\) in a form that clarifies the relationship to refined Chern-Simons theory,
\begin{equation}
\Delta_{q,t}(\mathcal{R}) =\frac{S_{0{\mathcal{R}}}S_{0{\mathcal{R}}}}{G_{\mathcal{R}}} 
\end{equation}
where \(S_{PQ}\) is the S-matrix for refined Chern-Simons theory, but analytically continued away from integer level, and \(G_{R}\) is the finite $N$ Macdonald metric.  These elements take the form,
\begin{eqnarray}
S_{\mathcal{R}0} = S_{00}\textrm{dim}_{q,t}(\mathcal{R}) & = & \prod_{m=0}^{\beta-1}\prod_{1\leq i<j \leq N} \lbrack \mathcal{R}_{i} - \mathcal{R}_{j} + \beta(j-i)+m \rbrack_{q} \label{eqn:qtdimbeta} \\
G_{\mathcal{R}} & = & \prod_{m=0}^{\beta-1}\prod_{1 \leq i<j \leq N}\frac{ \lbrack \mathcal{R}_{i}-\mathcal{R}_{j}+\beta(j-i)+m\rbrack_{q} }{\lbrack \mathcal{R}_{i}-\mathcal{R}_{j}+\beta(j-i)-m\rbrack_{q}}\label{eqn:metricbeta}
\end{eqnarray}
Putting this together, we can rewrite the entire partition function as,
\begin{equation}
Z_{\textrm{ref BH}}= \sum_{\mathcal{R}}\frac{(S_{\mathcal{R}0})^{2-2g}}{(G_{\mathcal{R}})^{1-g}}
 q^{\frac{p(\mathcal{R},\mathcal{R})}{2}}t^{p(\rho,\mathcal{R})}Q^{\sum_{i}\mathcal{R}_{i}} 
\end{equation}
in agreement with equation \ref{eqn:qtym11}.  
%

Finally, for comparison with refined topological string theory, it is helpful to include an overall \(Q\)-independent normalization factor, 
\begin{equation}
\alpha_{BH} = \exp\Bigg(-\frac{\epsilon_{2}^{2}}{\epsilon_{1}}\frac{\rho^{2}(p+2g-2)^{2}}{2p}+\frac{N\theta^{2}}{2p\epsilon_{1}} + \frac{\epsilon_{2}^{2}}{\epsilon_{1}}(2g-2)\rho^{2} \Bigg)\Big((t;q)_{\infty}(q;q)_{\infty}\Big)^{N(g-1)} \label{eqn:alphanorm}
\end{equation}
Therefore, our final result takes the form,
\begin{equation}
Z_{qt\textrm{YM}}(\Sigma) =\alpha_{BH} \sum_{\mathcal{R}}\frac{(S_{\mathcal{R}0})^{2-2g}}{(G_{\mathcal{R}})^{1-g}}
 q^{\frac{p(\mathcal{R},\mathcal{R})}{2}}t^{p(\rho,\mathcal{R})}Q^{\sum_{i}\mathcal{R}_{i}} 
 \label{eqn:qtym}
\end{equation}

In conclusion, \((q,t)\)-deformed Yang-Mills gives a precise prediction for the refined black hole partition function for D4 branes wrapping an arbitrary complex line bundle over \(\Sigma\).  From the discussion above, this can also be rephrased as a mathematical prediction for the \(\chi_{y}\) genus of instanton moduli space.  

It is important to notice that,  as written, \(Z_{qt\textrm{YM}}\) is an expansion in \(q=e^{-\epsilon_{1}}\) and \(t=e^{-\epsilon_{2}}\), while the original black hole index of equation \ref{eqn:rtrace} should be expanded in \(e^{-1/\epsilon_{1}}\) and \(e^{-\gamma}\).  Therefore, to extract D4/D2/D0 refined degeneracies, we must use TST duality to resum the partition function \(Z_{BH}\) so that it is written in the appropriate expansion.

\subsection{Example: $\mathcal{O}(-1)\to \mathbb{P}^{1}$ \label{sec:o1}}
In this section we focus on D4 branes wrapping the bundle
 \(\mathcal{O}(-1)\) over a genus \(g=0\) Riemann surface.  
 $${\cal D}_0\;\; = \;\;\mathcal{O}(-1)\to\mathbb{P}^{1}
 $$
 This geometry is simply given by blowing up \(\mathbb{C}^{2}\) at a point, which means that the corresponding instanton moduli space is especially simple.  In the mathematical work of Yoshioka and Nakajima \cite{Yoshioka:1996pd, Nakajima:2003uh}, the authors proved an explicit formula for the Hodge polynomial of \(U(N)\) instantons on the blow-up geometry,
\begin{eqnarray}
P_{\textrm{blow-up}}(\phi_{0},\phi_{2};x,y) & = & \sum_{ch_{2},c_{1},i}e^{-\phi_{0}ch_{2}-\phi_{2} c_{1}}(-1)^{i+j}x^{i}y^{j}h_{i,j}(\mathcal{M}_{ch_{2},c_{1}}) \\
& = & \frac{1}{\eta(e^{-\phi_{0}})^{N}}\sum_{\{n_{i}\}=-\infty}^{\infty}e^{-\phi_{0}\frac{(n,n)}{2}}(xy)^{(\rho,n)}e^{-\phi_{2}\sum_{i}n_{i}} \label{eqn:yoshioka}
\end{eqnarray}
Note that since this geometry is toric, \(h_{i,j}\) is only nonzero if \(i=j\).\footnote{From equation \ref{eqn:hodgespin}, it follows that all D4/D2/D0 BPS states in this geometry have zero R-charge (\(R=0\)).  Therefore, the ordinary and protected spin characters agree.}.  Setting \(x=1\) we obtain the \(\chi_{y}\) genus of interest,
\begin{equation}
P_{\chi_{y}; \textrm{blow-up}}(\phi_{0},\phi_{2},y) = \frac{1}{\eta(e^{-\phi_{0}})^{N}}\sum_{\{n_{i}\}=-\infty}^{\infty}e^{-\phi_{0}\frac{(n,n)}{2}}y^{(\rho,n)}e^{-\phi_{2}\sum_{i}n_{i}} \label{eqn:yoshioka}
\end{equation}
    We would like to compare this answer against the prediction of \((q,t)\)-deformed Yang-Mills,
\begin{equation}
Z_{\textrm{ref}}({\cal D}_0) = \alpha_{BH} \sum_{n_{1}\geq n_{2}\geq\cdots\geq n_{N}}q^{\frac{(n,n)}{2}}t^{(\rho,n)}Q^{\sum_{i}n_{i}}\frac{\textrm{dim}_{q,t}(\{n_{i}\})^{2}}{g(\{n_{i}\})} 
\end{equation}
This partition function can be rewritten using the remarkable identity,\footnote{An analogue of this identity for the \(SU(N)\) case has been proven by Cherednik in \cite{Ch:MacMehta, Ch:DAHA, Ch:Diff}.}
\begin{equation}
\sum_{n_{1}\geq n_{2}\geq\cdots\geq n_{N}}q^{\frac{(n,n)}{2}}t^{(\rho,n)}Q^{\sum_{i}n_{i}}\frac{\textrm{dim}_{q,t}(\{n_{i}\})^{2}}{g(\{n_{i}\})} = \Bigg(\prod_{k=1}^{N-1}\prod_{j=1}^{\infty}\frac{1-q^{j}t^{k}}{1-q^{j}}\Bigg)\sum_{n_{i}=-\infty}^{\infty}q^{\frac{(n,n)}{2}}t^{(\rho,n)}Q^{\sum_{i}n_{i}}
\end{equation}

Since the summation on the right is over all integers \(\{n_{i}\}\), we can shift the definition of \(n\) by any integer amount.  Using this freedom we can finally write the partition function as,
\begin{equation}
Z_{\textrm{ref}}({\cal D}_0)= \alpha_{BH}(q,t)\Bigg(\prod_{k=1}^{N-1}\prod_{j=1}^{\infty}\frac{1-q^{j}t^{k}}{1-q^{j}}\Bigg)\sum_{n_{i}=-\infty}^{\infty}q^{\frac{(n,n)}{2}}(t/q)^{(\rho,n)}Q^{\sum_{i}n_{i}} \label{eqn:blowup}
\end{equation}
The \(Q\)-independent prefactor in this expression is related to \(D4/D0\) degeneracies and is ambiguous because our geometry is non-compact.  For this reason, in subsequent formulas we will drop it.

To compare our result with the \(\chi_{y}\)-genus of instanton moduli space, we must remember that the partition function of \((q,t)\)-deformed Yang-Mills is an expansion in \(e^{-\epsilon_{k}}\), while D-brane degeneracies arise as coefficients of an expansion in \(e^{-\frac{1}{\epsilon_{1}}}\).  To relate these two expansions, we can rewrite \(Z_{\textrm{ref}}\) as a product of Jacobi theta functions,
\begin{equation}
Z_{\textrm{ref}}({\cal D}_0)\sim \prod_{j=1}^{N}\vartheta\Big(\frac{(\epsilon_{1}-\epsilon_{2})(N+1-2j)}{4\pi i} - \frac{\theta}{2\pi},-\frac{\epsilon_{1}}{2\pi i})
\end{equation}
where the Jacobi theta function is defined by \(\vartheta(z,\tau) = \sum_{n}e^{\pi i \tau n^{2} + 2\pi i n z}\).
It is helpful to recall that the Jacobi theta function has the modular property,
\begin{equation}
\vartheta\Big(\frac{z}{\tau},-\frac{1}{\tau}\Big) = (-i\tau)^{1/2}\exp\Big(\pi i z^{2}/\tau\Big)\vartheta(z,\tau)
\end{equation}
Applying this transformation to the black hole partition function we obtain,
\begin{equation}
Z_{\textrm{ref}}({\cal D}_0)\sim \sum_{n_{i}=-\infty}^{\infty}e^{-\phi_{0} \frac{(n,n)}{2}}e^{-\phi_{2}\sum_{i}n_{i}}e^{-2\gamma(\rho,n)} \label{eqn:goodo1}
\end{equation}
where \(\phi_{0}\), \(\phi_{2}\), and \(\gamma\) are given by the definitions in equation \ref{eqn:cov}.
As discussed above, physically this modular transformation arises from performing TST duality on the D-brane configuration.

This result precisely agrees with the expected \(\chi_{y}\)-genus in equation \ref{eqn:yoshioka}.  In section \ref{sec:wc}, we will give an independent physical derivation of equation \ref{eqn:yoshioka} by using the refined semi-primitive wall crossing formula.






\section{Large N Factorization and the Refined OSV Conjecture \label{sec:factor}}
Now that we have explained how to compute both the refined black hole partition function and the refined topological string, we would like to see how they are connected.  As explained in section \ref{sec:osv}, the refined OSV conjecture predicts that the refined partition function of \(N\) D4-branes wrapping,
\begin{equation}
\mathcal{O}(-p) \to \Sigma_{g}
\end{equation}
should be equal to the square of the partition function of refined topological string on
\begin{equation}
\mathcal{O}(-p)\oplus\mathcal{O}(2g-2+p) \to \Sigma_{g}
\end{equation}
to all orders in the \(1/N\) expansion.  In equations, this implies
\begin{equation}
Z_{qt\textrm{YM}}(\phi_{0},\phi_{2},\gamma) \sim \vert Z_{\textrm{ref top}}(\epsilon_{1},\epsilon_{2},k) \vert^{2}
\end{equation}
with the change of variables,
\begin{eqnarray}
k & = & \frac{2\pi^{2}}{\phi_{0}}\Big(\beta P + i \frac{\phi_{2}}{\pi}\Big)\nonumber \\
\epsilon_{1} & = & \frac{4\pi^{2} C}{\phi_{0}} \label{eqn:cov2} \\
\frac{\epsilon_{2}}{\epsilon_{1}} & = & \beta = 1 - \frac{\gamma}{2\pi i} \nonumber
\end{eqnarray}
Note that we have included the additional constant factor, \(C\).  It will become clear (see equation \ref{eqn:cov}) that in our example the value for \(C\) is \(1\) so that,
\begin{equation}
\phi_{0} = \frac{4\pi^{2}}{\epsilon_{1}}
\end{equation}
This is in contrast with the more symmetric choice of
\begin{equation}
\phi_{0} \sim \frac{4\pi^{2}}{\epsilon_{1} + \epsilon_{2}}
\end{equation}
that was used in our discussion of the refined OSV conjecture in section \ref{sec:osv}.  This apparent discrepancy can be resolved by postulating that \(D0\)-branes in our setup carry intrinsic charge under \((J_{3}-R)\), which will change their effective weighting in the refined partition function.  As explained in section \ref{sec:n4qt}, this is expected since in the dual refined Chern-Simons construction, \(D0\)-branes carry angular momentum which causes them to be weighted by either \(q\) or \(t\), but not \(\sqrt{qt}\).

We can make the change of variables in equation \ref{eqn:cov2} even more explicit for the geometries we are studying.   It might seem that the D4-brane charge, \(P\), is simply the number of \(D4\) branes, \(N\), that wrap the bundle, \(C_{4} = \mathcal{O}(-p) \to \Sigma_{g}\).  However, we should really measure this charge in electric \(D2\) brane units.  As explained in \cite{Aganagic:2004js} these charges differ because of the nontrivial intersection number of the Riemann surface \(\Sigma\) with the four-cycle wrapped by the D4 branes, \(C_{4}\),
\begin{equation}
\#(\Sigma \cap C_{4}) = 2g-2+p
\end{equation}
leading to the identification,
\begin{equation}
P = N(2g-2+p)
\end{equation}

Now we want to test the above predictions by using the results that we have built up in sections \ref{sec:tqft} and \ref{sec:branes}, where we solved for both the refined black hole partition function and the refined topological string on these geometries.  We found in section \ref{sec:branes}, that the refined brane partition function is computed by the two-dimensional \((q,t)\)-deformed Yang-Mills, whose partition function depends on two coupling constants \((\epsilon_{1},\epsilon_{2})\) and a theta-term, \(\theta\).  As explained in equation \ref{eqn:cov}, these gauge theory variables are related to the refined black hole chemical potentials by,
\begin{equation}
\phi_{0} = \frac{4\pi^{2}}{\epsilon_{1}}, \qquad \qquad \phi_{2} = \frac{2\pi \theta}{\epsilon_{1}}, \qquad \qquad \gamma = \frac{2\pi i(\epsilon_{2}-\epsilon_{1})}{\epsilon_{1}}
\end{equation}
Putting this together with the predictions of the refined OSV conjecture, we find that the large \(N\) limit of \((q,t)\)-deformed Yang-Mills should be equal to the square of the refined topological string with the K\"ahler modulus equal to,
\begin{equation}
k = 2\pi i \frac{X_{1}}{X_{0}} = \frac{1}{2}(2g-2+p)N\epsilon_{2}+i\theta \label{eqn:kahler}
\end{equation}
and the topological string couplings \((\epsilon_{1},\epsilon_{2})\) identified with the same variables in the Yang-Mills theory.


\begin{figure}[htp]
\centering
\includegraphics[scale=0.80]{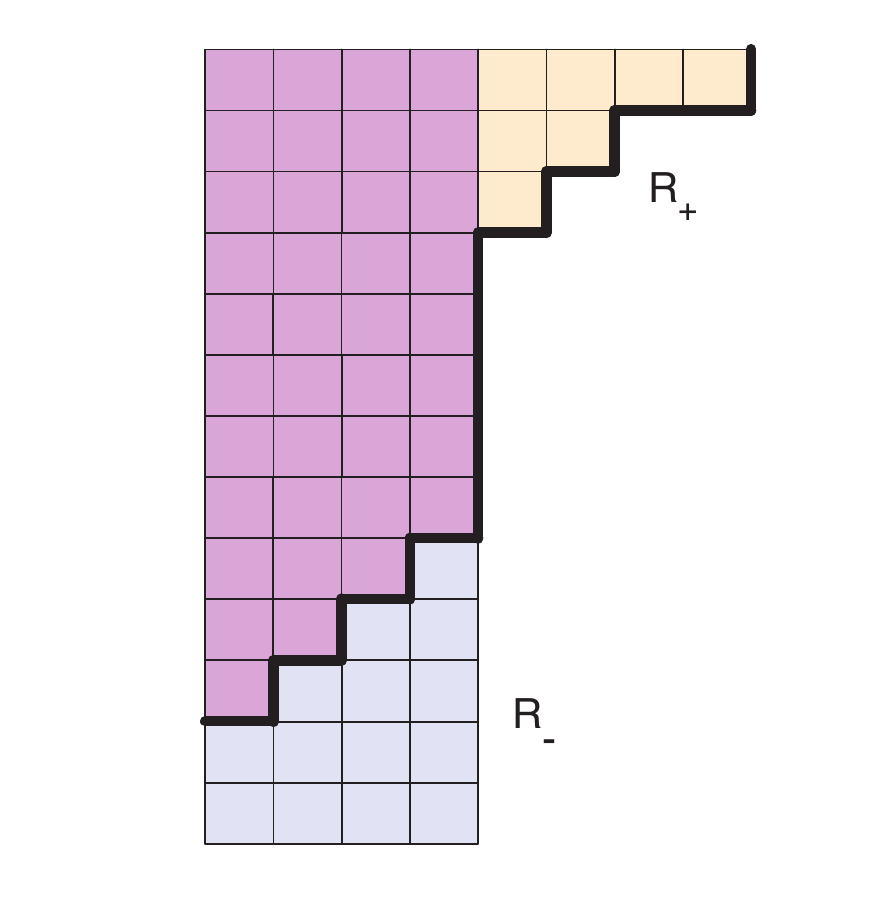}
\caption{The composite Young Tableau, \(R_{+}\overline{R}_{-}\), is shown for two \(SU(N)\) representations $R_{+}$ and $R_{-}$. \label{fig:composite}}
\end{figure}

To test this precise prediction, we must carefully study the large \(N\) limit of \((q,t)\)-deformed Yang-Mills. The key observation, first made for ordinary two-dimensional Yang-Mills in \cite{Gross:1993hu}, is that representations of \(SU(N)\) for large \(N\) can be viewed as composites of two Young Tableaux as shown in Figure \ref{fig:composite}.  This splitting can be thought of as a splitting of the Hilbert space as \(N\to \infty\) into two chiral pieces,
\begin{equation}
\mathcal{H} \sim \mathcal{H}_{+} \otimes \overline{\mathcal{H}}_{-}
\end{equation}
If the dynamics of the theory respect this splitting, then the partition function will also factorize into two pieces as predicted by the refined OSV conjecture.  In this section we will show that the partition function does indeed satisfy this factorization, after properly accounting for the sum over \(RR\) fluxes and asymptotic boundary conditions.  It is important to note that this splitting is valid to all orders in \(1/N\), but it is not valid nonperturbatively since the two Young Tableaux will interfere with each other at finite \(N\).  It would be interesting to compute the nonperturbative refined corrections to our result \cite{nonpertme}.

As in the unrefined case, it is helpful to split our discussion into the genus \(g\geq 1\) and the genus \(g=0\) cases.  We begin with the higher genus geometries.

\subsection{Genus $g\geq 1$ Case}
As explained in equation \ref{eqn:qtym}, the refined black hole partition function for \(N\) \(D4\) branes wrapping \(\mathcal{O}(-p) \to \Sigma_{g}\) is given by,
\begin{equation}
Z_{qt\textrm{YM}}(\Sigma_{g},p) =\alpha_{BH} \sum_{\mathcal{R}}\Bigg(\frac{(S_{\mathcal{R}0})^2}{G_{\mathcal{R}}}\Bigg)^{1-g}
 q^{\frac{p(\mathcal{R},\mathcal{R})}{2}}t^{p(\rho,\mathcal{R})}Q^{\sum_{i}\mathcal{R}_{i}} 
 \end{equation}
The sum is over all \(U(N)\) representations which we denote by \(\mathcal{R}\).  The normalization constant \(\alpha_{\textrm{BH}}\) defined in equation \ref{eqn:alphanorm} is included for convenience when making the connection with refined topological strings.

When we take the large \(N\) limit, it will be helpful to decompose each \(U(N)\) representation into an \(SU(N)\) and \(U(1)\) representation.  Recall that representations of \(U(N)\) are labeled by integers, \(\mathcal{R}_{i}\), such that \(\mathcal{R}_{1}\geq \mathcal{R}_{2} \geq \cdots \geq \mathcal{R}_{N}\) where the \(\mathcal{R}_{i}\) can take positive or negative values.  We can decompose this into a representation of \(SU(N)\) and a representation of \(U(1)\) by rewriting it as,
\begin{eqnarray}
\mathcal{R}_{i} & = & R_{i} + r \;\;\qquad i=1,\ldots,N-1 \\
\mathcal{R}_{N} & = & r \nonumber
\end{eqnarray}
where the \(R_{i}\) label an \(SU(N)\) representation.  Then the \(U(1)\) charge is given by
\begin{equation}
m= \vert R \vert + Nr
\end{equation}
where \(r \in \mathbb{Z}\).  We can rewrite the terms appearing in the partition function, as,
\begin{eqnarray}
\vert \mathcal{R} \vert & = & \sum_{k}\mathcal{R}_{k} = m \nonumber \\
(\mathcal{R}, \mathcal{R}) & = & \sum_{k}\mathcal{R}_{k}^{2} = \sum_{k}R_{k}^{2} + \frac{m^{2}}{N} - \frac{\vert R \vert^{2}}{N} \label{eqn:undecomp} \\
2(\rho, \mathcal{R}) & = & \sum_{k}\mathcal{R}_{k}(N+1-2k) = \sum_{k}R_{k}(1-2k)+N\vert R \vert \nonumber
\end{eqnarray}
For genus \(g\neq 1\), the partition function also involves the metric and \((q,t)\)-dimension.  These quantities are the same for the \(U(N)\) representation \(\mathcal{R}\) and its \(SU(N)\) part \(R\),
\begin{eqnarray}
\textrm{dim}_{q,t}(\mathcal{R}) & = & \textrm{dim}_{q,t}(R) \\
g_{\mathcal{R}} & = & g_{R} \nonumber
\end{eqnarray}
From equations \ref{eqn:metricbeta} and \ref{eqn:qtdimbeta}, this is true because both quantities can be written as functions only of the differences \(\mathcal{R}_{i}-\mathcal{R}_{j}\) which are independent of the \(U(1)\) charge.

Now we would like to study the large \(N\) limit of this theory.  As mentioned above, in this limit each \(SU(N)\) representation can be decomposed into a composite of two representations as depicted in Figure \ref{fig:composite}.  If we form the composite representation of \(R_{+}\) and \(R_{-}\), which we denote by  \(R_{+}\overline{R}_{-}\), then its row lengths are given by,
\begin{equation}
(R_{+}\overline{R}_{-})_{i} = (R_{-})_{1}+(R_{+})_{i}-(R_{-})_{N+1-i}
\end{equation}
Note that for this to be a good composite representation the two representations should not interact with each other, so that either \((R_{+})_{i}\) or \((R_{-})_{N+1-i}\) is equal to zero for each \(i\).  As explained above, we can neglect these interactions provided that we only study the perturbative expansion in \(1/N\).

Now we want to study how the quantities in equation \ref{eqn:undecomp} decompose for a \(U(N)\) representation \(\mathcal{R}\) whose \(SU(N)\) part consists of the composite representation \(R_{+}\overline{R}_{-}\).  In this case the \(U(1)\) charge is given by, \(m = Nl + \vert R_{+}\vert - \vert R_{-} \vert\), where we have defined \(l=r+(R_{-})_{1}\).  Then we find the decomposition,
\begin{eqnarray}
\vert \mathcal{R} \vert & = & \sum_{i}\mathcal{R}_{i} = \vert R_{+}\vert - \vert R_{-} \vert +Nl \nonumber \\
(\mathcal{R},\mathcal{R}) & = & \sum_{i}\mathcal{R}_{i}^{2} = \vert \vert R_{+} \vert \vert^{2} + \vert \vert R_{-} \vert \vert^{2} + Nl^{2} + 2l(\vert R_{+}\vert - \vert R_{-} \vert) \\
2(\rho, \mathcal{R}) & = & \sum_{k}\mathcal{R}_{k}(N+1-2k) = -\vert \vert R_{+}^{T} \vert \vert^{2} - \vert \vert R_{-}^{T} \vert \vert^{2} + N\vert R_{+} \vert + N \vert R_{-} \vert \nonumber
\end{eqnarray}
where \(\vert \vert R \vert\vert^{2} = \sum_{k}R_{k}^{2}\) and \(\vert \vert R^{T} \vert \vert^{2} = \sum_{k}(2k-1)R_{k}\).

Putting together these results, we can rewrite the refined black hole partition function as,
\begin{eqnarray}
Z_{\textrm{ref BH}} & = & \alpha_{\textrm{BH}} \sum_{l\in \mathbb{Z}}\sum_{R_{+},R_{-}}\Bigg(\frac{G_{R_{+}\overline{R}_{-}}}{\big(S_{0R_{+}\overline{R}_{-}}\big)^{2}}\Bigg)^{g-1}q^{\frac{p}{2}\big(\vert\vert R_{+}\vert\vert^{2} + \vert\vert R_{-} \vert \vert^{2}\big)} \\
& & t^{-\frac{p}{2}\big(\vert\vert R^{T}_{+}\vert\vert^{2} + \vert\vert R^{T}_{-} \vert \vert^{2}\big)} e^{-i\theta(\vert R_{+} \vert - \vert R_{-} \vert)}q^{lp(\vert R_{+} \vert - \vert R_{-} \vert)}t^{\frac{Np}{2}(\vert R_{-} \vert + \vert R_{+} \vert)}q^{\frac{N p l^{2}}{2}} e^{-i\theta N l} \nonumber
\end{eqnarray}
With the refined OSV relation in mind, we can use the formula for the K\"ahler modulus in equation \ref{eqn:kahler} to rewrite \(\alpha_{\textrm{BH}}\) as,
\begin{eqnarray}
\alpha_{\textrm{BH}} & = & \Big((t;q)_{\infty}(q;q)_{\infty}\Big)^{N(g-1)} \\
& & \cdot \exp\Bigg(-\frac{1}{\epsilon_{1}\epsilon_{2}}\frac{k^{3}+\overline{k}^{3}}{6p(p+2g-2) }+ \beta\frac{(k+\overline{k})(p+2g-2)}{24p} + \epsilon_{1}\beta^{2}\rho^{2}(2g-2) \Bigg) \nonumber
\end{eqnarray}

So far we have explained how the framing factor and \(\theta\)-dependent terms factorize, but we still need to understand the metric and \((q,t)\)-dimension.  Their factorization properties are derived in Appendix \ref{sec:qtfac}, with the result that,
\begin{eqnarray}
\frac{G_{R_{+}\overline{R}_{-}}}{q^{2\beta^{2}\rho^{2}} \big(S_{0R_{+}\overline{R}_{-}}\big)^{2}} & = &\Big(M(q,t)M(t,q)\Big)^{-1}\Big((t;q)_{\infty}(q;q)_{\infty}\Big)^{-N}T_{R_{+}}^{2}T_{R_{-}}^{2} Q^{\vert R_{+}\vert + \vert R_{-} \vert} \nonumber \\
& &  \cdot \frac{g_{R_{+}}g_{R_{-}}K_{R_{+}R_{-}}(Q\frac{q}{t})K_{R_{+}R_{-}}(Q)}{W_{R_{+}}^{4}W_{R_{-}}^{4}}
\end{eqnarray}
where we have defined \(Q = t^{N}\) and used the framing factor, \(T_{R} = q^{\vert \vert R \vert \vert^{2}/2}t^{-\vert\vert R^{T} \vert \vert^{2}/2}\), and where
\begin{equation}
K_{R_{+}R_{-}}(Q;q,t) = \sum_{P} \frac{1}{g_{P}}Q^{\vert P \vert}(t/q)^{\vert P \vert}W_{PR_{+}}(q,t)W_{PR_{-}}(q,t)
\end{equation}

From these factorization formulas, we can write the entire refined black hole partition function as a sum over chiral blocks,

\begin{equation}
Z_{\textrm{ref BH}} = \sum_{l \in \mathbb{Z}}\sum_{R_{1},\cdots R_{2g-2}}Z^{+}_{R_{1},\cdots, R_{2g-2}}(k+pl\epsilon_{1}) Z^{+}_{R_{1},\cdots, R_{2g-2}}(\overline{k}-pl\epsilon_{1}) \label{eqn:largenosv}
\end{equation}
where the chiral block is defined by,
\begin{eqnarray}
Z^{+}_{R_{1},\cdots, R_{2g-2}}(k) & = & Z_{0}(q,t) \cdot \big(t^{\frac{N}{2}}\big)^{\vert R_{1}\vert + \cdots \vert R_{g-1}\vert}\big(t^{\frac{N+1}{2}}q^{-\frac{1}{2}}\big)^{\vert R_{g}\vert + \cdots \vert R_{2g-2}\vert} \\
& & \sum_{R}(T_{R})^{p+2g-2}e^{-k\vert R\vert}\frac{W_{R_{1}R}(q,t) \cdots W_{R_{2g-2}R}(q,t)}{W_{0R}^{4g-4}}\frac{(g_{R})^{g-1}}{g_{R_{1}}\cdots g_{R_{2g-2}}} \nonumber
\end{eqnarray}
where \(k=\frac{1}{2}(p+2g-2)N\epsilon_{2}+i\theta\) and where we used the definition of the degree zero piece, \(Z_{0}\), defined in equation \ref{eqn:degzero}.

Now we come to the physical interpretation of these chiral blocks.  First notice that the chiral block is precisely the refined topological string amplitude for the geometry 
\begin{equation}
\mathcal{O}(2g-2+p)\oplus \mathcal{O}(-p) \to \Sigma_{g}
\end{equation}
with branes in the fibers over \(2g-2\) points, as explained in section \ref{sec:fiberbranes},
\begin{equation}
Z^{+}_{R_{1},\cdots, R_{2g-2}}(k) = Z_{\textrm{ref top, } R_{1},\cdots, R_{2g-2}}(k) 
\end{equation}
The factors, \(t^{\frac{1}{2}N\vert R_{i} \vert}\) appear because these branes have been moved in the fiber away from the origin.  These ``ghost branes'' were explained in \cite{Aganagic:2005dh} as parametrizing noncompact K\"ahler moduli in the fiber directions.  Naively, one might expect there to be noncompact moduli over every point on the Riemann surface.  However, these moduli can be localized by using the symmetries corresponding to meromorphic vector fields on \(\Sigma\).  Since these vector fields have \(2g-2\) poles generically, we find ghost branes over precisely \(2g-2\) points.  

This picture must be modified slightly in the refined case, since not all of these branes are the same.  They split into two groups of \(g-1\) branes, which differ only by their fiber K\"ahler parameters which are either 
\begin{equation}
k_{f} = \frac{1}{2}N\epsilon_{2}, \qquad  \textrm{or} \qquad k_{f} = \frac{1}{2}N\epsilon_{2} + \frac{1}{2}(\epsilon_{2}-\epsilon_{1})
\end{equation}
It would be interesting to derive this splitting from first principles.  Another important aspect of our formula for large N factorization is the sum over the \(U(1)\) charge, \(l\).  As in the unrefined case \cite{Vafa:2004qa}, we interpret this as arising from the sum over \(RR\)-flux through the Riemann surface.  Alternatively, this sum arises because the black hole partition function is trivially invariant under shifts \(\phi_{2} \to \phi_{2} + 2\pi i p n\).\footnote{The extra factor of \(p\) arises because the black holes have charges, \(Q_{2} \in \frac{1}{p}\mathbb{Z}\), as explained in \cite{Aganagic:2004js}.}  The sum over \(U(1)\) charge enforces this same periodicity on the topological string side of the correspondence.

\subsection{Genus $g=0$ Case \label{sec:genuszero}}
The genus zero case works much the same way as the higher genus case.  By decomposing the \(U(N)\) representations into \(U(1)\) and composite \(SU(N)\) representations, we can write the corresponding brane partition function as,
\begin{eqnarray}
Z_{\textrm{ref BH}} & = & \alpha  \sum_{l\in \mathbb{Z}}\sum_{R_{+},R_{-}} \frac{(S_{0R_{+}\overline{R}_{-}})^2}{G_{R_{+}\overline{R}_{-}}} q^{\frac{p}{2}\big(\vert\vert R_{+}\vert\vert^{2} + \vert\vert R_{-} \vert \vert^{2}\big)}t^{-\frac{p}{2}\big(\vert\vert R^{T}_{+}\vert\vert^{2} + \vert\vert R^{T}_{-} \vert \vert^{2}\big)} \nonumber \\
& & e^{-i\theta(\vert R_{+} \vert - \vert R_{-} \vert)}q^{lp(\vert R_{+} \vert - \vert R_{-} \vert)}t^{\frac{Np}{2}(\vert R_{-} \vert + \vert R_{+} \vert)}q^{\frac{N p l^{2}}{2}} e^{-i\theta N l}
\end{eqnarray}
However, since \(\textrm{dim}_{q,t}(R_{+}\overline{R}_{-})\) appears in the numerator, we will use the other set of identities from appendix \ref{sec:qtfac} to give,
\begin{eqnarray}
\frac{q^{2\beta^{2}\rho^{2}} (S_{0R_{+}\overline{R}_{-}})^2}{G_{R_{+}\overline{R}_{-}}} & = & M(q,t)M(t,q) \Big((t;q)_{\infty}(q;q)_{\infty}\Big)^{N}T_{R_{+}}^{-2}T_{R_{-}}^{-2} Q^{-\vert R_{+}\vert - \vert R_{-} \vert} \nonumber \\
& &  \cdot \frac{N_{R_{+}R_{-}}(Q\frac{q}{t})N_{R_{+}R_{-}}(Q)}{g_{R_{+}}g_{R_{-}}}
\end{eqnarray}
where,
\begin{equation}
N_{R_{+}R_{-}}(Q,q,t) := \sum_{P}\frac{1}{g_{P}}Q^{\vert P\vert}(t/q)^{\vert P\vert}\widetilde{W}_{RP}(q,t)W_{PS}(q,t)
\end{equation}
and where \(\widetilde{W}_{RP}\) is the cap amplitude for placing a brane in the base and an anti-brane in the fiber as explained in section \ref{sec:fiberbranes}.

Therefore, we can rewrite the black hole partition function as a sum over chiral blocks,
\begin{equation}
Z_{\textrm{ref BH}} = \sum_{l \in \mathbb{Z}}\sum_{R_{1}, R_{2}}Z^{+}_{R_{1},R_{2}}(k+pl\epsilon_{1}) Z^{-}_{R_{1}, R_{2}}(\overline{k}-pl\epsilon_{1}) \label{eqn:largenosvg0}
\end{equation}
where the K\"ahler parameter is given by,
\begin{equation}
k = \frac{1}{2}(p-2)N\epsilon_{2}+i\theta
\end{equation}
It is important to notice that in the genus zero case, the chiral and anti-chiral blocks are not precisely the same.  The chiral block is equal to,
\begin{equation}
Z^{+}_{R_{1},R_{2}}(k) = Z_{0}(q,t) t^{\frac{1}{2}N\vert R_{1}\vert}\big(t^{\frac{N+1}{2}}q^{-\frac{1}{2}}\big)^{\vert R_{2} \vert} \sum_{R}(T_{R})^{p-2}e^{-k\vert R \vert}\frac{W_{R_{+}R_{1}}(q,t) W_{R_{+}R_{2}}(q,t)}{g_{R}\; g_{R_{1}}\, g_{R_{2}}}
\end{equation}
while the anti-chiral block takes the slightly different form,
\begin{equation}
Z^{-}_{R_{1},R_{2}}(k) = Z_{0}(q,t) t^{\frac{1}{2}N\vert R_{1}\vert}\big(t^{\frac{N+1}{2}}q^{-\frac{1}{2}}\big)^{\vert R_{2} \vert} \sum_{R}(T_{R})^{p-2}e^{-k\vert R \vert}\frac{\widetilde{W}_{R_{+}R_{1}}(q,t) \widetilde{W}_{R_{+}R_{2}}(q,t)}{g_{R}\; g_{R_{1}}\, g_{R_{2}}}
\end{equation}

As in the higher genus case, the refined ghost branes split into two types depending on their fiber K\"ahler moduli and we obtain a sum over RR-flux.  The chiral block computes precisely the refined topological string amplitude on \(\mathcal{O}(p-2)\oplus \mathcal{O}(-p) \to \mathbb{P}^{1}\) with two ``ghost'' branes in the fiber.  

The main difference between the higher genus and genus zero cases is that here the anti-chiral amplitude is the refined topological string on the same geometry but with two anti-branes rather than branes in the fiber.  We can use the same argument as before for why there are precisely two ghost branes.  However, now localization by using a generic vector field on the \(\mathbb{P}^{1}\) will have two zeros rather than poles.

\section{Black Hole Entropy and Refined Wall Crossing \label{sec:wc}}
In section \ref{sec:o1}, we used \((q,t)\)-deformed Yang-Mills to compute the entropy of D-branes wrapping the blow-up geometry \(\mathcal{O}(-1) \to \mathbb{P}^{1}\) and found agreement with the mathematical result of \cite{Yoshioka:1996pd}.  In this section, we explain an alternative way to compute this black hole partition function by using refined wall-crossing.  We follow the approach of \cite{Nishinaka:2010qk, Nishinaka:2010fh}, giving a refined generalization of their unrefined computations.

\begin{figure}[htp]
\centering
\includegraphics[scale=1.15]{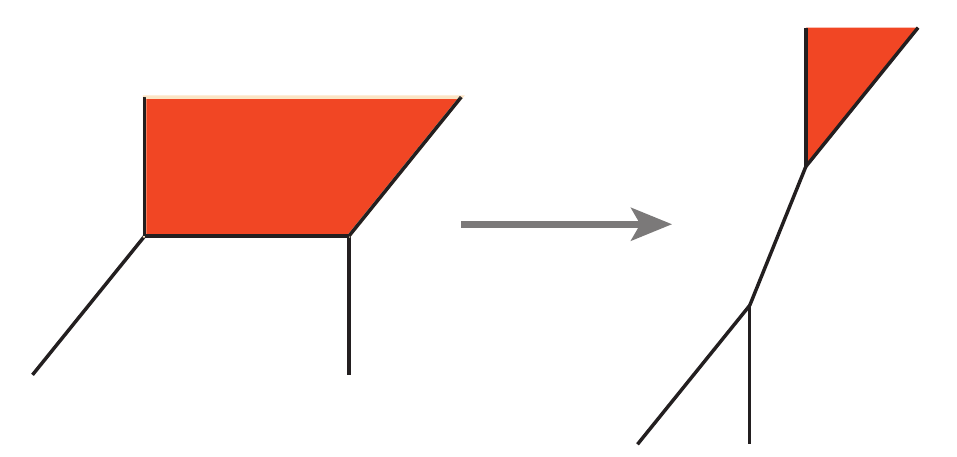}
\caption{The flop transition for the conifold.  $N$ D4-branes wrap the shaded four-cycle, which changes topology under the flop. \label{fig:conifoldflop}}
\end{figure}

We start by studying the resolved conifold, \(\mathcal{O}(-1)\oplus \mathcal{O}(-1) \to \mathbb{P}^{1}\), with \(N\) D4-branes wrapping the four cycle \(C_{4} = \mathcal{O}(-1)\to \mathbb{P}^{1}\).  We would like to compute the refined degeneracies of these \(D4\) branes bound to lower dimensional branes.  Note that since \(C_{4}\) contains the compact two-cycle, \(\mathbb{P}^{1}\), D2 branes can form bound states with the stack of \(D4\)s.

The key insight of \cite{Nishinaka:2010qk} is that by varying the K\"ahler modulus of the \(\mathbb{P}^{1}\), which we denote by \(z\),  the refined partition function will simplify in certain chambers.  Specifically, by sending \(z \to -\infty\), the conifold will undergo a flop transition, and the four-cycle, \(C_{4}\) will become a cycle wrapping only the fiber directions, as in shown in Figure \ref{fig:conifoldflop}.  Most importantly, this new four-cycle is topologically \(\mathbb{C}^{2}\) and does not have any compact two-cycles.  Thus, in this chamber, only D4 and D0 branes can bind, and the refined partition function simplifies dramatically.

Now by starting in this simple chamber, we can vary the K\"ahler parameter \(z\) and keep track of how the refined partition function jumps.  The partition function is locally constant, but along real codimension-one walls of marginal stability it will jump.  Since the conifold geometry is fairly simple, we can identify all walls of marginal stability and jumps as we take \(z\) from \(-\infty\) to \(\infty\).  This will allow us to explicitly compute the refined partition function at \(z=\infty\) and find agreement with our \((q,t)\)-deformed Yang-Mills computation in section \ref{sec:o1}.

To find these walls, we must first compute the central charge of BPS bound states.  Note that since our geometry is noncompact, the central charge of the \(D4\) branes is infinite and its phase is not well-defined.   As explained originally in \cite{Jafferis:2008uf}, this can be remedied by starting with a compact geometry and including a component of the complexified K\"ahler form along the direction that is becoming noncompact.  The result is that the central charge of the \(D4\) brane is given by,
\begin{equation}
Z(D4) = - \frac{1}{2}\Lambda^{2}e^{2i\phi}
\end{equation}
where  \(\Lambda \gg 1\), and ultimately we want to take \(\Lambda \to \infty\) to obtain the resolved conifold.  Note that \(\phi\) is still a free real parameter, but for our purposes we will keep it fixed and only vary the complexified K\"ahler parameter, \(z\).

Now we can consider a more general D4/D2/D0 bound state with charges, 
\begin{equation}
\Gamma = (P_{6},P_{4},Q_{2},Q_{0}) = (0,N,m,n)
\end{equation}
Its central charge is given by,
\begin{equation}
Z(\Gamma) = - \frac{1}{2}N\Lambda^{2}e^{2i\phi}+mz+n
\end{equation}
This state can decay as \(\Gamma \to \Gamma_{1} + \Gamma_{2}\) precisely when \(Z(\Gamma_{1})\) and \(Z(\Gamma_{2})\) are aligned.  Depending on the charges of \(\Gamma_{i}\), there are two types of walls that we must consider.  

First, \(\Gamma\) could decay into two states that each have nonzero \(D4\)-brane charge.  Such fragmentation of the stack of D4 branes was discussed in this context in \cite{Nishinaka:2010fh}.  Although the alignment of central charges seems to suggest that these fragments can form, we argue that fragmentation walls are not physical in our setup.  

The crucial point is that the \(D4\) branes remain noncompact throughout moduli space.  From a field theory perspective fragmentation corresponds to changing a scalar field's vev.  But for a field theory on noncompact spacetime, this vev is a background parameter of the theory and changing it would cost infinite energy.  Therefore, we conclude that because of the noncompactness of the \(D4\)-branes, there can be no binding or decay across these walls, and they can be consistently ignored.

The second type of wall is much more interesting for us and involves the decay \(\Gamma \to \Gamma_{1} + m\Gamma_{2}\) where \(\Gamma_{2}\) only has D2/D0 charge.  From the Gopakumar-Vafa invariants of the conifold, it follows that the only BPS D2/D0 bound states are,
\begin{eqnarray}
& & \Omega_{0}(0,0,0,n;y) = -2 \nonumber \\
& & \Omega_{0}(0,0,\pm 1,n;y) = 1
\end{eqnarray}
where \(\Omega_{j}(\Gamma)\) is the refined degeneracy for states with spin \(j\).

Now we would like to find the walls of marginal stability where \(\Gamma = (0,N,m,n)\) can decay into these bound states.  Since \(\Lambda \gg 1\), the phase of the central charge is equal to \(\arg(Z(\Gamma)) = 2\phi+ \pi\), which means that the walls of marginal stability for \(\Gamma\) will be independent of \(N\), \(m\), and \(n\), provided \(N>0\).

First we can consider the pure \(D0\)-brane decay channel where \(\Gamma_{2}=(0,0,0,n)\).  However, the central charge of \(\Gamma_{2}\) is always real, which means that for a generic fixed choice of \(\phi\), the central charges of \(\Gamma\) and \(\Gamma_{2}\) will never align.  This means that there are no walls of marginal stability associated with \(D0\) decay.

\begin{figure}[htp]
\centering
\includegraphics[scale=1.25]{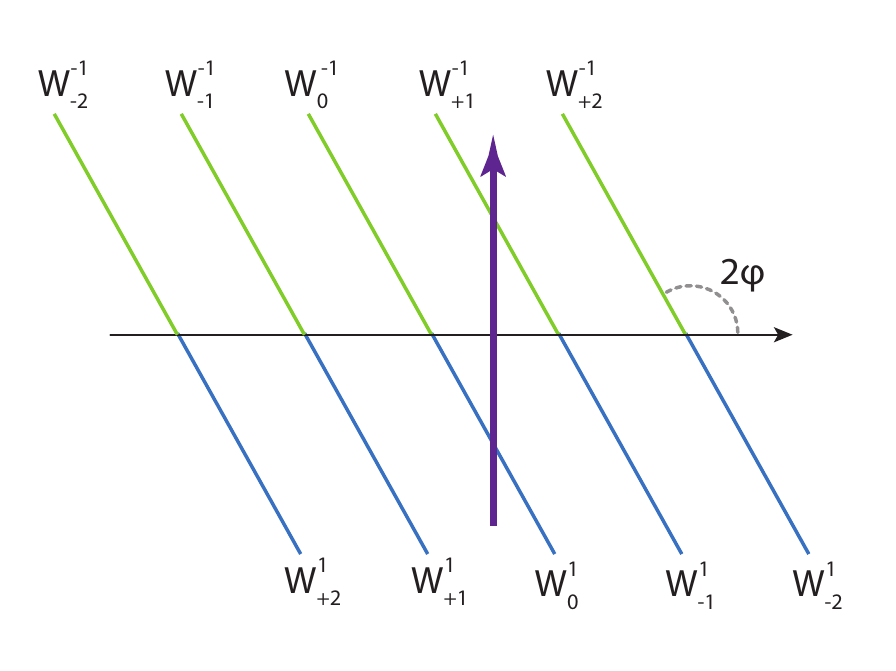}
\caption{Walls of marginal stability for the conifold moduli space.  The path we follow is shown by the arrow. \label{fig:conifoldwalls}}
\end{figure}

Next, we consider the second possibility of a decay involving \(\Gamma_{2}=(0,0,\pm 1,n)\).  In this case, the phase of the central charge is given by, \(\arg(Z(\Gamma_{2})) = \arg(\pm z + n)\), which implies that a wall of marginal stability can occur as we vary \(z\).  We will denote these walls of marginal stability by \(W^{\pm 1}_{n}\), and they are given explicitly by,
\begin{eqnarray}
W^{1}_{n}:\qquad \phi & = & \frac{1}{2}\arg(-z-n) \\
W^{-1}_{n}:\qquad \phi & = & \frac{1}{2}\arg(z-n)
\end{eqnarray}
These walls of marginal stability are shown in Figure \ref{fig:conifoldwalls}.

Now that we have identified all walls of marginal stability, we want to study how the partition function jumps across these walls.  For general decays, this jump can be quite complicated and is determined by the formula of Kontsevich and Soibelman \cite{KSoriginal}.  However, for our purposes we only must deal with semi-primitive decays of the form, \(\Gamma \to \Gamma_{1} + m\Gamma_{2}\).  The refined semi-primitive wall-crossing formula was computed in \cite{Dimofte:2009bv}, and is given by,
\begin{equation}
\sum_{n=0}^{\infty}\Omega(\Gamma_{1}+n\Gamma_{2};y)x^{n} = \Omega(\Gamma_{1};y)\prod_{k=1}^{\infty}\prod_{j=1}^{k\vert \langle \Gamma_{1},\Gamma_{2} \rangle \vert}\prod_{n}\Big(1+(-1)^{n}x^{k}y^{\frac{k\vert \langle \Gamma_{1},\Gamma_{2} \rangle \vert +1 - 2j}{2}}\Big)^{(-1)^{n}\Omega_{n}(k\Gamma_{2})}
\end{equation}
where we have used the intersection form \(\langle \Gamma_{1},\Gamma_{2}\rangle\).  In the mirror IIB geometry, this intersection form is simply the geometric intersection number of the corresponding lagrangian three-cycles.  In IIA, it is equal to,
\begin{equation}
\langle \Gamma, \Gamma' \rangle = P_{6}\cdot Q'_{0}-P'_{6}\cdot Q_{0}+P_{4}\cdot Q'_{2}-P'_{4}\cdot Q_{2}
\end{equation}
Finally, in the exponent of the semi-primitive wall-crossing formula we have used the refined degeneracies, \(\Omega_{n}(\Gamma)\), that compute the index of states with spin \(n\).

Now we that we have explained the refined semi-primitive wall-crossing formula, we want to apply this to our setup.  We start with some BPS state \(\Gamma = (0,N,m,n)\).  At the \(W^{1}_{n}\) wall, our state will decay into \(\Gamma_{1}+\Gamma_{2}\) where \(\Gamma_{2}=(0,0,1,n)\).  This means that the intersection is given by \(\vert \langle \Gamma_{1},\Gamma_{2} \rangle \vert = N\).  From our discussion above, the semi-primitive wall-crossing formula simplifies since \(\Omega(k\Gamma_{2})=0\) for \(k>1\), and \(\Omega_{n}(\Gamma_{2}) = \delta_{0,n}\).

Putting this together, we can write the refined black hole partition function using chemical potentials \(\tilde{q}\) and \(\widetilde{Q}\) for the D0 and D2-brane charge respectively.  Then we find that the partition function jumps across the wall \(W^{1}_{n}\) as,
\begin{equation}
Z_{\textrm{ref BH}}(\widetilde{Q},\tilde{q},y;z) \to Z_{\textrm{ref BH}}(\widetilde{Q},\tilde{q},y;z)\prod_{j=1}^{N}\Big(1+y^{\frac{N+1}{2}-j}\widetilde{Q} \tilde{q}^{n}\Big)
\end{equation}
Note that the wall-crossing factor takes the form of the Fock space character for a spin \( \frac{N-1}{2} \) multiplet.

Similarly, crossing the wall \(W^{-1}_{n}\) results in the jump,
\begin{equation}
Z_{\textrm{ref BH}}(\widetilde{Q},\tilde{q},y;z) \to Z_{\textrm{ref BH}}(\widetilde{Q},\tilde{q},y;z)\prod_{j=1}^{N}\Big(1 + y^{\frac{N+1}{2}-j}\widetilde{Q}^{-1}\tilde{q}^{n}\Big)
\end{equation}

Now having understood how to cross individual walls, we want to follow a path in moduli space that connects the flopped geometry to the the large volume limit of interest.  As shown in Figure \ref{fig:conifoldwalls}, we can take \(z=\frac{1}{2} + ir\) and follow the path from \(r=-\infty\) to \(r=+\infty\).  This path crosses all \(W^{1}_{n}\) and \(W^{-1}_{n}\) walls for \(n>0\), along with the wall \(W^{1}_{0}\).  This implies the relationship,
\begin{equation}
Z_{+\infty}(\tilde{q},\widetilde{Q},y)= Z_{-\infty}(\tilde{q},y) \prod_{j=1}^{N}\Bigg\{ \big(1 + y^{\frac{N+1}{2}-j}\widetilde{Q}\big)\prod_{n=1}^{\infty} \big(1 + y^{\frac{N+1}{2}-j}\widetilde{Q}\tilde{q}^{n}\big)\big(1 + y^{j-\frac{N+1}{2}}\widetilde{Q}^{-1}\tilde{q}^{n}\big) \Bigg\}
\end{equation}
Using the Jacobi Triple product, this can be rewritten as,
\begin{eqnarray}
Z_{+\infty}(\tilde{q},\widetilde{Q},y) & = & Z_{-\infty}(\tilde{q},y) \Bigg(\prod_{n=1}^{\infty}(1-\tilde{q}^{n})^{-N}\Bigg)\sum_{\{n_{i}\}}\tilde{q}^{\frac{1}{2}(n,n)}(\widetilde{Q}\tilde{q}^{-1/2})^{\sum_{i}n_{i}} y^{\sum_{i}(\frac{N+1}{2}-i)n_{i}} \\
& \sim & \sum_{\{n_{i}\}}e^{-\phi_{0}\frac{1}{2}(n,n)}e^{-\phi_{2}\sum_{i}n_{i}} y^{\sum_{i}(\frac{N+1}{2}-i)n_{i}}
\end{eqnarray}
where we have identified \(e^{-\phi_{0}} = \tilde{q}\) and \(e^{-\phi_{2}} = \widetilde{Q}\tilde{q}^{-1/2}\).\footnote{This shift in charges arises because of the Freed-Witten anomaly \cite{Freed:1999vc}, which implies that the spacetime D-brane charge, \(\Gamma\), is related to the chern character of a vector bundle, \(E\), over a four-cycle \(S\), by \(\Gamma = \textrm{ch}(E) e^{\frac{1}{2}c_{1}(S)}\).  This leads to the above nontrivial relationship between the D-brane charges seen by wall-crossing, and the instanton charges seen by \((q,t)\)-deformed Yang-Mills.}  Up to the D0/D4-brane bound states which are determined by the \(z=-\infty\) chamber, this formula agrees precisely with the partition function computed in equation \ref{eqn:goodo1}.  Thus, we have given two independent derivations of this mathematical formula, from gauge theory and from wall-crossing.

\acknowledgments{We thank Chris Beem, Emanuel Diaconescu, Tudor Dimofte, Ori Ganor, Lotte Hollands, Andrei Okounkov, Vasily Pestun, Ashoke Sen, Andy Strominger, Cumrun Vafa, and Kevin Wray for discussions.  We also thank the 2012 Simons Workshop in Mathematics and Physics for hospitality while this work was being completed.  The research of K.S. is supported by the Berkeley Center for Theoretical Physics and by the National Science Foundation (award number 0855653).  The research of M.A. is supported in part by the Berkeley Center for Theoretical Physics, by the National Science Foundation (award number 0855653), by the Institute for the Physics and Mathematics of the Universe, and by the US Department of Energy under Contract DE-AC02-05CH11231.
}

\appendix

\section{Macdonald Polynomials and Identities \label{sec:macs}}

In this appendix we fix our conventions and collect some useful results on Macdonald polynomials.  We refer the reader to \cite{macdonald_hall} for more details.

Macdonald polynomials form a special class of symmetric functions.  They are rational functions of \(q\) and \(t\), and are symmetric functions of N variables, \(x_{i}\).  The simplest way to understand Macdonald polynomials comes from defining an inner product on the space of symmetric functions,
\begin{equation}
\langle f, g \rangle = \frac{1}{N!}\oint dz_{1} \cdots \oint dz_{N}\Delta_{q,t}(z_{1},\cdots z_{N}) f(z_{1},\cdots, z_{N})g(z_{1}^{-1},\cdots z_{N}^{-1})
\end{equation}
where the measure is given by,
\begin{equation}
\Delta_{q,t}(z_{1},\cdots, z_{N}) = \prod_{1\leq i<j \leq N}\frac{\big(z_{i}/z_{j};q\big)_{\infty}\big(z_{j}/z_{i};q\big)_{\infty}}{\big(z_{i}t/z_{j};q\big)_{\infty} \big(z_{j}t/z_{i};q\big)_{\infty}}
\end{equation}
where \((x;q)_{\infty} = \prod_{m=0}^{\infty}(1-xq^{m})\).  

Then we can uniquely associate a Macdonald polynomial, \(M_{R}(z;q,t)\), to every \(SU(N)\) representation, \(R\), by requiring the following two properties,
\begin{eqnarray}
\langle M_{R}, M_{S} \rangle & = & 0 \qquad \textrm{if }\; R \neq S \\
M_{R} & = & m_{R} + \textrm{(lower order)}
\end{eqnarray}
where \(m_{R}\) is the monomial symmetric polynomial given by,
\begin{equation}
m_{R}(z_{1},\cdots,z_{N}) = \sum_{\sigma}z_{1}^{R_{\sigma(1)}}\cdots z_{N}^{R_{\sigma(N)}}
\end{equation}
where the sum is over all elements \(\sigma\) of the symmetric group.

Therefore, orthogonality and a condition on the leading behavior completely determine the Macdonald polynomials.  It is important to note that in the limit \(t=q\), Macdonald polynomials reduce to the more familiar Schur functions (which are independent of \(q\)),
\begin{equation}
M_{R}(x_{i};q,q) = s_{R}(x_{i})
\end{equation}

In this paper, we use Macdonald polynomials with either finitely many variables or infinitely many variables - the finite polynomials appear in \((q,t)\)-deformed Yang-Mills and the infinite polynomials appear in the closed refined topological string.  We begin by reviewing the finite case.

\subsection{$SU(N)$ Macdonald Polynomials}
For finitely many variables, the inner product of a Macdonald polynomial with itself can be written either in a combinatorial way
\begin{equation}
g_{R}:= \frac{G_{R}}{G_{0}} = \frac{ \langle M_{R}, M_{R} \rangle}{\langle M_{0}, M_{0} \rangle} = \prod_{(i,j)\in \lambda}\frac{1-q^{R_{i}-j+1}t^{R^{T}_{j}-i}}{1-q^{R_{i}-j}t^{R^{T}_{j}-i+1}}\frac{1-q^{j-1}t^{N+1-i}}{1-q^{j}t^{N-i}} \label{eqn:metric1}
\end{equation}
or for the case when \(\beta \in \mathbb{Z}_{\geq 0}\), in a Lie-theoretic way,
\begin{equation}
g_{R} = \prod_{m=0}^{\beta-1}\prod_{1 \leq i<j \leq N}\frac{ \lbrack R_{i}-R_{j}+\beta(j-i)+m\rbrack_{q} }{\lbrack R_{i}-R_{j}+\beta(j-i)-m\rbrack_{q}}\frac{\lbrack \beta(j-i) -m \rbrack_{q}}{\lbrack \beta(j-i) + m \rbrack_{q}}
\end{equation}
We will often refer to \(g_{R}\) as the Macdonald metric for \(R\).

We can also give an explicit formula for Macdonald polynomials evaluated at \(z_{k}=t^{\rho_{k}}\).  This gives a generalization of the quantum dimension of a representation, which we refer to as the \((q,t)\)-dimension,
\begin{equation}
\textrm{dim}_{q,t}(R) := M(t^{\rho}) = t^{\frac{(N+1)\vert R \vert}{2}}\prod_{(i,j)\in R}\frac{t^{i-N-1}-q^{j-1}}{1-q^{R_{i}-j}t^{R^{T}_{j}-i+1}}
\end{equation}
This formula can also be rewritten when \(\beta\in \mathbb{Z}_{\geq 0}\) in a Lie-theoretic way,
\begin{equation}
M(t^{\rho}) =  \prod_{m=0}^{\beta-1}\prod_{1 \leq i<j \leq N}\frac{ \lbrack R_{i}-R_{j}+\beta(j-i)+m\rbrack_{q} }{\lbrack \beta(j-i) + m\rbrack_{q}}
\end{equation}
Now that we have explained the finite \(N\) formulas, we would like to take the \(N\to \infty\) limit.  

\subsection{$SU(\infty)$ Macdonald Polynomials}
For studying the closed refined topological string, it is also useful to collect some formulas for Macdonald polynomials with infinitely many variables.  Starting with the metric in equation \ref{eqn:metric1}, and taking the \(N\to \infty\) limit naively gives,
\begin{equation}
g^{(mac)}_{R} = \prod_{(i,j)\in R}\frac{1-t^{R^{T}_{j}-i}q^{R_{i}-j+1}}{1-t^{R^{T}_{j}-i+1}q^{R_{i}-j}}
\end{equation}
which is the standard formula for the metric given in \cite{macdonald_hall}.  However, unlike its finite \(N\) counterpart, this expression is not symmetric under \((q,t) \to (q^{-1},t^{-1})\).  We can fix this by using a slightly different definition,
\begin{equation}
g_{R}(q,t) := \prod_{(i,j)\in R}\frac{t^{\frac{R^{T}_{j}-i}{2}}q^{\frac{R_{i}-j+1}{2}}-t^{-\frac{R^{T}_{j}-i}{2}}q^{-\frac{R_{i}-j+1}{2}}}{t^{\frac{R^{T}_{j}-i+1}{2}}q^{\frac{R_{i}-j}{2}} - t^{-\frac{R^{T}_{j}-i+1}{2}}q^{-\frac{R_{i}-j}{2}}} = (t/q)^{\frac{\vert R \vert}{2}}
\cdot g^{(mac)}_{R} \label{eqn:metricsym} \end{equation}
In order to preserve the relationship between the metric and Macdonald polynomials, \(g_{R} = \langle M_{R}, M_{R} \rangle\), we must also rescale \(M_{R}\) in the infinite-variable limit,
\begin{equation}
M_{R}(x_{1},x_{2},\ldots ;q,t) := (t/q)^{\frac{\vert R \vert}{4}} \cdot M^{(mac)}_{R}(x_{1},x_{2},\ldots;q,t)
\end{equation}
We can also give a formula for evaluating the infinite-variable Macdonald polynomial at a certain value,
\begin{equation}
M_{R}(t^{\rho};q,t) = q^{\frac{1}{4}\vert \vert R \vert \vert^{2}}t^{-\frac{1}{4}\vert \vert R^{T} \vert \vert^{2}} (t/q)^{\frac{\vert R \vert}{2}} \prod_{(i,j)\in R}\Big(q^{\frac{R_{i}-j}{2}}t^{\frac{R^{T}_{j}-i+1}{2}}-q^{-\frac{R_{i}-j}{2}}t^{-\frac{R^{T}_{j}-i+1}{2}}\Big)^{-1}
\end{equation}
where \((\rho)_{k}= -k + 1/2\)
%

%

There are also generalized Cauchy identities for Macdonald polynomials.  The most useful ones are,
\begin{eqnarray}
\sum_{R}\frac{1}{g_{R}}M_{R}(x;q,t)M_{R}(y;q,t) & = & \exp\Big(\sum_{n=1}^{\infty}\frac{1}{n}\frac{1-t^{n}}{1-q^{n}}p_{n}(x)p_{n}(y)\Big) \label{eqn:maccauchy} \\
\sum_{R}M_{R}(x;q,t)M_{R^{T}}(y;t,q) & = & \prod_{i,j}(1+x_{i}y_{j}) \label{eqn:cauchytranspose}
\end{eqnarray}
where \(p_{n}(x) = \sum_{i}x_{i}^{n}\) is the \(n\)'th power sum.

Finally, when understanding anti-branes in this paper, it is useful to use the operation that flips the sign of the power sums,
\begin{equation}
\iota \big(p_{n}(x)\big) = - p_{n}(x)
\end{equation}
This operation acts on Schur functions as,
\begin{equation}
\iota s_{R}(x) = (-1)^{\vert R \vert}s_{R^{T}}(x)
\end{equation}
Its action on Macdonald polynomials is more complicated.  However, from equation \ref{eqn:maccauchy} and the definition of \(\iota\), it is straightforward to write down a generalized Cauchy identity for \(\iota M\),
\begin{equation}
\sum_{R}\frac{1}{g_{R}}\iota M_{R}(x;q,t)M_{R}(y;q,t) = \exp\Big(- \sum_{n=1}^{\infty}\frac{1}{n}\frac{1-t^{n}}{1-q^{n}}p_{n}(x)p_{n}(y)\Big)
\end{equation}

\section{Gopakumar-Vafa Invariants \label{sec:gv}}

In section \ref{sec:tqft} we presented a TQFT that computes the refined topological string amplitudes for geometries of the form \(\mathcal{L}_{1}\oplus \mathcal{L}_{2} \to \Sigma_{g}\).  As explained originally by Gopakumar and Vafa in \cite{Gopakumar:1998ii, Gopakumar:1998jq}, the topological string partition function can be rewritten as counting spinning M2-branes which wrap a two-cycle in the Calabi-Yau and are free to move in the noncompact \(\mathbb{R}^{4,1}\).  Since these M2-branes are massive, they should transform in a definite representation of the (4+1)-dimensional little group, \(SO(4) = SU(2)_{l} \times SU(2)_{r}\).  BPS states will also fall in representations of the \(SU(2)_{R}\) R-symmetry.  As explained in section \ref{sec:tqft}, when tracing over BPS states we should study the diagonal subgroup \(SU(2)_{r}' = \textrm{diag}(SU(2)_{r}\times SU(2)_{R})\).  It is necessary to use this diagonal subgroup \(SU(2)_{r}'\) rather than \(SU(2)_{r}\) in order to guarantee that the Gopakumar-Vafa invariants will be invariant under complex structure deformations.

If we denote these spins as \(J_{l}\) and \(J^{'}_{r}\), then the refined free energy (defined by \(F = -\log Z\)) takes the general form,

\begin{equation}
F_{\textrm{ref top}}(X;q,t) = \sum_{d=1}^{\infty}\sum_{\beta, J_{l},J_{r}'}\frac{N^{\beta}_{J_{l},J_{r}'}\Big((qt)^{dJ_{l}}+\cdots+(qt)^{-dJ_{l}}\Big)\Big((q/t)^{dJ_{r}'}+\cdots + (q/t)^{-dJ_{r}'}\Big)Q^{d \beta}}{d(q^{d/2}-q^{-d/2})(t^{d/2}-t^{-d/2})}
\end{equation}
where the \(N^{\beta}_{J_{l},J_{r}'}(X)\) are the Gopakumar-Vafa (GV) invariants associated to the Calabi-Yau.  Since these invariants count M2-branes, they are expected to be integers on general grounds.

One nontrivial check of our proposed TQFT is the integrality of the corresponding GV invariants.  We have found that the invariants arising from the TQFT are always integral as expected, but that they do not sit in full representations of \(SU(2)_{l} \times SU(2)_{r}' \).  Instead, they generally sit in representations of \(SU(2)_{l} \times U(1)_{r}'\).  This is generally true for any Calabi-Yau manifold that engineers a field theory with a nonzero five-dimensional Chern-Simons level, as has been observed previously in \cite{Awata:2008ed, Chuang:2012dv}.  It would be interesting to understand this breaking more precisely from a spacetime perspective.

Since the symmetry group has been reduced to \(SU(2)_{l} \times U(1)_{r}'\), it is more convenient to write the free energy as,
\begin{equation}
F_{\textrm{ref top}}(X;q,t) = \sum_{d=1}^{\infty}\sum_{\beta, J_{l},m_{r}'}\frac{N^{\beta}_{J_{l},m_{r}}\Big((qt)^{dJ_{l}}+\cdots+(qt)^{-dJ_{l}}\Big) (q/t)^{d m_{r}'} Q^{d \beta}}{d(q^{d/2}-q^{-d/2})(t^{d/2}-t^{-d/2})}
\end{equation}
where \((J_{l},m_{r}')\) labels an irreducible representation of \(SU(2)_{l} \times U(1)_{r}'\).  Some examples of these invariants are listed below,

\begin{table}[htp]
\centering
\begin{tabular}[c]{| c | | c |}
\hline
$\beta$ & $\sum_{J_{l}, m_{r}'}N^{\beta}_{(j_{l},m_{r}')}(j_{l},m_{r}')$ \\
\hline
& \\
1 & $-(\frac{1}{2},1)+(0,1)+(0,0)$ \\
& \\
\hline
& \\
2 & $-(1,\frac{3}{2})+(\frac{1}{2},2)+2(\frac{1}{2},1)-2(0,\frac{3}{2}) -2(0,\frac{1}{2})$ \\
& \\
\hline
& \\
3 & $-(2,3)+(\frac{3}{2},\frac{7}{2})+(\frac{3}{2},\frac{5}{2})+(1,2)-(\frac{1}{2},\frac{7}{2})-3(\frac{1}{2},\frac{5}{2})-(\frac{1}{2},\frac{3}{2})+2(0,3)+2(0,2)$ \\
& \\
\hline
\end{tabular}
\caption{Genus 1 Gopakumar-Vafa Invariants for the geometry $\mathcal{O}(1)\oplus \mathcal{O}(-1) \to T^{2}$.}
\end{table}

\begin{table}[htp]
\centering
\begin{tabular}[c]{| c | | c |}
\hline
$\beta$ & $\sum_{J_{l}, m_{r}'}N^{\beta}_{(j_{l},m_{r}')}(j_{l},m_{r}')$ \\

\hline
& \\
1 & $-(1,1)+2(\frac{1}{2},\frac{3}{2})+2(\frac{1}{2},\frac{1}{2})-(0,2)-3(0,1)-(0,0) $ \\
& \\
\hline
& \\
2 & $-(3,\frac{7}{2})+2(\frac{5}{2},4)+2(\frac{5}{2},3)-(2,\frac{9}{2})-2(2,\frac{7}{2})-2(\frac{3}{2},4)-4(\frac{3}{2},3)-(\frac{3}{2},2)$ \\
& $+2(1,\frac{9}{2})+7(1,\frac{7}{2})+(1,\frac{5}{2})-(1,\frac{3}{2})-2(\frac{1}{2},4)+3(\frac{1}{2},3)+8(\frac{1}{2},2)+(\frac{1}{2},1)$ \\
& $-(0,\frac{9}{2})-4(0,\frac{7}{2})-11(0,\frac{5}{2})-5(0,\frac{3}{2})$ \\
& \\
\hline
& \\
3 & $-(\frac{13}{2},\frac{15}{2})+2(6,8)+2(6,7)-(\frac{11}{2},\frac{17}{2})-2(\frac{11}{2},\frac{15}{2})-2(5,8)-3(5,7)-(5,6)$ \\
& $+2(\frac{9}{2},\frac{17}{2})+5(\frac{9}{2},\frac{15}{2})-(\frac{9}{2},\frac{11}{2})-(4,8)+3(4,7)+4(4,6)$ \\
& $-(\frac{7}{2},\frac{17}{2})-2(\frac{7}{2},\frac{15}{2})-(\frac{7}{2},\frac{13}{2})+3(\frac{7}{2},\frac{11}{2})+3(\frac{7}{2},\frac{9}{2})  - 5(3,7)-7(3,6)-7(3,5)-4(3,4)$ \\
& $ +2(\frac{5}{2},\frac{15}{2})+3(\frac{5}{2},\frac{13}{2})+3(\frac{5}{2},\frac{11}{2})+8(\frac{5}{2},\frac{9}{2})+2(\frac{5}{2},\frac{7}{2})$ \\
& $ +(2,8)+3(2,7)+3(2,6)-6(2,5)-3(2,4)+2(2,3)$ \\
& $ -2(\frac{3}{2},\frac{15}{2})-2(\frac{3}{2},\frac{13}{2})+5(\frac{3}{2},\frac{11}{2})-2(\frac{3}{2},\frac{9}{2})-6(\frac{3}{2},\frac{7}{2})-2(\frac{3}{2},\frac{5}{2})$ \\
& $ -5(1,6)+4(1,5)+8(1,4)-3(1,3)-(1,2) $ \\
& $ +3(\frac{1}{2},\frac{13}{2})+2(\frac{1}{2},\frac{11}{2})-3(\frac{1}{2},\frac{9}{2})+19(\frac{1}{2},\frac{7}{2})+13(\frac{1}{2},\frac{5}{2})+(\frac{1}{2},\frac{3}{2}) $ \\
& $ -(0,7)-4(0,6)-3(0,5)-18(0,4)-24(0,3)-6(0,2) $ \\
& \\
\hline
\end{tabular}
\caption{Genus 2 Gopakumar-Vafa Invariants for the geometry $O(3)\oplus O(-1) \to \Sigma_{g=2}$.}
\end{table}

\section{Factorization of the $(q,t)$-Dimension and Metric \label{sec:qtfac}}
In this appendix we derive formulas for the \((q,t)\)-dimension and metric of a composite representation.  As a warm-up exercise, it is helpful to recall how this works for the quantum dimension of an \(SU(N)\) representation, \(R\),
\begin{equation}
\textrm{dim}_{q}(R) = \prod_{1\leq i<j\leq N}\frac{\lbrack R_{i}-R_{j}+j-i\rbrack}{\lbrack j-i \rbrack}
\end{equation}
where we are using the definition, \(\lbrack n \rbrack = q^{n/2}-q^{-n/2}\).  For the composite representation coming from \(R\) and \(S\), we obtain,
\begin{equation}
\textrm{dim}_{q}(R\overline{S}) = \textrm{dim}_{q}(R)\textrm{dim}_{q}(S)\prod_{i=1}^{c_{R}}\prod_{j=1}^{c_{S}}\frac{\lbrack S_{j}+R_{i}+N+1-j-i \rbrack\lbrack N+1-i-j\rbrack}{\lbrack S_{j}+N+1-i-j\rbrack\lbrack R_{i}+N+1-j-i \rbrack}
\end{equation}
where \(c_{R}\) and \(c_{S}\) are the number of rows in \(R\) and \(S\).  

Now we can perform the same calculation for the \((q,t)\)-dimension.  For computational convenience, we will specialize to \(t=q^{\beta}\) where \(\beta \in \mathbb{Z}_{>0}\), but our final results will be valid for any \(q\) and \(t\).

Recall that the \((q,t)\)-dimension is given by the expression,
\begin{equation}
\textrm{dim}_{(q,t)}(R) = \prod_{m=0}^{\beta-1}\prod_{1\leq i<j\leq N}\frac{\lbrack R_{i}-R_{j}+\beta(j-i)+m \rbrack}{\lbrack \beta(j-i)+m\rbrack}
\end{equation}
A short calculation shows that we obtain the same type of splitting,
\begin{eqnarray}
\textrm{dim}_{(q,t)}(R\overline{S}) & = & \textrm{dim}_{(q,t)}(R)\cdot\textrm{dim}_{(q,t)}(S)\cdot\prod_{m=0}^{\beta-1}\prod_{i=1}^{c_{R}}\prod_{j=1}^{c_{S}}\frac{\lbrack R_{i}+S_{j}+\beta(N+1-i-j)+m \rbrack }{\lbrack R_{i}+\beta(N+1-i-j)+m \rbrack} \nonumber \\
& & \cdot \frac{\lbrack \beta(N+1-i-j)+m \rbrack }{\lbrack S_{j}+\beta(N+1-i-j)+m \rbrack} \label{eqn:qtfactor}
\end{eqnarray}

Now we would like to understand the additional factors that appear in this formula, which are interpreted in section \ref{sec:factor} as arising from ghost branes.  To do so, it is helpful first to convert the \(SU(N)\) \((q,t)\)-dimension into an expression for the \(SU(\infty)\) \((q,t)\)-dimension, since it is the \(N\to \infty\) formulas that appear in the refined topological string.  We find,
\begin{eqnarray}
\textrm{dim}_{q,t}(R) & = & \prod_{m=0}^{\beta-1}\prod_{1\leq i<j\leq N}\frac{\lbrack R_{i}-R_{j}+\beta(j-i)+m\rbrack}{\lbrack \beta(j-i)+m\rbrack} \\
& = & W_{R}(q,t)T_{R}^{-1}(-1)^{\vert R \vert}(q/t)^{\vert R \vert/4}Q^{-\frac{\vert R \vert}{2}}\prod_{i=1}^{c_{R}}\prod_{j=1}^{R_{i}}(1-t^{-i+1}q^{j-1}Q) \label{eqn:wr}
\end{eqnarray}
where \(Q=t^{N}\) and we have used the definitions,
\begin{eqnarray}
W_{R}(q,t) & = & (q/t)^{\frac{\vert R \vert}{2}}M_{R}(t^{\rho};q,t) \\
T_{R} & = & q^{\frac{1}{2}\vert\vert R \vert \vert^{2}}t^{-\frac{1}{2}\vert\vert R^{T}\vert \vert^{2}}
\end{eqnarray}
where \((\rho)_{i}=-i+1/2\).  


Next, it is helpful to rewrite the factors in \ref{eqn:qtfactor} and \ref{eqn:wr} in exponential form,
\begin{equation}
\prod_{m=0}^{\beta-1}\prod_{i=1}^{c_{R}}\prod_{j=1}^{c_{S}}\Big(1-q^{R_{i}+S_{j}+\beta(N+1-i-j)+m}\Big) = \exp\Big(-\sum_{n=1}^{\infty}\frac{1}{n}Q^{n}g_{1}(q^{n},t^{n})\Big)
\end{equation}
where \(g_{1}(q,t) = \sum_{m=0}^{\beta-1}\sum_{i=1}^{c_{R}}\sum_{j=1}^{c_{S}}q^{R_{i}+S_{j}+\beta(1-i-j)+m}\).  Similarly,
\begin{equation}
\prod_{m=0}^{\beta-1}\prod_{i=1}^{c_{R}}\prod_{j=1}^{c_{S}}\Big(1-q^{R_{i}+\beta(N+1-i-j)+m}\Big) = \exp\Big(-\sum_{n=1}^{\infty}\frac{1}{n}Q^{n}g_{2}(q^{n},t^{n})\Big)
\end{equation}
where \(g_{2}(q,t) = \sum_{m=0}^{\beta-1}\sum_{i=1}^{c_{R}}\sum_{j=1}^{c_{S}}q^{R_{i}+\beta(1-i-j)+m}\).  
\begin{equation}
\prod_{m=0}^{\beta-1}\prod_{i=1}^{c_{R}}\prod_{j=1}^{c_{S}}\Big(1-q^{S_{j}+\beta(N+1-i-j)+m}\Big) = \exp\Big(-\sum_{n=1}^{\infty}\frac{1}{n}Q^{n}g_{3}(q^{n},t^{n})\Big)
\end{equation}
where \(g_{3}(q,t) = \sum_{m=0}^{\beta-1}\sum_{i=1}^{c_{R}}\sum_{j=1}^{c_{S}}q^{S_{j}+\beta(1-i-j)+m}\).  
\begin{equation}
\prod_{m=0}^{\beta-1}\prod_{i=1}^{c_{R}}\prod_{j=1}^{c_{S}}\Big(1-q^{\beta(N+1-i-j)+m}\Big) = \exp\Big(-\sum_{n=1}^{\infty}\frac{1}{n}Q^{n}g_{4}(q^{n},t^{n})\Big)
\end{equation}
where \(g_{4}(q,t) = \sum_{m=0}^{\beta-1}\sum_{i=1}^{c_{R}}\sum_{j=1}^{c_{S}}q^{\beta(1-i-j)+m}\).  
\begin{equation}
\prod_{i=1}^{c_{R}}\prod_{j=1}^{R_{i}}\Big(1-t^{-i+1}q^{j-1}Q\Big) = \exp\Big(-\sum_{n=1}^{\infty}\frac{1}{n}Q^{n}f_{R}(q^{n},t^{n})\Big)
\end{equation}
where \(f_{R}(q,t) = \sum_{i=1}^{c_{R}}\sum_{j=1}^{R_{i}}q^{\beta(-i+1)+j-1}\).  Putting all of these results together, we can rewrite the composite \((q,t)\)-dimension as,
\begin{eqnarray}
\textrm{dim}_{q,t}(R\overline{S}) & = & W_{R}(q,t)W_{S}(q,t) T_{R}^{-1}T_{S}^{-1} (-1)^{\vert R\vert + \vert S \vert}(q/t)^{\frac{1}{4}(\vert R \vert + \vert S \vert)} \nonumber \\
& & \cdot Q^{-\frac{\vert R\vert +\vert S \vert}{2}}\exp\Bigg(-\sum_{n=1}^{\infty}\frac{1}{n}Q^{n}M_{RS}(q^{n},t^{n})\Bigg)
\end{eqnarray}
where we have defined 
\begin{equation}
M_{RS}(q,t) = g_{1}(q,t)+g_{4}(q,t)-g_{2}(q,t)-g_{3}(q,t)+f_{R}(q,t)+f_{S}(q,t) \label{eqn:mrs}
\end{equation}
Next, we would like to simplify \(M_{RS}\).  To do so, it is helpful to notice that \(f_{R}\) can be rewritten as,
\begin{equation}
f_{R}(q,t) = \frac{t}{q-1}\sum_{i=1}^{c_{R}}\Big(q^{R_{i}-\beta i}-q^{-\beta i}\Big)
\end{equation}
Then a short calculation reveals that \(M_{RS}\) can be rewritten as,
\begin{equation}
M_{RS}(q,t) = \frac{(1-q)(1-t)}{t}f_{R}(q,t)f_{S}(q,t)+f_{R}(q,t)+f_{S}(q,t)
\end{equation}

We would like to use this rewriting to make contact with the refined topological vertex amplitudes, \(W_{RQ}\) and \(\widetilde{W}_{RQ}\).  As discussed in section \ref{sec:tqft}, these amplitudes can be computed from the large \(N\) limit of the refined Chern-Simons S-matrix and are equal to,
\begin{eqnarray}
W_{RQ} & = & (q/t)^{\frac{\vert R \vert + \vert Q \vert}{2}}M_{R}(t^{\rho};q,t)M_{Q}(t^{\rho}q^{R};q,t) \\
\widetilde{W}_{RQ} & = & (q/t)^{\frac{\vert R \vert + \vert Q \vert}{2}}M_{R}(t^{\rho};q,t)\iota M_{Q}(t^{\rho}q^{R};q,t)
\end{eqnarray}
We define the following refined quantity,
\begin{eqnarray}
K_{RS}(Q,q,t) & := & \sum_{P}\frac{1}{g_{P}}Q^{\vert P\vert}(t/q)^{\vert P \vert} W_{PR}(q,t)W_{PS}(q,t) \\
& = & W_{R}(q,t)W_{S}(q,t)\sum_{P}\frac{1}{g_{P}}Q^{\vert P\vert} M_{P}(t^{\rho}q^{R};q,t)M_{P}(t^{\rho}q^{S};q,t) \nonumber \\
& = & W_{R}(q,t)W_{S}(q,t)\exp\Bigg(\sum_{n=1}^{\infty}\frac{1}{n}\frac{1-t^{n}}{1-q^{n}}Q^{n} p_{n}(x)p_{n}(y)\Bigg)
\end{eqnarray}
where in the last line we have used the generalized Cauchy identity of \ref{eqn:maccauchy} and where \(x_{i}=q^{R_{i}}t^{-i+1/2}\) and \(y_{j} = q^{S_{j}}t^{-j+1/2}\).  A straightforward evaluation shows that,
\begin{eqnarray}
K_{RS}(q,t) & = & K_{\cdot\cdot}(q,t)W_{R}(q,t)W_{S}(q,t)\exp\Bigg(\sum_{n=1}^{\infty}\frac{1}{n}Q^{n}M_{RS}(q^{n},t^{n})\Bigg) 
\end{eqnarray}
This means that we can rewrite the composite \((q,t)\)-dimension as,
\begin{equation}
\textrm{dim}_{q,t}(R\overline{S}) = T_{R}^{-1}T_{S}^{-1}(-1)^{\vert R \vert + \vert S \vert}Q^{-\frac{\vert R \vert + \vert S \vert}{2}}(q/t)^{\frac{1}{4}(\vert R \vert + \vert S \vert)} \frac{K_{\cdot\cdot}(Q,q,t)W_{R}(q,t)^{2}W_{S}(q,t)^{2}}{K_{RS}(Q,q,t)}
\end{equation}
In section \ref{sec:factor}, we use this result to understand the genus \(g>1\) geometries.

It will also be useful to rewrite the \((q,t)\)-dimension in another way.  Define the refined quantity,
\begin{eqnarray}
N_{RS}(Q,q,t) & := & \sum_{P}\frac{1}{g_{P}}Q^{\vert P\vert}(t/q)^{\vert P\vert}\widetilde{W}_{RP}(q,t)W_{PS}(q,t) \\
& = & W_{R}(q,t)W_{S}(q,t)\sum_{P}\frac{1}{g_{P}}Q^{\vert P\vert}\iota M_{P}(t^{\rho}q^{R};q,t)M_{P}(t^{\rho}q^{S};q,t) \nonumber \\
& = & W_{R}(q,t)W_{S}(q,t)\exp\Bigg(-\sum_{n=1}^{\infty}\frac{1}{n}\frac{1-t^{n}}{1-q^{n}}Q^{n} p_{n}(x)p_{n}(y)\Bigg) \nonumber
\end{eqnarray}
where as before, \(x_{i}=q^{R_{i}}t^{-i+1/2}\) and \(y_{j} = q^{S_{j}}t^{-j+1/2}\).  Then using the same analysis as above, we can rewrite this expression as,
\begin{equation}
N_{RS}(Q,q,t) = N_{\cdot\cdot}(q,t)W_{R}(q,t)W_{S}(q,t)\exp\Bigg(-\sum_{n=1}^{\infty}\frac{1}{n} Q^{n}M_{RS}(q^{n},t^{n})\Bigg) 
\end{equation}
Therefore, we can alternatively rewrite the composite \((q,t)\)-dimension as,
\begin{equation}
\textrm{dim}_{q,t}(R\overline{S}) = T_{R}^{-1}T_{S}^{-1}(-1)^{\vert R \vert + \vert S \vert}(q/t)^{\frac{1}{4}(\vert R \vert + \vert S \vert)}Q^{-\frac{\vert R \vert + \vert S \vert}{2}}\frac{N_{RS}(Q,q,t)}{N_{\cdot\cdot}(Q,q,t)}
\end{equation}
This formula is useful in section \ref{sec:factor} for studying the genus \(g=0\) geometries.  It is also helpful to notice that \(N_{\cdot\cdot}(Q,q,t) = (K_{\cdot\cdot}(Q,q,t))^{-1}\).

It is important to note here that by using the above definition of \(\textrm{dim}_{q,t}(R)\), we are rewriting the expression in equation \ref{eqn:qtym} as,
\begin{equation}
\frac{S_{0R}^{2}}{G_{R}} = \frac{S_{00}\widetilde{S}_{00}\textrm{dim}_{q,t}(R)^{2}}{g_{R}}
\end{equation}
where we used the definition from above that \(g_{R} = G_{R}/G_{0}\) and have also introduced the normalization factors, \(S_{00}\widetilde{S}_{00}\).  In the large \(N\) limit, these additional factors can be rewritten as,
\begin{eqnarray}
q^{2\beta^{2}\rho^{2}} S_{00}(q,t)\widetilde{S}_{00}(q,t) & := & q^{2\beta^{2}\rho^{2}} \prod_{m=0}^{\beta-1}\prod_{1\leq i<j\leq N}\lbrack \beta(j-i)+m \rbrack \lbrack \beta(j-i)-m \rbrack \\
& = & \Big(K_{\cdot\cdot}(Q)K_{\cdot\cdot}(Q\frac{q}{t})\Big)^{-1} \prod_{k=0}^{\infty}\Big(1-tq^{k}\Big)^{N}\Big(1-q^{k+1}\Big)^{N} \\
& & \cdot \prod_{j,k=1}^{\infty}\Big(1-t^{k}q^{j-1}\Big)^{-1}\Big(1-q^{k}t^{j-1}\Big)^{-1} \nonumber \\
& = & \Big(K_{\cdot\cdot}(Q)K_{\cdot\cdot}(Q\frac{q}{t})\Big)^{-1}  \Big((t;q)_{\infty}(q;q)_{\infty} \Big)^{N} M(q,t)M(t,q)
\end{eqnarray}
where we have written the expression in terms of the refined MacMahon function, 
\begin{equation}
M(q,t) = \prod_{j,k=1}^{\infty}\Big(1-t^{k}q^{j-1}\Big)^{-1}
\end{equation} 
that appears in the refined topological vertex \cite{Iqbal:2007ii}.

Now we move on to studying the large \(N\) factorization of the metric, \(g_{R}\).  Recall that the metric is given by the formula,
\begin{equation}
g_{R}^{(N)} = \prod_{m=0}^{\beta-1}\prod_{1\leq i<j\leq N}\frac{\lbrack R_{i}-R_{j}+\beta(j-i)+m \rbrack}{\lbrack R_{i}-R_{j}+\beta(j-i)-m \rbrack}\frac{\lbrack \beta(j-i)-m \rbrack}{\lbrack \beta(j-i)+m \rbrack}
\end{equation}
As a first step, we must express this in terms of the \(N\to\infty\) metric,
\begin{eqnarray}
g_{R}^{(N)} & = & (q/t)^{\frac{\vert R \vert}{2}}g_{R}^{(\infty)} \prod_{m=0}^{\beta-1}\prod_{i=1}^{N}\prod_{j=1}^{\infty}\frac{\lbrack R_{i}+\beta(N+j-i)-m \rbrack}{\lbrack R_{i}+\beta(N+j-i)+m \rbrack}\frac{\lbrack \beta(N+j-i)+m \rbrack}{\lbrack \beta(N+j-i)-m \rbrack} \label{eqn:metriclargen}
\end{eqnarray}
where \(g^{(\infty)}_{R}\) is the symmetrized metric defined by \ref{eqn:metricsym}.  
We also find that the metric for the composite representation \(S\overline{R}\) is given by,
\begin{eqnarray}
g_{S\overline{R}}^{(N)}& = & g_{R}^{(N)}\cdot g_{S}^{(N)} \cdot \prod_{m=0}^{\beta-1}\prod_{i=1}^{c_{S}}\prod_{j=1}^{c_{R}}\frac{\lbrack S_{i}+R_{j}+\beta(N+1-j-i)+m \rbrack}{\lbrack S_{i}+R_{j}+\beta(N+1-j-i)-m \rbrack}\frac{\lbrack S_{i}+\beta(N+1-i-j)-m \rbrack}{\lbrack S_{i}+\beta(N+1-i-j)+m \rbrack} \nonumber \\
& & \cdot  \frac{\lbrack R_{j}+\beta(N+1-i-j)-m \rbrack}{\lbrack R_{j}+\beta(N+1-i-j)+m \rbrack}\frac{\lbrack \beta(N+1-j-i)+m \rbrack}{\lbrack \beta(N+1-j-i)-m \rbrack} \label{eqn:metricfac}
\end{eqnarray}
Now as with the analysis of the \((q,t)\)-dimension, we want to write all of the factors in \ref{eqn:metriclargen} and \ref{eqn:metricfac} in terms of exponentials,
\begin{equation}
\prod_{m=0}^{\beta-1}\prod_{i=1}^{c_{R}}\prod_{j=1}^{c_{S}}\frac{1-q^{R_{i}+S_{j}+\beta(N+1-i-j)+m}}{1-q^{R_{i}+S_{j}+\beta(N+1-i-j)-m}} = \exp\Big(-\sum_{n=1}^{\infty}\frac{1}{n}Q^{n}h_{1}(q^{n},t^{n})\Big)
\end{equation}
where \(h_{1}(q,t) = \sum_{m=0}^{\beta-1}(q^{m}-q^{-m})\sum_{i=1}^{c_{R}}\sum_{j=1}^{c_{S}}q^{R_{i}+S_{j}+\beta(1-i-j)}\). 
\begin{equation}
\prod_{m=0}^{\beta-1}\prod_{i=1}^{c_{R}}\prod_{j=1}^{c_{S}}\frac{1-q^{S_{j}+\beta(N+1-i-j)+m}}{1-q^{S_{j}+\beta(N+1-i-j)-m}} = \exp\Big(-\sum_{n=1}^{\infty}\frac{1}{n}Q^{n}h_{2}(q^{n},t^{n})\Big)
\end{equation}
where \(h_{2}(q,t) = \sum_{m=0}^{\beta-1}(q^{m}-q^{-m})\sum_{i=1}^{c_{R}}\sum_{j=1}^{c_{S}}q^{S_{j}+\beta(1-i-j)}\). 
\begin{equation}
\prod_{m=0}^{\beta-1}\prod_{i=1}^{c_{R}}\prod_{j=1}^{c_{S}}\frac{1-q^{R_{i}+\beta(N+1-i-j)+m}}{1-q^{R_{i}+\beta(N+1-i-j)-m}} = \exp\Big(-\sum_{n=1}^{\infty}\frac{1}{n}Q^{n}h_{3}(q^{n},t^{n})\Big)
\end{equation}
where \(h_{3}(q,t) = \sum_{m=0}^{\beta-1}(q^{m}-q^{-m})\sum_{i=1}^{c_{R}}\sum_{j=1}^{c_{S}}q^{R_{i}+\beta(1-i-j)}\). 
\begin{equation}
\prod_{m=0}^{\beta-1}\prod_{i=1}^{c_{R}}\prod_{j=1}^{c_{S}}\frac{1-q^{\beta(N+1-i-j)+m}}{1-q^{\beta(N+1-i-j)-m}} = \exp\Big(-\sum_{n=1}^{\infty}\frac{1}{n}Q^{n}h_{4}(q^{n},t^{n})\Big)
\end{equation}
where \(h_{4}(q,t) = \sum_{m=0}^{\beta-1}(q^{m}-q^{-m})\sum_{i=1}^{c_{R}}\sum_{j=1}^{c_{S}}q^{\beta(1-i-j)}\). 
\begin{equation}
\prod_{m=0}^{\beta-1}\prod_{i=1}^{N}\prod_{j=1}^{\infty}\frac{1-q^{R_{i}+\beta(N+j-i)-m}}{1-q^{R_{i}+\beta(N+j-i)+m}}\frac{1-q^{\beta(N+j-i)+m}}{1-q^{\beta(N+j-i)-m}} = \exp\Big(-\sum_{n=1}^{\infty}\frac{1}{n}Q^{n}h_{R}(q^{n},t^{n})\Big)
\end{equation}
where 
\begin{eqnarray}
h_{R}(q,t) & = & \sum_{m=0}^{\beta-1}(q^{-m}-q^{m})\sum_{i=1}^{c_{R}}\sum_{j=1}^{\infty}\Big(q^{R_{i}+\beta j - \beta i}-q^{\beta j -\beta i}\Big) \\
& = & \frac{q-t}{1-q}\sum_{i=1}^{N}\Big(q^{R_{i}- \beta i}-q^{-\beta i}\Big) \\
& = & \Big(1-\frac{q}{t}\Big)f_{R}(q,t)
\end{eqnarray}
Putting this all together we can write the composite metric as,
\begin{equation}
g^{(N)}_{R\overline{S}} = (q/t)^{\frac{\vert R \vert + \vert S \vert}{2}}g_{R}^{(\infty)}g_{S}^{(\infty)}\exp\Bigg(-\sum_{n=1}^{\infty}\frac{1}{n}Q^{n}H_{RS}(q^{n},t^{n})\Bigg)
\end{equation}
where \(H_{RS}\) is given by,
\begin{equation}
H_{RS} = h_{1}(q,t)+h_{4}(q,t)-h_{2}(q,t)-h_{3}(q,t)+\Big(1-\frac{q}{t}\Big)\Big(f_{R}(q,t)+f_{S}(q,t)\Big)
\end{equation}
A little algebra shows,
\begin{eqnarray}
H_{RS} & = & \Big(1-\frac{q}{t}\Big)\frac{(1-q)(1-t)}{t}f_{R}f_{S}+\Big(1-\frac{q}{t}\Big)\Big(f_{R}+f_{S}\Big) \\
& = & \Big(1-\frac{q}{t}\Big)M_{RS}(q,t)
\end{eqnarray}
where \(M_{RS}\) is the quantity that we defined in equation \ref{eqn:mrs} while studying the \((q,t)\)-dimension.  Therefore,
\begin{equation}
g_{R\overline{S}}^{(N)}= (q/t)^{\frac{\vert R \vert + \vert S \vert}{2}}g_{R}^{(\infty)}g_{S}^{(\infty)}\exp\Bigg(-\sum_{n=1}^{\infty}\frac{1}{n}Q^{n}M_{RS}(q^{n},t^{n})+\sum_{n=1}^{\infty}\frac{1}{n}Q^{n}\frac{q^{n}}{t^{n}}M_{RS}(q^{n},t^{n})\Bigg)
\end{equation}
Using the results from above, this means that we can write the composite metric in two ways,
\begin{eqnarray}
g_{R\overline{S}}^{(N)} & = & (q/t)^{\frac{\vert R \vert + \vert S \vert}{2}} g_{R}^{(\infty)}g_{S}^{(\infty)}\frac{K_{RS}(Q\frac{q}{t},q,t)}{K_{\cdot\cdot}(Q\frac{q}{t},q,t)}\frac{K_{\cdot\cdot}(Q,q,t)}{K_{RS}(Q,q,t)} \\
g_{R\overline{S}}^{(N)} & = & (q/t)^{\frac{\vert R \vert + \vert S \vert}{2}} g_{R}^{(\infty)}g_{S}^{(\infty)}\frac{N_{\cdot\cdot}(Q\frac{q}{t},q,t)}{N_{RS}(Q\frac{q}{t},q,t)}\frac{N_{RS}(Q,q,t)}{N_{\cdot\cdot}(Q,q,t)}
\end{eqnarray}

\section{Refined S-Duality \label{sec:sduality}}

From string theory we can study the action of TST duality on the D4/D2/D0 system.  Our goal is to understand how the chemical potentials in the index transform under this duality.  If we simply wrapped our \(D4\)-branes on a euclidean time circle of circumference \(\lambda\), we would be computing the trace,
\begin{equation}
\textrm{Tr}(-1)^{F}e^{-\lambda H}
\end{equation}
The contribution of a D4/D2/D0-brane bound state to this index is given by, \(e^{-\lambda M}\), where \(M\) is the mass of the bound state and is equal to the magnitude of the central charge, \(M = \vert Z \vert\).

We would like to identify the \(D0\) and \(D2\) chemical potentials with the corresponding \(D0\) and \(D2\)-brane masses.  However, for a generic choice of the K\"ahler parameter, \(k=B+iJ\), we are studying genuine D4/D2/D0 bound states whose mass is not equal to the sum of the constituent masses.  This can be fixed by studying the special limit when \(J=0\), so that the D2-branes get all of their mass from a background B-field.\footnote{A similar limit was used to derive wall-crossing formulas from M-theory in \cite{Aganagic:2009kf, Aganagic:2009cg}.}  In this limit the central charges of the D0 and D2-branes align, which means that they form marginal bound states.  In this case, the total mass of the bound state is simply equal to the sum of the D2/D0 masses.\footnote{More precisely, the central charge for a bound state of \(N\) D4-branes, \(m\) D2-branes, and \(n\) D0-branes is given by \(Z = N\Lambda^{2}e^{2i\phi}+mk+n\) (see section \ref{sec:wc} for details).  Then using the fact that \(\Lambda \gg 1\), the mass is given by, \(M = \vert Z \vert \approx \frac{1}{2}N\Lambda^{2} - (mk+n)\cos(2\phi)\).}  Therefore, we can identify the D-brane masses with the chemical potentials,\footnote{Throughout this section we set \(2\pi \sqrt{\alpha'} = 1\).}
\begin{equation}
\phi_{0} \leftrightarrow \frac{2\pi \lambda}{g_{s}}, \qquad \phi_{2} \leftrightarrow \frac{2\pi \lambda k}{g_{s}}
\end{equation}
To determine their transformation properties under TST duality, it is helpful first to recall how the coupling, \(g_{s}\), transforms,
\begin{eqnarray}
g_{s} \stackrel{T}{\longrightarrow} \frac{g_{s}}{\lambda} & \stackrel{S}{\longrightarrow} & \frac{\lambda}{g_{s}}  \stackrel{T}{\longrightarrow}\frac{\lambda^{2}}{g_{s}}
\end{eqnarray}
Therefore,
\begin{equation}
\frac{2\pi \lambda}{g_{s}} \stackrel{TST}{\longrightarrow} 4\pi^{2} \Big(\frac{g_{s}}{2\pi \lambda}\Big)
\end{equation}
which implies that the D0-brane chemical potential transforms as,
\begin{equation}
\phi_{0} \to \frac{4\pi^{2}}{\phi_{0}}
\end{equation}
To determine the transformation of the \(D2\)-brane charge, we must simply follow the background B-field, \(k\), which transforms as,
\begin{eqnarray}
(B_{ij}=k) \stackrel{T}{\longrightarrow} B_{ij} & \stackrel{S}{\longrightarrow} & C_{ij} \stackrel{T}{\longrightarrow} \Big(C_{ijt} =  \frac{k}{ \lambda}\Big)
\end{eqnarray}
Therefore, the \(TST\) duality converts the background B-field into a background RR three-form which couples to the D2-brane as,
\begin{equation}
2\pi i \int C_{(3)} = 2\pi i \frac{k}{ \lambda} \int dt = 2\pi i k
\end{equation}
Therefore, we have the transformation property,
\begin{equation}
\frac{2\pi \lambda k}{g_{s}} \stackrel{TST}{\longrightarrow} 2\pi i k
\end{equation}
which implies that the D2-brane chemical potential transforms as,
\begin{equation}
\phi_{2} \to 2\pi i \frac{\phi_{2}}{\phi_{0}}
\end{equation}

Now that we have explained how the unrefined potentials transform, we want to study the transformation properties of the spin character chemical potential, \(\gamma\).  First, recall that the inclusion of \(e^{-\lambda \gamma J_{3}}\) in the trace is equivalent to considering the geometry,
\begin{equation}
(\mathbb{R}^{2} \times S^{1})_{\gamma} \times \cdots
\end{equation}
where we rotate the \(\mathbb{R}^{2}\) plane by \(\gamma\) as we go around the thermal circle.  The metric for this geometry can be written explicitly as,
\begin{equation}
ds^{2} = \frac{\lambda^{2}}{4\pi^{2}}dt^{2} + (dx^{i}+\Omega^{i}dt)(dx^{i}+\Omega^{i}dt)
\end{equation}
where \(\Omega\) is given by \(\Omega = \gamma r^{2}d\theta\) and \((r,\theta)\) are the coordinates on the \(\mathbb{R}^{2}\) plane.  

Applying T-duality, this metric is converted to first order into a background B-field,
\begin{equation}
B_{\theta t} = \alpha' \frac{g_{\theta t}}{g_{tt}} = \frac{\gamma r^{2}}{\lambda^{2}}
\end{equation}
Applying S-duality converts this into a RR two-form, \(C_{\theta t}\), and after the final T-duality we are left with a RR one-form,
\begin{equation}
C_{\theta} = \frac{\gamma r^{2}}{\lambda}
\end{equation}
Now it is important to remember that the RR one-form couples to D0-branes as,
\begin{equation}
2\pi i \int C_{(1)} = 2\pi i \int \frac{\gamma r^{2}}{\lambda}d\theta = \frac{2\pi i}{\lambda} \int \gamma r^{2}\frac{d\theta}{dt}dt
\end{equation}
Therefore, we can rewrite this coupling as,
\begin{equation}
\frac{2\pi i \gamma}{\lambda m_{\textrm{D0}}} \int J_{3} dt = \frac{g_{s} i}{\lambda} \gamma \int J_{3} dt = \frac{g_{s} i}{\lambda} \gamma \lambda J_{3} 
\end{equation}
This implies that under \(TST\) duality, \(\gamma\) transforms as \(\gamma \to \gamma \frac{g_{s}}{\lambda} \).  In terms of chemical potentials, this implies,
\begin{equation}
\gamma \to 2\pi i \frac{\gamma}{\phi_{0}}
\end{equation}  
It might seem that after the TST duality, \(\gamma\) only couples to the angular momentum of D0-branes.  However, the RR one-form also couples to D2-branes in the presence of a background B-field, and the same arguments go through,
\begin{equation}
2\pi i \int B_{(2)}\wedge C_{(1)} = 2\pi i k \int \frac{\gamma r^{2}}{\lambda}d\theta = \frac{2\pi i k\gamma}{\lambda m_{\textrm{D2}}} \int J_{3} dt = \frac{g_{s} i}{\lambda} \gamma \int J_{3} dt
\end{equation}
%

%

%

So far, we have only studied the transformation properties of the \(J_{3}\) chemical potential, and have not studied the full combination \((J_{3}-R)\).  However, the \(R\)-symmetry can be realized as a geometric rotation in the Calabi-Yau, so the same arguments apply.  Further, by supersymmetry we know that the chemical potential for \(R\) must transform in precisely the same way as the chemical potential for \(J_{3}\).

Combining all of these results together we find the potentials transform under TST duality as,
\begin{equation}
\phi_{0} \to \frac{4\pi^{2}}{\phi_{0}}, \qquad \phi_{2} \to 2\pi i \frac{\phi_{2}}{\phi_{0}}, \qquad \gamma \to 2\pi i \frac{\gamma}{\phi_{0}}
\end{equation}
This general result agrees nicely with the explicit transformation properties found in section \ref{sec:o1}.

\bibliographystyle{utphys}
\bibliography{ROSV}{}

\end{document}